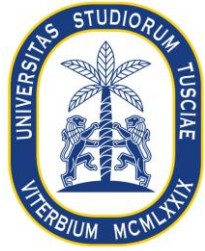

**Dipartimento di Economia, Ingegneria, Società e Impresa (DEIM)**

**Corso di Dottorato di Ricerca in**

**Economia, Management e Metodi Quantitativi (EMMQ)**

**Curriculum: Economia Agro-Alimentare - XXXIII Ciclo**

**TESI DI DOTTORATO DI RICERCA**

**The role of Common Agricultural Policy (CAP) in enhancing and stabilising farm income: an analysis of income transfer efficiency and the Income Stabilisation Tool**

AGR/01

**Tesi di dottorato di:**
Dott. Luigi Biagini
Firma _______________________

| | |
|---|---|
| **Coordinatore del corso** | **Tutore** |
| Prof. Alessandro Sorrentino | Prof. Simone Severini |
| Firmer ____________________ | Firma ____________________ |

A.A. 2019/2020

<div style="text-align: right"><em>To my family</em></div>

*"The only thing greater than the power of the mind is the courage of the heart."*
<div style="text-align: right"><em>John Nash</em></div>

*"It is not knowledge, but the act of learning, not possession but the act of getting there, which grants the greatest enjoyment.*
*When I have clarified and exhausted a subject, then I turn away from it, in order to go into darkness again; the never satisfied man is so strange if he has completed a structure, then it is not in order to dwell in it peacefully, but in order to begin another."*
<div style="text-align: right"><em>Carl Friedrich Gauss</em></div>

# CONTENTS

# Abstract

Since its inception, the E.U.'s Common Agricultural Policy (CAP) aimed at ensuring an adequate and stable farm income. While recognizing that the CAP pursues a larger set of objectives, this thesis focuses on the impact of the CAP on the level and the stability of farm income in Italian farms. It uses microdata from a high standardized dataset, the Farm Accountancy Data Network (FADN), that is available in all E.U. countries. This allows, if perceived as useful, to replicate the analyses to other countries.

The thesis first assesses the Income Transfer Efficiency (i.e., how much of the support translate to farm income) of several CAP measures. Secondly, it analyses the role of a specific and relatively new CAP measure (i.e., the Income Stabilisation Tool - IST) that is specifically aimed at stabilising farm income.

The first issue is investigated considering the dynamic dimension of income: the previous year has an undoubted impact on the current economic result. This aspect is a crucial point in this analysis because, differently from other studies, it is hypothesized that the present choice is fundamental sticky. The tools used for static modelling cannot be used under these circumstances because they remain affected by endogeneity, simultaneity bias, and omitted variables. To overcome these issues, the Generalised Method of Moments (GMM) is used together with a system of Instrumental Variables (IV) taken in level and first difference (*à la* Blundell-Bond).

The outcomes are in line with previous studies and economic expectations. Decoupled direct payments provide the highest contribution to agricultural incomes, followed by agri-environmental payments and on-farm investment subsidies. Coupled payments, indeed, have no significant impacts on farmers' income. For the first time, we found the efficiency of CAP measures for three different economic dimensions of farms, to the best of our knowledge. We have obtained that large farms benefit from greater transfer efficiency than medium and small farms. These differences among instruments and across farms suggest that policy-participation costs may play a pivotal role, together with farms' economic structure, in determining the income transfer efficiency of CAP policies.

The second part of this thesis investigates the potential impact of the IST, currently introduced in the European CAP, to reduce farmers' income risk. The analysis is motivated by the fact that this tool, while potentially useful, is not implemented yet apart few cases. One of the reasons is that, as is the case of all new insurance schemes, it is not easy to define the structure of the premiums (i.e., to develop a correct ratemaking). The main objectives are to assess the potential effects of introducing this tool on the Italian farms and assess whether Machine Learning procedures can develop adequate ratemaking.

The first objective has been pursued by developing a simulation based on a FADN panel data set of 3421 farms over seven years. This allowed investigating the effects of the introduction of the IST on (a) farm-level income variability, (ii) the expected level and variability of indemnifications at the level of mutual funds and (iii) the distribution of net benefits from this policy instrument across the farm population. We find that IST's introduction would lead to a significant reduction of income variability in Italian agriculture. Our results support establishing a national M.F. due to the high volatility of indemnification levels at more disaggregated (e.g. regional or sectoral) levels. Besides, our results propose that farmers' contribution to mutual funds, i.e. premiums paid, should be modulated according to farm size. This reduces the inequality of the distribution of benefits of such tool within the farm population.

The assessment of the potential use of Machine Learning procedures to develop an adequate ratemaking is based on the following. These are used to predict indemnity levels because this is an essential point for a similar insurance scheme. The assessment of ratemaking is challenging: indemnity distribution is zero-inflated, not-continuous, right-skewed, and several factors can potentially explain it. We address these problems by using Tweedie distributions and three Machine Learning procedures. The objective is to assess whether this improves the ratemaking by using the prospective application of the Income Stabilization Tool in Italy as a case study. We look at the econometric performance of the models and the impact of using their predictions in practice. Some of these procedures efficiently predict indemnities, using a limited number of regressors, and ensuring the scheme's financial stability.

This thesis fills some gaps in the analysis of "farm problem" and, in particular, by assessing the role of agricultural policies in enhancing and stabilizing farm income. While the first part of the analysis refers to past measures, the second refers to an innovative not-yet implemented instrument. Thus, the thesis uses a relatively large set of methodologies that have not been applied so far in these areas of analysis, providing preliminary results regarding their possible pros and cons. Furthermore, because the analyses rely on a widely available database, the proposed approaches could be used in future similar applications and other E.U. countries and consider different insurance tools.

# Acknowledgements

I am deeply indebted to my tutor, Prof. Simone Severini, for his advice at different writing stages of this dissertation but, first of all, for his wisdom in my defects, madness, dark moments, and my impetuousness. Like a scientific mentor, Prof. Severini supported me, especially in the despair's moments that all the researcher is destined to face during the course.

Prof. Simone Severini supervised this dissertation all along with its variegated development resulting from three years of work. The numerous and lengthy conversations I had with DAFNE, DEIM and DIBAF economic professors, particularly the ex-Dear membership, were a constant source of inspiration and broad perspectives. I am grateful to the PhD board, particularly Prof. Alessandro Sorrentino, for having impeccably managed this course even though it was the first time it was held.

I am in debt toward Dr Federico Antonioli, my office mate in the department, and Dr Cinzia Zinnanti, current and valid successor; my gratitude towards them went beyond the simple relationship and turned into a solid friendship. I am also immensely grateful to classmates of the XXXIII PhD cycle, especially with those I spent most of the time, who have collected my outbursts, supported me in difficult times and have never left me alone in some painful moments: Dr Eleonora Rapiti, Dr Giorgio Monfeli and, last but not least, Dr Chiara Grazini. My debt to them goes far beyond mere thanks for this dissertation.

I must necessarily spend more than a few words about Prof. Robert Finger (ETH - Zurich) and Prof. Nadja El Benni (Agroscope - Switzerland): to them, I am indebted for having accompanied and supported me for much of this thesis. Their help was not limited to the thesis itself but has also concerned constructive criticisms and how to think and develop research. Their skills have also been shown to make me feel at ease in a dynamic and scientific environment like ETH and Agroscope: I will never be able to thank them entirely for their hospitality.

During my three years in the PhD, I have been lucky to meet influential scholars who improve my knowledge with constructive criticisms and substantial suggestions. I remember (in order of when I met them, not of importance) Dr Alessandro Di Nola ( University of Konstanz), Dr Luciano Lavecchia ( Banca d'Italia), Prof. Martina Bozzola (Queen's University Belfast), Prof. Martin Odenin (Humboldt university Zu Berlin), Prof. Oliver Mußhoff (University of Gottingen), Prof. Mauro Vigani (University of Gloucestershire), Prof. Alberto Bisin (New York University), Dr Nadia Garbellini (Università Cattolica di Piacenza), Marco Veronese Passarella (Leeds University Business School, Leeds) Stefano Ciliberti (Università di Perugia), Angelo Frascarelli (Università di Perugia), Dr Maria Marino (Università degli Studi di Firenze), Dr Elisa Giampietri (Università degli Studi di Padova). His always available assistance was essential, and his influence is visible throughout all the essays.

I want to thank the participants in the EAAE, AIEAA and SIDEA seminars and congresses for allowing the development of all the articles included in this thesis: their advice has allowed me to make tremendous improvements and fill my weak experience in Research activities.

I am very in debt to Prof. Thomas Heckled (Rheinische Friedrich Wilhelms Universität Bonn) and Prof. Mauro Vigani (University of Gloucestershire) having acted as reviewers in this dissertation. Their advice highlights some of my shortcomings and certainly inspires me to improve

In Machine Learning, I am immensely thankful for suggestions and criticisms to Prof. Matteo M. Pelagatti (University of Milano-Bicocca), Prof. Shishir Shakya (Shippensburg University of Pennsylvania), and Dr Giovanni Millo (Generali Investments) that also help to me to develop my knowledge in statistics with R, without them, most of the results were impossible to achieve.

I am in deep debt with Prof. Stephen Bond (University of Oxford); without his explanations, some remarkable steps in my knowledge in the theory of dynamic panel data regression were not feasible.

I wish to acknowledge a particular debt I owe to my teachers in the University of Tuscia, particularly ex-DEAR, for my graduate course, who helped the formation of my ideas on economic theory and policy matters. Only the author of this dissertation knows how important their contribution is.

This dissertation is dedicated to my family that supported me: the words all always few in this case; I hope that this result partially reimburses their patience and support.

All credit is the extraordinary people I mentioned above; meanwhile, the mistake is just my responsibility.

# Chapter 1. Introduction

A considerable amount of public resources are used in the E.U. to support the farm sector (Massot 2020a). The E.U.'s Common Agricultural Policy (CAP) supports agricultural income through a wide range of measures (Matthews 2017; World Bank, 2017).

Most of the support is channelled through Pillar 1 Direct Payments (D.P.), whose primary goal is to increase farm income (Massot 2020a, 2020b).

Although the CAP objectives have been broadened compared to the past, securing income support continues to be one of its nine specific goals, strengthened in the CAP legislative proposal beyond 2020. (European Commission, 2018; Matthews, 2020). CAP has also placed a growing emphasis on decreasing the variation of farm income (Diaz-Caneja et al. 2008; Meuwissen, van Asseldonk, and Huirne 2008). Consequently, assessing the degree to which the various CAP measures affect farm income in terms of magnitude and variability/volatility remains essential to evaluate the policy's effectiveness.

This introduction reports a review of the literature on the "farm problem", followed by the definition of the topic and research questions. The third section describes the thesis's data, and the methodologies implemented to answer the research questions. Finally, the structure of the thesis is reported.

## 1.1 The "Farm problem"

The past two centuries have been characterised by epochal transition from an economy based primarily on the agricultural sector to one where it currently has a limited contribution to total GDP.

A tremendous increase in agricultural productivity initiated the economic revolution in the 19$^{th}$ century through an outflow of the workforce from the agricultural towards other sectors. In particular, according to Zellner et al. (Zellner, Kmenta, and Dreze 1966), technology has been the factor that can explain how output cannot be seen as a simple sum of inputs. The value-added increased by technology compared to the simple number of factors has represented, and represent today, the pave which economic development is relied on.

In the first moment, the manufacturing sector utilising this cheap labour has had a vibrant development.

After this initial change, the tertiary sector also incredibly increased its economic performance. This evolution has fostered significant changes in society, with an increase in well-being that has never been seen in history (Hill 2012; Tweeten and Thompson 2002).

On the other side of the coin, the massive, not regulated, and tumultuous use of technology has many problems. Jevons (Jevons 2001; York 2006) reported that some input (in the specific case the coil) is dramatically increased with tremendous consequences in the environment after the first technology revolution. Particular painful to London remember the "big-smoke" event due to excessive use of colin in industrial production and domestic warming that caused over than 4000 deaths. Undoubtedly a no-scientific evaluation of development impacts has been caused some social disorder and significant social change followed to technology revolutions that characterised the last centuries, with very dramatic consequence.

This essay only regards only a tiny part of the economic consequence of policy implication in the technological impact, particularly the agricultural economic aspects (Marino, Rocchi, and Severini 2021; Rocchi, Marino, and Severini 2020). I hope that it contributes to paving the way to highlight the importance of policy evaluation for social dynamics and environmental effects.

*1.1.1   Definition of the "farm problem."*

The farm problem refers to the economic troubles in the agricultural sector. In general, scholars left this definition deliberately ambiguous. In most cases, this term translates to low and variable income generated by a peculiar return of assets in agricultural, caused by exogenous and uncontrollable factors such as disease, weather or markets. (Tweeten and Thompson 2002).

Hopkins and Morehart (2002) highlight that the differences in operating profit margins explain most of the divergence in profit rates for entrepreneurs in the agricultural and non-agricultural sectors. These changes have affected small farms' income significantly (Gardner 1992).

According to the fundamental paper of Gardner (1992), in which he investigates this issue, it is crucial to focus on two main aspects of agricultural income, starting from or basing his theory on the studies by Tweeten (1970) and Brandow (1977). Agricultural income is characterised by lower levels when compared to the whole economy may not allow covering living expenses. This result is aggravated by the low level of rates return of capitals (Tweeten 1970). On the other hand, Brandow (1977) puts the spotlight on unfair low income and, at the same time, on the number of farms that have fallen dramatically in recent years. These studies revealed that the instability of agricultural incomes is a key factor enhancing/exacerbating the farm problem.

*1.1.2   Causes of the "farm problem."*

According to Gardner (1992), Tweeten (1970) and Brandow (1977), the "farm problem" is the result of soaring agricultural productivity and technical efficiency. On the one hand, this change has indisputably led to the improvement of society's welfare; on the other hand, it has depressed the agricultural sector.

To explain this issue, many economists have focused on the supply-demand model by observing a substantial decrease in the price level and an increase in instability. The agricultural sector is characterized by demand and supply of inelastic products and by having a supply growth that is lower than that of demand.

Moreover, after WWII, technological progress has generated only a slightly larger increase in supply than demand, causing prices substantially, given the inelastic supply and demand. Furthermore, the relatively low capacity to absorb exogenous shocks causes substantial price fluctuations (Gardner 1992).

Recently, globalization, coupled with a vibrant increase in internet and communication technologies, transformed the agricultural problem from local to global and from slow to fast. Globalisation increases the elasticity, particularly on the side of demand, implying that the production-control approach to raising farm incomes is less effective than low elasticities (Alston and Pardey 2014). The world market, mainly for commodities, has suffered further profound changes caused by globalisation. New players appeared in the global market after the 90s, such as China and India, characterised by lower labour and product costs and dumping measures. These events have altered world economics and have required new policy tools to avoid strong repercussion caused by external and sudden shocks. Without policy regulation, the impact could have been shocking for all the economic sectors (Alston and Pardey 2014).

Trends and oscillations of the farm income may entail a systemic impact in both food and other value-chains (Bhat and Jõudu 2019; Hernandez et al. 2017). Eventually, a crisis in the agricultural sector can propagate its effect quickly and uncontrolled compared to the past.

*1.1.3    Empirical evidence of the "farm problem."*

Most empirical studies have emphasised the relationship between "farm problem" (*lato sensu*) with labour and society (Banerjee and Newman 1993), investments and capital (Foster and Rosenzweig 1996; Hu 1972), degree of openness of an economy (Matsuyama 1992)

According to Gollin, Lagakos and Waugh (2014), it is necessary to hold the farm sector at the equilibrium to reach a state of health in order to guarantee the economic balance, especially during the bust and contraction phases characterising economic cycles, also keeping into consideration the consequence on labour, finance and natural resources. Recently Bustos, Garber and Ponticelli (2020) have stressed the strict linkage between the rural and urban environment, proving the systemic effect of the changes in the agricultural sector.

Hopkins and Morehart (2000) have studied another strand of the farm problem regarding the low returns, comparing expenditures, household incomes, and net worth to a minimum threshold or poverty line. These scholars have measured the incidence, intensity, and inequality of poverty in agriculture. This study confirms the presence of the farm problem and considers the relevance of off-farm revenue to stabilise and maintain the level of the farm's household income.

Notwithstanding these changes, the economic results are the more critical factors used to evaluate the efficiency of policy measures, particularly regarding the impact on both level and variation of the farm income.

Recently this strand of research in "Farm Problem" has received a new life regarding the comparison between agricultural and non-agricultural income (Marino, Rocchi, and Severini 2021; Rocchi, Marino, and Severini 2020).

### 1.1.4   Agricultural policy and Farm Problem

The farm problem is the principle supporting public intervention in agricultural economics, with the E.U. considering the agricultural sector as a cornerstone of its policy interventions since its constitution. After the first years since its inception, the E.U. has created the Common Agricultural Policy (CAP) to coordinate the policies of its Member States (M.S.). In the beginning, the CAP was characterised by having a supporting pricing policy and regulating imports-exports.

Substantial changes to the E.U.'s CAP was introduced in 1992 with the `MacSharry' reforms: cut institutional support prices for some major agricultural commodities and farmers compensated by other direct payments. The main objectives were the increase in competitiveness and agricultural income, also introducing essential goals regarding the environment, landscape, society and food quality.

The "Agenda 2000" package in 1999 represented a fundamental breaking point with the introduction of the cross-compliance strategy. In particular, the objective of agricultural policy did no longer only focus on economic support. Still, it started pursuing environmental and social goals, especially via the Rural Development Program (RDP) (the second pillar of CAP). After 2005, most direct payments transformed into the Single Payments Scheme, leading to a vigorous shift towards increasingly decoupled CAP forms. The renovation of CAP has followed in 2007 and 2013 and have posed the basis for the next CAP (Bozzola and Finger, 2020).

This thesis tries to fill the gap in the connection of farm problems with the E.U.'s CAP in Italy.

## 1.2 Topic and Research Questions

This thesis's topic is pivoted on the CAP's impact on the level and stability of income. In particular, it refers to the Income Transfer Efficiency (hereinafter ITE) and the role of a specific CAP measure like the Income Stabilisation Tool (IST).

ITE can be defined as the ratio of income gain of the targeted beneficiaries, i.e., farmers and the volume of the associated public expenditures and consumer costs (OECD 1995). Strictly speaking, in this thesis, ITE is defined as the ratio of the gain in farm income to the monetary value of public transfers to support farmers (Dewbre, Antón, and Thompton 2001) hence the ability of certain subsidies to enhance the farmers' income.

The scant literature has shown that ITE varies according to the CAP measures analysed. This variability can be explained by the impacts of various subsidies on the outputs, inputs, production function, and the costs incurred by farmers' participation in these policies. According to economic size, differentiating farms can have potentially significant impacts on ITE since eligibility criteria for support payments have become stricter with successive policy reforms. To the best of my knowledge, the literature has not considered evaluating ITE, considering that it may vary according to the farm's economic size. Assuming that income could have a dynamic component, its variation, due to ITE, can extend overtime with a geometric progression. The effect can be investigated in the short and long term. All these effects may be evaluated simultaneously to grasp the different aspects of ITE.

The second topic of this thesis concerns the instability of income which dramatically characterises the agricultural sector (E.C. 2009). The E.U. has looked at many instruments in order to deal effectively with this issue (Cafiero et al. 2007; Diaz-Caneja et al. 2008; E.C. 2001, 2017b; Meuwissen, Assefa, and van Asseldonk 2013; Meuwissen, van Asseldonk, and Huirne 2008) the focus was on income stabilising insurance schemes such as IST - Income Stabilisation Tool (E.C. 2010b, 2011b, 2011a, 2013a, 2013b). IST is a CAP measure devoted to stabilising the income of farmers.

Being an innovative measure, many Members State has decided not to applicate it till more information is made available to policymakers who will then decide whether to adopt it or not. A substantial amount of exploratory research is now underway on the IST's agricultural, sectoral, and country-level effects. The emphasis of this literature is on actuarial assessments of potential income insurance, its government costs, potential beneficiaries within the farm community, as well as conceptual analyses of adverse selection and moral hazard issues with specific full-farm income insurance instruments (El Benni, Finger, and Meuwissen 2016; Dell'Aquila and Cimino 2012; Finger and El Benni 2014b, 2014a; Liesivaara et al. 2012; Mary, Santini, and Boulanger 2013; Pigeon, Henry de Frahan, and Denuit 2014).

However, the research carried out remained at the conceptual level, with no explicit implementation issues. This observation is also due to the fact that considering the possible appeal of this method, IST has not yet been introduced in Italy, thus creating a vicious circle that can only be broken by analyzing the real impact of this tool.

Significant measures towards implementation that have not previously been discussed in empirical research are i) the specification of aspects relating to the structure of mutual funds (M.F.) across sectors and space and (ii) the design of IST regarding the modality of participation at M.F. by the farmers. These decisions could have an effect on income-stabilising properties, the profitability of mutual funds, income inequality in the agricultural sector and the distribution of benefits through land and farm types.

Regarding point ii), a first hypothesis has made concerns the participation of M.F. through contributions. Farms can contribute to the IST by applying flat-rate premium levels (Finger and El Benni 2014a) or by income or risk. Different designs could have cost implications and potentially could reduce the capacity of IST to stabilise the income.

Another exercise concerns the implementation in IST, replacing the contribution used in the first hypothesis of an insurance-like system with a premium depending on the probability and extent of the farm's damage.

This approach is very challenging because it needs to define what are the variables that influence the indemnities. Another interesting aspect concerns the economic sustainability of the insurance-like system. Whereby "economic sustainability" is defined both on the side of the "policyholder" who must have the availability to acquire the premium and on the other side of the "insurer" (in this case the M.F.), which must, following the payment of the premiums, to support the agricultural impact of extreme adverse events without incurring insolvency problems.

This thesis attempts to answer the question of how the CAP affects the level and stability of income.

## 1.3 Data and Methodologies

The analyses conducted in this thesis are based on uniform farm-level data of the Italian Farm Accountancy Data Network (FADN) for the period 2008 to 2018[1]. FADN is a stratified panel dataset specifically designed to assess Common Agricultural Policy's impacts, embracing over 100.000 observations gathered yearly via surveys. The stratified characteristic represents all the farm's representativeness in all the years analysed, while the farm-level data allows micro-founded analysis. FADN represents a powerful source of data, considering that the information is standardised for all the country in the European Union, even allowing a comparison between different economy types.

The amount of data extracted from the FADN is vast, and to manage it, it is necessary to have a statistical language to exploit all the potential of the dataset. Considered this challenging goal, it has been used R (R Core Team 2020), a specific language designed for statistics, econometrics and deep learning.

---

[1] Considering the change in the CAP in 2014, the analysis of the CAP was divided into two periods from 2008 to 2014 and from 2015 to 2018.

Because the topics differ, several methodologies are applied. Born in mind that the FADN is a panel dataset, it was exploited this feature to evaluate the ITE. Considering these characteristics, it has been likely to reduce the economic problem's dimensionality analysed through time and individual fixed-effects: all fixed variables have been eliminated, including those omitted. The FADN panel data also allows evaluating the dynamic effect of ITE. For this reason, the lagged income has been used. However, this approach highlighted many endogeneity problems to reduce, which, it was necessary to introduce the Cochrane-Orcutt procedure and an econometric estimation method such as Generalized Method of Moments (GMM) with Instrumental Variables (IV) in the form of a system (i.e., the system use IV in level and first difference).

The Income Stabilisation Tool assessment to maintain an unchanging income level is made with a simulation of this measure's application (the design is specified in chapter 3). In order to analyse the effect of IST on income stability, considering that we do not have actual data, we need to simulate the application. Also, for this evaluation, we use the FADN dataset that allows assessing many aspects of IST. The challenge in this operation relates preliminarily to the searching and tidying up of the dataset to make the simulation representative of IST's eventually future adoption. The simulations have also regarded possible M.F. structures, considering different aggregations and two kinds of contribution: fixed and variable. Using the information of FADN, this assessment has been referred to all Italian farm population.

After having refined the methodologies to simulate IST and analysed different typologies of contributes to M.F., it was change point of view using an approach like the insurance. In this case, the objective is to predict the levels of indemnity. The indemnities are linked to the characteristics of the company and the farmers. To estimate them, it is necessary to use a large dataset, such as the FADN. Suppose many variables can be an advantage to make the evaluations to understand which of these are essential, considering that there is a cost associated with the information. In that case, it is necessary to implement a methodology able to select only the relevant variables. Another challenge regards the peculiar probability density function of IST indemnities.

This study can lead to considerable increases in the analysis of this topic as, i) the mandatory condition is relaxed (generally assumed in almost all papers on IST), ii) it goes to understand which the variables are directly connected with the compensation through machine learning tools while guaranteeing excellent forecasts and finally iii) the estimation of the peculiar type of pdf that characterizes the indemnities will be made possible through the use of the Tweedie distribution. To the best of our knowledge, this is the first time that this probability density function is used in agricultural economics.

Finally, in order to evaluate several hypotheses, it is required to check the performance both in a statistical and economic sense. These three objectives, i.e., estimating the minimum set of information necessary to predict the indemnities, evaluate which model is the best in terms of statistical and economic performance, are analysed in the second part of chapter 3. The goal is to assess whether this improves the ratemaking by using the Income Stabilization Tool's prospective Italy application.

## 1.4 Structure of the thesis

Because the two topics differ, the thesis is structured into two main chapters the first addresses the effect of CAP on income, the second how the IST can improve the stability of farm income and how it should be designed and managed to do so.

The first topic faced a critical issue regarding the CAP measure's capacity to be transferred in income (ITE). This evaluation is necessary for the policymaker to know what measure better design to guarantee the level of income. For example, some losses in efficiency are already taken into account for the policymaker when a specific policy is designed. This is the case of the indemnity for the losses incurred in following agri-environmental constraints such as organic. This measure is specifically designed, among other things, to avoid the reduction of the impact deriving from the use of synthesised chemical products and not to increase the level of income.

Other analyses were carried out to understand if the ITE of the CAP differs according to the various economic size of the company. Based on our scientific knowledge, this level of study has not yet been conducted for this specific topic. The rationale behind this share-out stems from considerations about information which in our research is considered a commodity. Following this, in order to have a greater amount of information, which allows a better economic result, it is necessary to have a high level of wealth to purchase this particular productive factor.

The second main topic of the thesis regarding the effect of CAP on income variation, in particular, concerning an innovative policy instrument such as IST. It splits the chapter into two parts: first, regarding IST's simulation regarding the stabilisation of income, second, concerning IST's design.

The first study is necessary to understand if IST can stabilise the income and reduce this farm problem. The simulation also concerns different levels of aggregation: the IST regulation does not provide anything on this aspect. It is necessary to give policymakers a vision of the best level of grouping or the best parameter to make this subdivision.

The second part of this chapter is about designing IST with an approach similar to insurance, with a premium established based on farm risk factors. Considering that the M.F. has not the objective of profit maximisation but a null profit goal, we need to design a very similar premium to indemnity.

However, because both topics are two sides of the farm problem, a final chapter concludes the thesis to summarise the findings of the two parts of the thesis that affect the two essential dimensions of the farm problem. More in detail, CAP policy measures the role of fostering the level of farm income and reducing the instability of farm income.

# Chapter 2. The role of Common Agricultural Policy in enhancing farm income: A dynamic panel analysis accounting for farm size in Italy

As a multi-objective policy, the E.U. Common Agricultural Policy continues to secure significant income support for farmers as one of the specific objectives. We estimate the income transfer efficiency of a broad set of pivotal policy measures, focusing on the effects of farm structure on income transfer efficiency. We use dynamic modelling based on a micro-data panel of Italian farms for the period 2008–2014, allowing for endogeneity, simultaneity bias, and omitted variables. In line with previous studies and economic expectations, we find that decoupled direct payments provide the highest contribution to agricultural incomes, followed by agri-environmental payments and on-farm investment subsidies. Coupled payments have no significant impacts on farmers' income. Generally, for all analysed Common Agricultural Policy measures, large farms benefit from greater transfer efficiency than medium and small farms. These differences among instruments and across farms suggest that policy-participation costs may play a pivotal role, together with farms' economic structure, in determining the income transfer efficiency of CAP policies.

This chapter reports a paper published in Journal of Agricultural Economics, written in collaboration with Simone Severini and Federico Antonioli (Biagini L, Antonioli F, Severini S. The Role of the Common Agricultural Policy in Enhancing Farm Income: A Dynamic Panel Analysis Accounting for Farm Size in Italy. J Agric Econ 2020;71:652–75).

## 2.1 Introduction

The $EU$ Common Agricultural Policy ($CAP$) supports farm income via a broad set of different measures (Matthews 2017; The World Bank 2017). Ensuring a fair standard of living for farmers is one of the multiple specific objectives of the $CAP$, (Treaty on the Functioning of the European Union ($TFEU$) (art. 39, point b)), and recent analysis shows that the $CAP$ reduces poverty in some $EU$ regions (World Bank 2017). Over the last decades, the $CAP$ has addressed a broader set of societal concerns, adding new dimensions, such as animal welfare, environment, food quality, and countryside management (European Parliament 2017). Despite the broadening of the $CAP$'s objectives, securing income support has remained one of its nine specific objectives, reinforced by the legislative proposal on the $CAP$ beyond 2020 (European Commission 2018a). As a consequence, evaluation of the extent to which the different $CAP$ measures affect farm income (income transfer efficiency, hereafter $ITE$) remains relevant, to assess the gain in farm income from the monetary value of $CAP$ support directly provided to farmers (Dewbre, Antón, and Thompson 2001).

We measure the $ITE$ of different $CAP$ subsidies as the marginal effect of one euro of support on farm income. The five measures analysed are: i) coupled direct payments ($CDP$), ii) decoupled direct payments ($DDP$), rural development support in the form of iii) agri-environmental payments ($RDP_{aes}$), iv) on-farm investments ($RDP_{inv}$), and v) other measures ($RDP_{other}$). The present study has two main objectives. The first is to assess whether $ITE$ differs across the different types of payments and, consequently, the relative $ITE$ performance of the different payment types. The second is to investigate whether the $ITE$ varies with farm size (small, medium and large). Differentiating farms according to economic size has not been explored in terms of income transfer efficiency, albeit having potentially significant impacts on $ITE$ since, eligibility criteria for support payments have become stricter with successive policy reforms. In this context, participation costs associated with specific policy support play an important role in determining the level of $ITE$. These costs may differ by economic size, since fixed costs bear a diverse weight on the cost structure: small farms rely upon limited information, economic and physical resources and often on limited technical competence for administrative tasks, encouraging them to outsource (Vernimmen et al., 2000). The existence of positive economies of scale – characterising large farms – may reduce the cost of participation.

The $EU$ currently allocates approximately 45 billion euros per year to support farmers through Pillar 1, of which more than 90% are direct payments (hereafter $DP$) (European Commission 2017a). Most of these are decoupled from production ($DDP$) and serve the objective of enhancing and stabilising farm income (European Commission 2018b; European Parliament 2012; Matthews 2018a). The remaining share of $DP$ is coupled ($CDP$), not directly to production levels, but to the amount of cropped land or number of livestock (European Commission 2017a). Rural development policies ($RDP$) offer further support (approximately 14 billion euros of commitment appropriations per year), to which national resources are added, following the co-financing requirement. Four axes characterise the $RDP$, with Axes 1 and 2 accounting for the largest share of the overall $RDP$ budget (European Commission, 2013). Axis 1 aims to improve the competitiveness of the agricultural and forestry sector, with most of the resources devoted to support farm investment ($RDP_{inv}$); Axis 2 intends to improve the environment and countryside conditions, mainly via agri-environmental scheme payments ($RDP_{aes}$). Axis 3 supports economic diversification and the quality of life in rural areas, while Axis 4 fosters local activities intended for local development. Axis 3 and 4 account for a small share of the budget, roughly 12% of the overall $RDP$ budget granted in Italy at the end of 2014 (CREA 2015). These two Axes, together with all the remaining $RDP$-related measures other than $RDP_{inv}$ and $RDP_{aes}$, are accounted for in the variable $RDP_{other}$. A substantial share of this category is represented by Less Favoured Areas ($LFA$ hereafter) payments, a significant source of income support for farmers settled in these areas.

The $ITE$ for each measure is estimated in a dynamic setting, employing the System-Generalised Method of Moments ($SYS-GMM$) model on the entire sample of the Italian Farm Accountancy Data Network ($FADN$) for the period 2008–2014, ensuring a constant application of the $CAP^2$. Models are estimated for the total sample (all farms), as well as for three subsamples defined according to the farms' economic size (i.e., small, medium, and large farms).

This analysis adds to the current literature in four main areas: i) it accounts for a recent period with a policy setting that has not been investigated in the literature; ii) it considers both short- and long-run ($SR$ and $LR$, respectively) effects of $CAP$ payments by taking advantage of the dynamic nature of the model; iii) it disaggregates $RDP$ support into its main components, offering a more in-depth analysis of the very different measures included within the axes of $RDP$; iv) it analyses the impact of the policy measures on subgroups of farms, according to their economic size.

The paper is structured as follows: Section 2 describes the application of the $CAP$ in Italy, whereas Section 3 offers a review of the literature on $ITE$. Section 4 presents the econometric model and estimation method, Section 5 describes the data used in the analysis, and Section 6 presents and discusses the results. Section 7 concludes and provides some policy implications.

## 2.2 Main characteristics of the CAP in Italy during the studied period

Agricultural subsidies directly transferred to Italian farmers amount to approximately one-third of the national value-added generated by the agricultural sector at the end of the considered period (CREA 2017). Approximately three-quarters of this support comes from $EU$ policies, with the remaining share granted through national and regional policies. The European Agricultural Guarantee Fund (i.e. $CAP$ Pillar 1) provides the most significant amount of subsidies (i.e. 4.5 billion euros per year), mostly via $DP$ (i.e. 3.9 billion euro per year). Community support from the European Agricultural Fund for Rural Development (i.e. $CAP$ Pillar 2) reaches approximately 1.5 billion euro per year, with the national co-financing accounting for 49.3% of the overall support from $RDP$ in the period 2007–13 (CREA 2017).

Italy implemented the historical model for allocating $DP$ to its 1.16 million recipients (approximately 72.5% of the total farm population) (European Commission 2015). Hence, the monetary amount of $DP$ support is determined on a per-farm basis according to the amount received during the reference period (i.e. 2000–2002), and further determined by the entitlements owned by each farmer. It is important to mention that at the national level there are less entitlements than eligible land (Ciaian, Kancs, and Swinnen 2013; Guastella et al. 2018), and such entitlements are freely traded between farmers without restriction (European Commission 2017b). Hence, the availability of entitlements represents the most relevant constraint on $DP$ support, which may explain the limited degree of capitalization of $DPs$ into land values observed in Italy (Guastella et al., 2018; Valenti et al., 2020). Approximately 10% of the national ceiling for $DP$ has been devoted to $CDP$ as specific support, which benefits a limited number of farmers (52,000 in 2014, according to the European Commission (2015)) and specific commodities, such as cattle, dairy, tobacco, and sugar beet (CREA 2017). It is worth underlining that none of the support measures are linked to product prices - as for deficiency payments and countercyclical support measures implemented in other non-$EU$ countries.

---
[2] A new *CAP* reform was implemented in 2015, generating a policy break that makes the comparison of results before and after the reform problematic.

Member States are allowed to set minimum thresholds, in terms of both land and value, under which the support is not granted, to avoid an excessive administrative burden. For example, $RDP_{aes}$ are not granted when the area for which the payments are requested is smaller than a Regional set level (generally one hectare). Finally, the aid the farmers receive under the direct support scheme is subject to cross-compliance (Matthews 2013). This encourages compliance with statutory requirements since a reduction or elimination of $DP$ is risked when cross-compliance is not met[3].

As mentioned, Axes 1 and 2 of Pillar 2 accounted for 38.3% and 48.2%, respectively, of the $RDP$ resources granted in Italy in 2014 (CREA 2017). Within Axis 1, the modernisation of agricultural holdings has represented approximately 53.2% of the total resources in the 2007–2013 programming period. This measure is aimed at supporting farm investments ($RDP_{inv}$) by providing public resources covering between 40% and 60% of the overall investment cost. $RDP_{aes}$ represent 37.9% of the total budget allocated to Axis 2, reaching a relatively high penetration in Italy[4]. They provide annual support to farmers who, on a voluntary basis, subscribe to environmental commitments. Hence, they are designed to encourage agricultural producers to adopt environmentally friendly practices beyond mandatory obligations (European Commission 2013). $RDP_{aes}$ have been granted over at least 5 consecutive years. $RDP_{other}$ reach farmers through many further policy instruments (e.g. $LFA$ and Natura 2000), and although such aggregated residual categories of measures are included in the analysis, its substantial heterogeneity and possible correlation effects amongst the many specific subsidies, hampers the identification of their separate effects.

---

[3] This refers to the Statutory Management Requirements in the areas of environmental, public, animal and plant health, animal welfare, and the minimum Good Agricultural and Environmental Conditions.
[4] In 2013, the share of agricultural land in Italy enrolled in agri-environmental measures was approximately 23.1% of the national utilised agricultural area (data retrieved from Eurostat, 2019).

## 2.3 Background on the $ITE$ of farm subsidies

Dewbre et al. (2001) and Dewbre and Short (2002) explored the impact of a set of policy measures in terms of $ITE$, trade distortion, and competitiveness. Guyomard et al. (2004) further extended the scope of the analysis, accounting for the reduction in negative externalities. These analyses used static, single-output partial equilibrium models for the farm sector and key input parameters estimated in previous studies (e.g., agricultural commodities' demand and supply elasticities, the elasticity of substitution between factors, and relative cost shares of the considered inputs). These analyses provided interesting results regarding the impact of output, land, and input subsidies, and decoupled income transfers with and without mandatory production. Specifically, payments without mandatory production were most efficiently transferred to farm income, followed by payments with mandatory production or land subsidies (Guyomard, Le Mouël, and Gohin 2004), with both output and input subsidies being the least efficient.

However, the CAP has now changed substantially since these studies. Specifically, most of the payments are now decoupled from production, with few measures now linked to specific crops and livestock while Pillar 2 measures are now important. Second, the policy programme ranking obtained in previous studies depended on prior given assumptions on key parameters (Guyomard, Le Mouël, and Gohin 2004). Third, to ensure the interpretability of their results, these studies assumed that all farms are identically affected.

A more recent strand of literature focuses on the empirical estimation of the $ITE$ of $CAP$ measures using a dynamic panel approach. Michalek et al. (2011) and Ciaian et al. (2015) applied econometric techniques to individual farm data, accounting for several $EU$ policy measures, including $RDP$. However, they analysed the latter as a single aggregated measure, ignoring the very different nature of the measures therein. Both these studies employed the $DIFF - GMM$ estimator (Arellano and Bond, 1991) to account for endogeneity problems and to address the issues of omitted variables and selection biases. Furthermore, their analyses included several explanatory variables that could contribute to explaining the observed level and evolution of farm income and, thus, to better identifying the relationship between farm income and the policy measures (Ciaian et al. (2015)). In Michalek et al. (2011) and Ciaian et al. (2015), these variables included – proxies - output and input prices, the relative amount of family labour used on-farm, the amount of available capital (differentiating among land and other assets, such as buildings and machinery), and the liabilities-to-total assets ratio. They both found that, depending on the considered $CAP$ measure, the $ITE$ differs, with the $RDP$ being the most efficient, followed by decoupled and coupled $DP$.

Differences in the $ITE$ of the different measures can be explained by the impact of subsidies on output and input prices and production decisions and to the costs incurred by farmers participation in these policies. Regarding input prices, the literature stresses the importance of land, since policy support increases the demand for land (Guyomard, Le Mouël, and Gohin 2004), thus capitalising (part of) the subsidy into land value and rents (see e.g. Ciaian and Kancs, 2012; Graubner, 2018; Guastella et al., 2018; O'Neill and Hanrahan, 2016).

Particularly, a situation in which entitlements are abundant when compared to the eligible hectares favours the full capitalisation of policy support into land value and rent (Ciaian, Kancs, and Swinnen 2013; Ciaian and Swinnen 2008). This is further cemented in Kilian and Salhofer (2008) and Kilian et al. (2012), whose findings show how the historical model leads to a partial capitalisation of subsidies into the land price, and in Klaiber et al. (2017), describing an increase in capitalisation when moving from the historical to regional model. Quoting Courleux et al. (2008, p. 8) on the application of the historical model, 'the degree of capitalization of the support into the land rental price can be zero, partial or total depending on the scarcity of entitlements relative to the number of hectares and on the land supply elasticity.'

However, in Italy, the $CAP$ Health Check of 2008 made all agricultural land uses eligible for entitlements (excepted forests), generating of abundancy of eligible hectares when compared to the volume of entitlements. As a result, Guastella et al. (2018) and Valenti et al. (2020) conclude that the degree of capitalisation of both $CDP$ and $DDP$ into land values is negligible.

The amount of *effectively fully decoupled* support (Cahill, 1997) should entirely translate into farm profits since, in a static environment, it does not affect production decisions (i.e. a unitary $ITE$ is expected)[5]. Conversely, as the other forms of support trigger changes in farmer behaviour, mainly concerning input use and production patterns, one should expect an $ITE$ lower than unity (see Minviel and Latruffe (2017) and Nilsson (2017) for some examples). For example, $CAP$ support has been found to influence the amount of labour force employed in the agricultural sector (see, e.g., Olper et al., 2014; Petrick and Zier, 2011, 2012), and on-farm investments (see, e.g., Latruffe et al., 2010; O'Toole and Hennessy, 2015; Sckokai and Moro, 2009; Traill, 2008).

Finally, the enlargement of the scope of $EU$ agricultural policy envisaged in the last decades has increased the effort required for farmers to become eligible for support (Mack et al. 2019; Mettepenningen, Verspecht, and Van Huylenbroeck 2009; Vernimmen, Verbeke, and van Huylenbroeck 2000). Such efforts incur participation costs, namely, additional costs, foregone income, and transaction costs[6]. While additional costs refer to the direct cost caused by fulfilling the subsidy requirements (causing additional working time or physical workload (Beckmann, 1996)), forgone income is related to the impact such requirements can have on production choices and yield levels (e.g., the introduction of a crop rotation or land set-aside). Finally, private transaction costs refer to the burden farmers face while completing the application, including both administrative and contact costs (Mettepenningen, Verspecht, and Van Huylenbroeck 2009). This suggests that the *ITE* of the measure is affected by the extent of participation costs, which can diverge across measures (Rørstad, Vatn, and Kvakkestad 2007) and farms. For example, coupled payments provide farmers with an incentive to adjust their farming system by adopting a less profitable enterprise in order to benefit from these payments. Furthermore, cross-compliance may produce a negative effect on economic results that could differ across farms of different size (Bennett et al. 2006; Cortignani, Severini, and Dono 2017) Differences in transaction costs arise across measures according to the complexity of the administrative process related to each measure (Mack et al. 2019). For example, while *DP* entail only the fulfilment of cross-compliance commitments, agri-environmental schemes require adherence to production practices that extend beyond cross-compliance. It is worth noting that cross-compliance costs vary according to the farm specialisation. Indeed, farms specialised in animal production usually face higher costs in the attempt of meeting the required cross-compliance. Rete Rurale Nazionale (2010) illustrates that this is due to both the complexity - requiring further consultancy- and service-related costs - and the monetary disbursement required for fulfilling the relative measures, particularly the treatment for animal effluents. Furthermore, farmers receiving an amount of direct payments lower than 1,250 Euro per year are allowed to submit a simplified demand (Henke et al., 2015)[7]. Similarly, $RDP_{inv}$ require higher participation costs, because of the need to involve external professionals[8]. The participation costs can differ across farms, even regarding the same measure and within the same region. For example, Ducos et al. (2009) indicate that the fixed component of participation costs in agri-environmental schemes explains why larger farms enrol more easily than smaller ones. Indeed, limited resources and technical competences in performing administrative tasks add to the reasons why small farms often resort to outsourcing (Vernimmen, Verbeke, and van Huylenbroeck 2000).

---

[5] Cahill (1997) defines a policy to be effectively fully decoupled if 'it results in a level of production and trade equal to what would have occurred if the policy were not in place' (Guyomard et al., 2004). Under uncertainty, fully decoupled support could affect the optimal production decisions of risk-averse farmers too, because of insurance and wealth effects (Hennessy, 1998).

---

[6] These three categories are derived from Reg. (EC) No 1698/2005. Farmers may face further costs related to uncertainties concerning the effects on the production levels - hence on the economic results of the farm – due to the change in production practices that some policy measures require (e.g., agri-environmental schemes) (Mettepenningen et al., 2009).

[7] The simplified scheme has been introduced by the Regulation (EC) No 1244/2001 of 19 June 2001.

[8] This requires an ex-ante economic evaluation of the investment and, for some types of investments, to provide investment design and planning. Finally, with this kind of measure, the quality of the process strongly influences the likelihood of a farmer becoming a beneficiary of the available resources that are granted on a competitive basis.

## 2.4 Methods and Data

### 2.4.1 Model Specification

The economic performance of a price-taking farm in the presence of government support can be represented by the following realised short-run farm profit ($\pi$) (Moro and Sckokai 2013):

$$\pi = pq - wx + G$$

where $q = f(x, k)$ is the output produced according to a production function using variable inputs $x$ (e.g. labour) and fixed inputs $k$ (e.g. capital, including owned land and other farm assets, such as machinery and buildings). $p$ and $w$ are the prices of output and variable inputs, respectively and $G$ is the government support directly granted to farmers.

$G$, placed directly in the profit function, may influence production decisions and takes different forms: coupled or decoupled to production, with or without mandatory production, and linked to the use of specific factors such as land (see Guyomard et al. (2004) for a thorough description of the diverse types of $CAP$ subsidies).

Farmers maximise their expected profit conditional to the information they have at the planning period: $E(\pi) = E(pf(x, k) - wx + G) \mid \Omega$, where $\Omega$ refers to all available information affecting farmers' choices (Arrow 1996)[9].

We assume similarly to Rizov et al., (2013), that a farmer's expectation for his/her profit ($E(\pi)$) is a First-Order Markov process with transition probability $Pr(\pi_{i,t} \mid \pi_{i,t-1})$. Therefore, the realised profit can be expressed as $\pi_t = E(\pi_t) + \varepsilon$, where $\varepsilon$ represents a generic deviation of the realised profit from the expected profit (i.e. defined when production decisions take place). Such deviation can be generated by several factors, including the fact that realised output prices differ from their expected levels, that production can be affected by exogenous and unpredictable events, and that exogenous factors could constrain the decisions regarding the use of production inputs.

Exploiting these theoretical assumptions and introducing farm net income ($FNI$)[10] that represents the remuneration for a farm's own factors of production (family work, land, and capital) and remuneration for the entrepreneurs' risks (loss/profit), net of the cost for external factors, including rented land, as the profit ($\pi$), the model is specified as follows

$$FNI_t = \alpha_1 FNI_{i,t-1} + \alpha_2 FNI_{i,t-2} + \sum_{k=1}^{5} \beta_k G_{k,i,t} + \sum_{k=1}^{5} \gamma_k G_{k,i,t-1} + \sum_{j=1}^{7} \delta_j X_{j,i,t} + \sum_{j=1}^{7} \zeta_j X_{j,i,t-1} + \tau_t^* + \eta_i^* + \varepsilon_{i,t}'' \quad (1)$$

---

[9] This formulation refers to the simplest risk-neutrality case, that is however unlikely in the real world (Iyer et al. 2020). As in many other studies, is not possible to retrieve any information about the risk profile of each farmer, although different risk behaviours may entail a different use of resources (for a recent example of how a different risk profile affects resources' allocation see Vigani and Kathage, 2019).

[10] FNI is defined as variable $SE420$ in the $FADN$. For further details, see the FADN guide, available at http://ec.europa.eu/agriculture/rica/pdf/site_en.pdf.

where $\alpha_1$ and $\alpha_2$ are the autoregressive coefficients for the lagged values of the dependent variable[11] and $G$ refers to the considered policy variables. $X$ represents other explanatory variables related to farm characteristics, including the value of farm assets ($Land\ Value$ and $NonLand\ Value$), the ratio between rented and total land, the amount of family labour, and output and input prices as explained below. $\tau_t^*$ is the year-specific intercept, $\eta_i^*$ is the time-invariant farm-specific fixed effect, and $\varepsilon_{i,t}''$ is the idiosyncratic error term[12].

### 2.4.2 Data

The dataset used in the analysis is an unbalanced panel of farms belonging to the Italian Farm Accountancy Data Network ($FADN$), covering the period 2008–2014[13]. Of the original 78,310 observations, only farms with a negligible amount of labour and land have been excluded from the analysis since, generally, they are not recipients of $CAP$ support. Furthermore, observations falling outside 0.05% of the highest and lowest percentile, with respect to the median of the dependent variable (farm income), were treated as outliers and eliminated, for a more homogeneous panel dataset. This resulted in the elimination of 4,300 observations (i.e. 5.5% of the original sample), leading to a final total of 74,010 observations.

Table 1 describes the variables used in the analysis. The first set of variables refers to the policy measures considered in the analysis and extensively described in previous sections. The second variable set refers to other farm-specific characteristics that, according to both the theoretical model reviewed above and the background literature detailed in section 3 (e.g. Ciaian et al., 2015; Michalek et al., 2011), are expected to affect $ITE$ and account for farm heterogeneity. These characteristics can influence the level of farm income: for example, higher use of unpaid (family) labour reduces explicit costs, whereas a high liabilities-to-assets ratio can affect the costs associated with financial activities. Given the impact that subsidies may have on output and input prices, as well as on land value, these variables are included to better estimate the effect of the policy support on income.

Thus, the model accounts for the family labour ($FAML$), the relative amount of rented land ($Land\ ratio$), the value of land ($LV$) and non-land ($Non\ LV$) assets, and the relative importance of farm liabilities ($Leverage$). For purely empirical reasons, the relative amount of rented land is not included in the estimation model presented in section 5[14]. The farm leverage is defined as the share of liabilities on total farm assets, hence providing an important dimension of the farm-specific financial conditions.

---

[11] The inclusion of the second lag of the dependent variable is motivated by the fact that such AR(2) specification results in the best fitting lag order (for a similar specification, see Baldoni et al., 2017).

[12] For further econometrics details, see Appendix (on-line).

[13] Data are managed by the Consiglio Nazionale per la Ricerca in Agricoltura e l'analisi dell'Economia Agraria (CREA) in Rome. Data and background materials are available at https://bancadatirica.crea.gov.it/Default.aspx.

[14] While results were very similar to those presented in section 5 for the total sample, the system proved not invertible when estimating the sub-samples for small, medium and large farms. For the sake of comparability among the four estimated models, the authors opted for the model omitting this variable.

Finally, additional price-shares variables are included reflect changes in output and input prices. The variable *Price ratio*[15] is defined as the ratio between the total output – excluding policy support - and the corresponding total intermediate consumption per farm as proxy for the evolution of output and input prices over time. However, this only reflect a farm-specific price ratio if the composition of output and input on each farm does not change over the estimation period and if the volume of output and input changes by the same proportion each year. To overcome these drawbacks, two further sector-specific price ratios are included as proxies of price differentials, namely the ratio of cereals and fruit and vegetable outputs over the total output each farm generates each year ($CER\ ratio$ and $F\&V\ ratio$, respectively). These ratios reflect systematic differences in the production structure of the different groups of farms (defined as the composition of outputs and inputs), which may otherwise lead to systematic differences in the evolution of farm profits for each group[16]. That is, proxies of price differentials are expected to reduce the potential correlation between price movements and payments' size received by farmers. However, one may expect the extent of these correlations to be very limited, since the unitary level of all considered support measures are not explicitly linked to prices[17].

All variables, except for $Price\ ratio,\ CER\ ratio,\ F\&V\ ratio$ and $Leverage$ are standardised by the total amount of utilised labour (expressed in annual work units, hereafter $AWU$), enabling comparability among different farm sizes and reducing heteroscedasticity since labour is ubiquitous amongst all our farms.

The three subsamples referring to the economic size of farms have been identified on the basis of farm total revenue[18]: the farm-specific median value of total revenue across all the years is used to provide a stable farm classification in terms of the economic size, namely small, medium, and large farms. All economic variables are deflated using the Eurostat Harmonised Consumer Price Index[19].

---

[15] Because the profit function is homogeneous of degree zero in prices, including them as a price ratio is advisable. The authors thank an anonymous referee for this useful suggestion.

[16] The two sector-specific price ratios have been selected by first analysing the trend of both output and input prices at national level to spot whether divergences exist with respect to a common movement. While any divergence is displayed with regards to input prices, both the fruit&vegetables and cereals sectors depict a singular trend with respect to the general agricultural output category. The authors are grateful to an anonymous referee for this suggestion.

[17] CDP represents the only category of policy support that may change in this regard, being linked to the amount of land (or livestock units) benefitting from this support. However, the support devoted to a specific production (e.g., durum wheat) is fixed by national ceilings, and distributed among producers according to the amount of planted land (or number of heads). Hence, even if a price increase induces an increase in planted area (or raised livestock units) at the national level, this would result in a decline of the unitary value of the support per hectare (or per head of livestock), limiting the correlation between prices and the size of support.

[18] The variable total revenue reported in the Italian FADN is determined as Total output + Total subsidies (excluding those on investments) (i.e., variable SE131+SE605 of the EU FADN).

[19] Eurostat, Table 'prc_hicp_aind'; retrieved from: https://ec.europa.eu/eurostat. Base 100 = 2015.

**Table 1 – Description of the variables used in the analysis.**

| Variable's category | Label | Description |
|---|---|---|
| Policy variables ($G$) | $FNI$ | Farm Net Income/Total Annual Work Units (AWU) |
| | $CDP$ | Coupled Direct Payments/AWU |
| | $DDP$ | Decoupled Direct Payments/AWU |
| | $RDPaes$ | Rural Development Program support for Agri-Environmental Schemes/AWU |
| | $RDPinv$ | Rural Development Program support for farm investments/AWU |
| | $RDPother$ | Other Rural Development Program payments/AWU |
| Other variables ($X$) | $LV$ | Land Value /AWU |
| | $Non\ LV$ | (Total Assets – Land Value) /AWU |
| | $FAML$ | Family labour expressed in annual work unit /AWU |
| | $Leverage$ | Liabilities/Total Assets (Liabilities+ Own Capital) |
| | $Price\ ratio$ | Total Output / Total Intermediate Consumption |
| | $CER\ ratio$ | Total Output of Cereals/Total Output |
| | $F\&V\ ratio$ | Total Output of Fruits & Vegetables/Total Output |
| Variables not used in model estimation | $Land\ ratio$ | Rented land/Total land (included in the theoretical model) |
| | $Total\ revenue$ | Total revenue (used for the characterisation of the three farm-subsamples) |

*Source: Authors' elaboration*

Farms differ considerably in terms of some important characteristics when grouped by size (Table 2). Farm income per unit of total farm labour increases when moving from small to large farms, whereas the importance of family work tends to decrease with size. A similar pattern is observed regarding the amount of land and non-land capital per unit of total labour while the opposite is true for $Leverage$. In contrast, $Price\ ratio$ remains substantially at the same level in all three subsamples. Regarding the five policy measures analysed, the largest share of the $CAP$ support is provided by $DDP$. $RDP_{inv}$ is the most varying measure, with large farms exhibiting the smallest variability, as expected, and small farms the greatest. This is also true for $CDP$ and $RDP_{aes}$. In contrast, $DDP$ is quite stable over all farm size groups.

**Table 2 – Descriptive statistics of the considered variables, 2008–2014.**

|  | Mean | Sd | Skewness | Kurtosis | Mean | Sd | Skewness | Kurtosis |
|---|---|---|---|---|---|---|---|---|
|  | **Total Sample (74,010 Obs.)** | | | | **Small Farms (24,664 Obs.)** | | | |
| FNI | 25699.12 | 48054.08 | 49.29 | 6479.17 | 10211.96 | 20132.61 | 26.43 | 1781.74 |
| CDP | 696.31 | 3530.72 | 18.32 | 601.19 | 149.44 | 644.61 | 12.10 | 288.02 |
| DDP | 6132.66 | 10854.12 | 6.15 | 88.88 | 2980.44 | 4638.67 | 5.83 | 78.38 |
| $RDP_{aes}$ | 466.02 | 2345.31 | 13.71 | 322.13 | 251.20 | 1873.22 | 27.65 | 1078.95 |
| $RDP_{inv}$ | 388.85 | 5349.35 | 32.52 | 1811.23 | 309.74 | 6188.65 | 41.98 | 2527.26 |
| $RDP_{other}$ | 614.05 | 2177.50 | 7.63 | 105.63 | 440.21 | 1769.42 | 9.80 | 158.96 |
| LV | 234667 | 475846 | 9.49 | 197.83 | 161436.56 | 268094 | 6.67 | 114.61 |
| NO LV | 154281 | 1020873 | 208.29 | 49395.06 | 75090.88 | 101726 | 6.12 | 80.24 |
| FAML | 0.839 | 0.259 | -1.505 | 1.084 | 0.951 | 0.135 | -3.333 | 11.649 |
| Leverage | 0.032 | 0.173 | 54.506 | 5475.827 | 0.026 | 0.123 | 14.096 | 406.023 |
| Price Ratio | 2.994 | 4.409 | 80.705 | 12401.561 | 2.917 | 5.901 | 90.225 | 11141.363 |
| CER Ratio | 0.151 | 1.209 | 252.899 | 67202.614 | 0.183 | 2.067 | 151.565 | 23540.962 |
| F&V Ratio | 0.021 | 0.133 | 6.668 | 43.656 | 0.004 | 0.051 | 15.598 | 257.566 |
|  | **Medium Farms (24,673 Obs.)** | | | | **Large Farms (24,673 Obs.)** | | | |
| FNI | 21519.20 | 52921.03 | 97.78 | 12862.64 | 45360.56 | 55473.27 | 9.18 | 271.79 |
| CDP | 457.89 | 1601.88 | 8.61 | 121.24 | 1481.38 | 5782.81 | 11.91 | 241.53 |
| DDP | 5620.32 | 8504.20 | 6.42 | 103.26 | 9796.07 | 15360.82 | 4.63 | 50.54 |
| $RDP_{aes}$ | 512.94 | 2245.39 | 10.30 | 171.33 | 633.83 | 2806.07 | 10.41 | 186.30 |
| $RDP_{inv}$ | 302.51 | 4006.78 | 25.78 | 958.82 | 554.26 | 5608.71 | 18.84 | 501.91 |
| $RDP_{other}$ | 858.45 | 2572.12 | 6.24 | 75.01 | 543.43 | 2093.48 | 8.01 | 115.50 |
| LV | 235945 | 451146 | 10.52 | 248.77 | 306592 | 627142 | 7.71 | 121.89 |
| NO LV | 132127 | 196899 | 32.67 | 2161.83 | 255595 | 1749317 | 123.96 | 17154.55 |
| FAML | 0.875 | 0.210 | -1.658 | 1.825 | 0.690 | 0.321 | -0.549 | -1.110 |
| Leverage | 0.032 | 0.239 | 55.939 | 4343.307 | 0.038 | 0.132 | 23.900 | 1317.917 |
| Price Ratio | 3.104 | 3.135 | 9.959 | 230.367 | 2.961 | 3.695 | 37.916 | 2861.211 |
| CER Ratio | 0.145 | 0.234 | 1.640 | 1.601 | 0.126 | 0.230 | 0.663 | 35.894 |
| F&V Ratio | 0.013 | 0.100 | 8.616 | 75.599 | 0.046 | 0.199 | 4.304 | 16.917 |

*Source: Authors' elaboration on FADN data.*

Finally, note that for all the policy variables and both farmland and non-land values, the skewness is always positive as expected, since their distributions are truncated at zero, and some (e.g. $RDP_{inv}$) are zero-inflated. In the total sample, the support provided by agricultural policy is approximately 32% of the farm income (on average) (Table 3). Approximately three-quarters of this support is provided as $DDP$, whereas $CDP$, $RDP_{aes}$, $RDP_{other}$ and $RDP_{inv}$ account for lower shares of the overall support, ranging from 11% to 4%.

**Table 3 – Relative importance of the support provided by the different policy measures in terms of farm net income and as a share of the overall $CAP$ support.**

**Average values, 2008–2014[20].**

| Type of Measure | % Support/FNI | % Support/CAP | % Support/FNI | % Support/CAP |
|---|---|---|---|---|
| | Total Sample (74,010 Obs.) | | Small Farms (24,664 Obs.) | |
| $CDP$ | 2.71% | 8.39% | 1.46% | 3.62% |
| $DDP$ | 23.86% | 73.91% | 29.19% | 72.15% |
| $RDP_{aes}$ | 1.81% | 5.62% | 2.46% | 6.08% |
| $RDP_{inv}$ | 1.51% | 4.69% | 3.03% | 7.50% |
| $RDP_{other}$ | 2.39% | 7.40% | 4.31% | 10.66% |
| $CAP$ | *32.29%* | *100.00%* | *40.45%* | *100.00%* |
| | Medium Farms (24,673 Obs.) | | Large Farms (24,673 Obs.) | |
| $CDP$ | 2.13% | 5.91% | 3.27% | 11.39% |
| $DDP$ | 23.12% | 72.50% | 21.60% | 75.30% |
| $RDP_{aes}$ | 2.38% | 6.62% | 1.40% | 4.87% |
| $RDP_{inv}$ | 1.41% | 3.90% | 1.22% | 4.26% |
| $RDP_{other}$ | 3.99% | 11.07% | 1.20% | 4.15% |
| $CAP$ | *36.02%* | *100.00%* | *28.68%* | *100.00%* |

*Source: Authors' elaboration on Italian FADN data.*

The relative importance of the overall $CAP$ support markedly declines when moving from small to large farms, indicating a high dependency of small farms to policy support. $DDP$ represent, on average, 24% of farm income, reaching almost 30% for small, and 22% for large farms. Furthermore, the measures' relative importance slightly differs according to farm size. $RDP_{aes}$ are more important for small and medium farms (2.5% and 2.4%, respectively) than for large farms (1.4%). $RDP_{other}$ represent 2.4% of the income in the total sample and are limited in large farms (1.2%). $CDP$ amounts to only 2.7% of farm income in the total sample and even less in small farms (1.5%). Finally, $RDP_{inv}$ amounts to only 1.5% of farm income on average, and its relative importance decreases when moving from small to large farms.

---

[20] A Wilcoxon test was used to evaluate the significance of differences among measures and subset ratios. Only $CDP/FNI$ between Total and Medium farms; $DDP/FNI$ between Small and Large, and Medium and Large farms; and, $RDP_{aes}/CAP$ between Small and Large farms were non-significant. All the results are available upon request.

### 2.4.3 Estimation Method

Several reasons justify the use of the $SYS-GMM$ estimator (Blundell and Bond 1998), both from theoretical and empirical point of view. This model is designed for situations where there exists an individual fixed effect (hereafter $FE$), as in our case, which is removed by a first-difference transformation (Roodman 2009), accounting for all time-invariant farm-specific characteristics (e.g. farm specialisation, farmers' managerial abilities, and different natural characteristics of the soil) (Olper et al. 2014; Ooms and Peerlings 2005). Moreover, the inclusion of one and more lags of the dependent variable accounts for the dynamic nature of farm income (as detailed previously, decisions are taken at $t-1$, and income persistence is explained by the time required for adjusting to any changes occurred in the economic and biophysical environments and the degree of income volatility) and for omitted variables (Dithmer and Abdulai 2017). The inclusion of time dummies removes time-specific effects (i.e. shocks) common to all farms (see, e.g., Dithmer and Abdulai, 2017; Zhengfei and Oude Lansink, 2006), preventing contemporaneous correlation among farms, which is the most likely form of cross-individual correlation (Roodman 2009). Finally, the $SYS-GMM$ approach permits using the own-lagged variable as an instrument, offering a suitable solution for endogeneity issues. In the present analysis, $CAP$ subsidies are not assigned randomly and could sometimes depend on farmers' choices, making them potentially correlated to the error term (Mary 2013a; Michalek et al. 2011).

## 2.5 Results and Discussion

### 2.5.1 The overall validity of the estimation results

Four groups of specification tests served as the basis for a robust and comprehensive application of the $SYS-GMM$ model and its structure[21] to demonstrate the overall econometric validity of the results (Table 4, lower panel).

First, the autoregressive coefficient of the dependent variable in the $SYS-GMM$ estimation lies between the $OLS$ (upward-biased) and the $FE$ (downward-biased) estimated coefficients (Deconinck and Swinnen 2015; Haile, Kalkuhl, and Von Braun 2016), indicating a satisfactory goodness of fit for the model[22]. Second, the $AR(1)$ test indicates that autocorrelation of order 1 exists, while this is not true for the $AR(2)$. This is required for the correct specification of the $SYS-GMM$ (Deconinck and Swinnen 2015; Roodman 2009). Third, the Sargan test for the suitability of the instruments fails to reject the null hypothesis of exogeneity of instruments (over-identifying restrictions), showing that the set of chosen instruments is valid. Fourth, the results from the Wald tests suggest the satisfactory specification of the model. Finally, the $R^2$ values are also quite satisfactory (see Bloom et al. (2001), and Windmeijer (1995) for the calculation of the $R^2$), with the lowest value found in the Small farm model. To further show the unbiasedness of the estimated results, potential under-identification problems of the $SYS-GMM$ models due to weak instruments have been investigated (Vigani and Dwyer 2019). This was done by utilising the Sanderson and Windmeijer (2016) test[23], which, in general, rejects the null hypothesis of under-identification for all the variables.

---

[21] See the on-line Appendix for more details on the econometric specification of the model.

[22] $OLS$ estimation is biased due to the correlation between the lagged dependent variable and the fixed effects in the error term, which is known as the dynamic panel bias (see Nickell, 1981 for more details). The logical solution, applying the $FE$ estimator, does not overcome the dynamic panel bias, since the regressors and errors are still correlated, generating a downward bias in the estimates (Roodman 2009). Thus, as indicated by Bond (2002), a good rule of thumb for establishing the reasonableness of the results from a $SYS-GMM$ estimation is that estimates from the latter should lie between the two bounds. Pooled $OLS$ and Two-way $FE$ results are available upon request.

### 2.5.2 *Estimates from the model based on the total sample*

In the model based on the total sample, the estimated coefficients, when significant, have the expected sign (Table 4). First, the autoregressive coefficients are both positive and significant, indicating a moderate-income persistency[24]. The first autoregressive coefficient is equal to 0.183, indicating to what extent the farm income at $t-1$ persists at time $t$. Such limited income persistency is consistent with the diverse nature of farm shocks, which increase the profit variability hence uncertainty. Indeed, the magnitude of the estimate is in line with what Hirsh and Gschwandtner (2013) find regarding Italian food industries, but larger than the value in Michalek et al. (2011), estimated on the entire EU instead of one specific country.

Likewise, coefficients related to non-policy variables generally exhibit the expected signs. In particular, the model returns positive estimates for the amount of both land and non-land capital. The policy-related coefficients at time $t$ are all significant and positive, ranging from 0.312 (for $RDP_{inv}$) to 1.102 (for $RDP_{other}$). The relative magnitude of the standard error of each policy coefficient indicates the existence of differences among policy measures. $CDP$ present large standard errors, consistent with the nature of the measure: $CDP$ support is provided only for specific activities and farms, and the level can change over time within the same farm, according to the change in production patterns. In contrast, $DDP$ is relatively more stable over time, being affected only by changes in the amount of eligible land and entitlements, but not by changes in the production mix.

---

[23] See Windmeijer (2018) for the implementation of the first test. Windmeijer (2019) criticised the validity of Bazzi and Clemens's (2013) approach for under-identification, recommending the Sanderson and Windmeijer (2016) test for dynamic panel models. Results are available upon request.

[24] Other scholars (see, e.g., Hirsch and Gschwandtner, 2013) use the term persistency when relying on dynamic models, as the estimate of the lagged dependent variable(s) indicates what part of the economic performance at stake (farm income in the present analysis) is passed-through from previous time lags.

**Table 4 – Estimation results of the SYS-GMM model. Total sample and small, medium, and large farm subsample models.**

|  | Total Sample | Small Farms | Medium Farms | Large Farms |
|---|---|---|---|---|
| $FNI_{t-1}$ | 0.183*** | 0.076** | 0.065** | 0.191*** |
|  | [0.032] | [0.035] | [0.021] | [0.048] |
| $FNI_{t-2}$ | 0.033*** | 0.013 | 0.005* | 0.025 |
|  | [0.013] | [0.008] | [0.002] | [0.020] |
| $CDP_t$ | 0.470* | 0.203 | 1.142 | 0.547* |
|  | [0.268] | [0.673] | [2.373] | [0.294] |
| $CDP_{t-1}$ | -0.209* | -0.292 | 0.052 | -0.355*** |
|  | [0.121] | [0.213] | [0.680] | [0.133] |
| $DDP_t$ | 0.928*** | 0.893** | 0.342 | 1.100*** |
|  | [0.128] | [0.452] | [0.342] | [0.167] |
| $DDP_{t-1}$ | -0.204** | -0.112 | 0.189 | -0.432*** |
|  | [0.092] | [0.167] | [0.195] | [0.131] |
| $RDP_{aes_t}$ | 0.600** | 0.666** | 0.198 | 0.976*** |
|  | [0.257] | [0.273] | [0.312] | [0.296] |
| $RDP_{aes_{t-1}}$ | -0.197 | -0.507*** | -0.111 | -0.498** |
|  | [0.194] | [0.185] | [0.172] | [0.248] |
| $RDP_{inv_t}$ | 0.312*** | 0.016 | 0.327*** | 0.426*** |
|  | [0.092] | [0.126] | [0.095] | [0.105] |
| $RDP_{inv_{t-1}}$ | 0.056 | -0.002 | 0.096 | 0.088 |
|  | [0.061] | [0.077] | [0.075] | [0.065] |
| $RDP_{other_t}$ | 1.102*** | 0.935*** | 1.241*** | 1.015*** |
|  | [0.203] | [0.310] | [0.287] | [0.288] |
| $RDP_{other_{t-1}}$ | 0.132 | -0.087 | 0.025 | 0.543* |
|  | [0.143] | [0.172] | [0.180] | [0.305] |
| $FAML_t$ | 4404.060 | -1142.919 | 15179.547* | 59393.341*** |
|  | [5734.300] | [8140.302] | [6595.720] | [15736.966] |
| $FAML_{t-1}$ | -9186.579*** | -3966.141 | 2820.252 | -28259.251*** |
|  | [3202.034] | [3540.357] | [3028.357] | [10687.842] |
| $LV_t$ | 0.023*** | 0.020*** | 0.023* | 0.010 |
|  | [0.008] | [0.008] | [0.010] | [0.011] |
| $LV_{t-1}$ | -0.001 | -0.005* | -0.004* | 0.008 |
|  | [0.002] | [0.003] | [0.002] | [0.011] |
| $Non\ LV_t$ | 0.017* | -0.011 | -0.013 | -0.004 |
|  | [0.010] | [0.017] | [0.014] | [0.011] |
| $Non\ LV_{t-1}$ | 0.002 | 0.020*** | 0.005 | 0.001 |
|  | [0.002] | [0.006] | [0.005] | [0.002] |
| $Price\ Ratio_t$ | 493.185 | 98.231** | 2563.059*** | 5526.357*** |
|  | [314.612] | [49.811] | [320.917] | [987.668] |
| $Price\ Ratio_{t-1}$ | 26.387 | 6.376 | -369.218* | -1865.361 |
|  | [40.233] | [14.769] | [170.564] | [1154.712] |
| $CER\ Ratio_t$ | -13410.958*** | -8227.088** | 4760.887 | 419.922 |
| | [4717.069] | [3599.189] | [4325.481] | [9265.048] |
| $CER\ Ratio_{t-1}$ | -9630.778*** | -2532.172 | 1587.628 | -12050.310** |
| | [2494.971] | [1612.817] | [2526.822] | [5014.529] |
| $F\&V\ Ratio_t$ | 6819.487** | 5690.012** | 7069.816 | 4555.486 |
| | [3142.037] | [2660.655] | [4463.530] | [9156.319] |
| $F\&V\ Ratio_{t-1}$ | -2250.426 | 7.967 | -179.891 | -2822.769 |
| | [1940.591] | [1520.780] | [2438.921] | [5870.981] |
| $Leverage_t$ | -752.323 | -2906.029 | -366.762 | 3499.460 |
| | [578.947] | [2449.650] | [609.696] | [4404.325] |
| $Leverage_{t-1}$ | -943.467 | -2387.542 | 96.245 | 3799.848 |
| | [676.885] | [3006.030] | [440.019] | [2695.491] |
| $Constant$ | 12879.006** | 8808.512 | -12493.149* | -7624.822 |
| | [5313.889] | [9643.683] | [5742.336] | [9319.996] |
| Individual FE | Yes | Yes | Yes | Yes |
| Time FE | Yes | Yes | Yes | Yes |
| Collapsed IV[b] | Yes | Yes | Yes | Yes |
| $R^2$ | 0.341 | 0.226 | 0.380 | 0.386 |
| Sargan[c] | 37.882(0.153) | 37.244(0.458) | 22.711(0.360) | 25.441(0.280) |
| AR(1) | -6.601(0.000) | -2.770(0.006) | -6.627(0.000) | -4.540(0.000) |
| AR(2) | 1.429(0.153) | -1.536(0.124) | -0.241(0.810) | 0.264(0.190) |
| Wald for coefficients[d] | 638.532(0.000) | 98.629(0.000) | 214.021(0.000) | 235.767(0.000) |
| Wald for Time dummies[e] | 38.956(0.000) | 10.923(0.027) | 19.237(0.001) | 0.815(0.041) |

**Notes:** Significance codes for p-values: *** ≤ 0.01; ** ≤ 0.05; * ≤ 0.10. Robust standard errors in square brackets, p-values in round brackets.
[a] The IV Lag is specific for each model and all the tests for the correct setting of Dynamic-GMM must be satisfied. In general, the IV lag is for FNI from 2 to 5 and for independent variables from 1 to 5. The data are available on request. [b] See Roodman (2009) for the thorough explanation of collapsed instrument matrix. [c] Degrees of freedom are 30 for Total, 37 for Small, 21 for Medium, 38 for Large [d] Degrees of freedom are 22 for all models. [e] Degrees of freedom are 4 for all models

*Source: Authors' elaboration on Italian FADN data.*

To ensure a more thorough and statistically-robust analysis regarding the effect of the measures, both estimated coefficients at time $t$ and $t_{-1}$ should be jointly considered for a more comprehensive assessment of the $ITE$, describing short and long-run effects ($SR$ and $LR$). These are calculated (Table 5) accounting for the dynamic nature of the model (see, e.g., Dithmer and Abdulai, 2017; Olper et al., 2014). The $SR$ effect refers to the effect of the policy instrument in time $t$ and $t-1$ and, using the notation of formula (1), is specified as $\beta_k + \gamma_k$; the $LR$ is specified as $\frac{\beta_k + \gamma_{k'}}{1 - \alpha_1 - \alpha_2}$, characterising cumulated effects of a change in support across all subsequent years (Pesaran, 2015). Being both $SR$ and $LR$ are a nonlinear combination of estimates, the delta method provides an approximation of standard errors through a Taylor approximation (Croissant and Millo, 2018; Weisberg, 2005)[25].

---

[25] This explains why the statistical significance of policy estimates in Table 4 is not directly comparable with that of SR and LR estimates in

**Table 5 – Short and long-run income effects of $CAP$ measures from SYS-GMM estimation (Total sample and small, medium, and large farm subsample models).**

| | \multicolumn{4}{c}{**Short-Run**} | | | |
|---|---|---|---|---|
| | *Total* | *Small* | *Medium* | *Large* |
| $CDP$ | 0.261 | -0.089 | 1.194 | 0.193 |
| $DDP$ | 0.725*** | 0.781** | 0.531** | 0.668*** |
| $RDP_{aes}$ | 0.403** | 0.159 | 0.087 | 0.478 |
| $RDP_{inv}$ | 0.369*** | 0.014 | 0.424*** | 0.514*** |
| $RDP_{other}$ | 1.235*** | 0.848** | 1.266*** | 1.558*** |
| | \multicolumn{4}{c}{**Long-Run**} | | | |
| | *Total* | *Small* | *Medium* | *Large* |
| $CDP$ | 0.333 | -0.097 | 1.284 | 0.246 |
| $DDP$ | 0.924*** | 0.858** | 0.571** | 0.852*** |
| $RDP_{aes}$ | 0.514** | 0.175 | 0.093 | 0.610* |
| $RDP_{inv}$ | 0.470*** | 0.015 | 0.456*** | 0.655*** |
| $RDP_{other}$ | 1.575*** | 0.931** | 1.362*** | 1.987*** |

Significance codes for p-values: *** ≤ 0.01; ** ≤ 0.05; * ≤ 0.01

*Source: Authors' elaboration on Italian FADN data.*

The resulting $SR$ effects imply that the measures have a positive marginal impact on farm income in all cases except for $CDP$, for which the estimated coefficient is not significant (Table 5). Nevertheless, levels smaller than one suggest leakages, as extensively shown in the literature (see section 2.3).

When concerning the total model, the discussion is based on the $SR$ only. Indeed, the $LR$ effect is merely a transformation of the $SR$ effect as it depends on the estimated lagged coefficient of $FNI$. Interestingly, $ITE$ differs according to the considered policy measure, and such differentials seem to depend on the characteristics of each measure, especially participation costs. As expected, the highest efficiency is found for $RDP_{other}$ and, to a lesser extent, $DDP$. The level of $ITE$ declines for the remaining measures, with $RDP_{aes}$ and $RDP_{inv}$ and, finally, $CDP$, which is not significant (Table 5). The very high level of $ITE$ for $RDP_{other}$ reflects the fact that the main component of this $RDP$ element is represented by $LFA$ payments. These payments are paid on top of $DDP$, since every beneficiary of $LFA$ payments must already have qualified for $DDP$, once the participation costs related to $DDP$ are incurred, no further additional costs exist for $LFA$ payments[26]. The high degree of $ITE$ of $DDP$ payments is also consistent with the limited capitalisation of such subsidies into land values and rents. As described above (section 2 and 3), the adoption of the historical model, together with the abundancy of eligible hectares compared to the volume of entitlements, determines the low capitalisation rate recently found by Guastella et al. (2018) and Valenti et al.

---

Table 5.

[26] We are grateful to an anonymous referee for suggesting this sensible explanation.

(2020). Furthermore, generally, participation costs concerning this subsidy are lower than those of other measures. The fact that the $SR$ effect of $RDP_{aes}$ is lower than that of $DDP$ seems consistent with the fact that participation costs differ, with stricter requirements for receiving $RDP_{aes}$ if compared to those imposed by $DP$-related cross-compliance. The non-significant $SR$ effect of $CDP$ may relate to the additional costs a farmer incurs for growing the crop or raising the type of livestock that allows him or her to receive the subsidy.

Due to its diverse nature, the limited $ITE$ of $RDP_{inv}$ requires a more complex explanation. On the one hand, the effect may stem from the fact that this measure is associated with relatively high transaction costs that cause a low-efficiency level. However, on the other hand, one can consider the effect this measure exerts on the farm's investment decisions: for certain farms, investments become profitable only because of the provision of $RDP_{inv}$. Therefore, in such cases, a part of the support is used to cover the (negative) difference between investment benefits and costs, and the higher the number of farms in this situation, the lower the average $ITE$ of this measure will be.

Comparing these results with those found in the literature, some similarities, as well as differences, arise. Michalek et al. (2011) and Ciaian et al. (2015) find a high-efficiency level for an aggregated $RDP$. Similarly, $RDP_{other}$ shows the highest $ITE$ level in our results, Michalek et al. (2011) disaggregated $RDP$ into three different measures, namely, an $RDP$-environment-related variable (similar to $RDP_{aes}$ in the present analysis), $RDP$ measures intended to support investments, and $LFA$ payments, which in our case is the main component of the aggregated variable $RDP_{other}$ (i.e., 76%). Interestingly, they found a similar ranking: $LFA$ payments showing the highest efficiency, followed by environment-related payments and, finally, $RDP$ measures supporting investments. On the contrary, when analysing the effects of $DDP$ and $CDP$, the literature shows mixed results. Although, generally speaking, one may expect larger efficiency for $DDP$ compared to $CDP$, (see Dewbre et al., 2001 and Guyomard et al., 2004), Michalek et al. (2011) and Ciaian et al. (2015) find a significant and similar $ITE$ for these two types of payments. However, the results presented here are in line with the theory that $CDP$ may have a limited $ITE$ due to leakage to non-farm agents. This supports the idea that, despite being no longer directly linked to production levels, $CDP$ still entail large adaptation costs, and potentially reduce output prices (Guyomard et al., 2004). Finally, one should bear in mind that Michalek et al. (2011) and Ciaian et al. (2015) referred to a larger group of Member States and to an earlier period (i.e. before 2008).

### 2.5.3 *Estimates of the models for small, medium and large farms*

Differences exist between the three models in terms of $ITE$ in the $SR$ and, since the level of income persistence of the three sub-groups differs, also in the $LR$ (Table 5). For $RDP$ measures, the general pattern is that $ITE$ increases with farm size. This pattern does not hold in the case of $DDP$, where the $SR$ effect for small farms the highest. However, in the long run ($LR$), the efficiency level for $DDP$ between small and large farm disappears. Interestingly, medium farms show the lowest level of efficiency. The $SR$ effects of $RDP_{aes}$ is significant only when the total sample is concerned, while large farms display a significant effect in the $LR$. In contrast, both the $SR$ and $LR$ effects of $RDP_{inv}$ are significant only for medium and large farms, with the latter exhibiting the highest $ITE$. Although results from the total sample ranks $RDP_{aes}$ as slightly more income transfer efficient than $RDP_{inv}$, the opposite is true when considering the three groups of farms, where the estimate of the $RDP_{aes}$ is non-significant in almost all specifications.

These results are interesting, as the $ITE$ of the same measure can strongly differ according to farm size. In particular, $DDP$ and $RDP_{others}$ seem able to transfer support to small farms efficiently, with both $RDP_{aes}$ and $RDP_{inv}$ showing no significant effects. The relatively higher level of $DDP$ $ITE$ in small farms concerning $DDP$ suggests that these farms face lower participation costs than the other two farm categories. Cross-compliance costs have been found to be relatively greater for farms specialized in livestock, generally large in economic size and, hence, not well represented within the small farm group (Rete Rurale Nazionale, 2010). A further explanation of the high $ITE$ of $DDP$ in the small farms might be the reduced participation costs linked to the simplified procedures for direct payments.

Economies of scale seem to satisfactorily explain the lower efficiency featured by medium farms concerning $DDP$. Such group shares a more similar composition in terms of farm specialisation of that of larger farms, but, on the other hand, the extent of scale economies is reduced, lowering the efficiency of decoupled direct payments.

A quite divergent situation is observed when considering $RDP_{aes}$ and $RDP_{inv}$. $RDP_{aes}$ exhibits a non-significant $ITE$ in the short run, turning significant for large farms in the long-run. Environmental requirements impose costs that offset the benefits derived from the provided payments for both small and medium farms, but not for large farms.

Similar considerations apply to $RDP_{inv}$: the non-significant effect found for small farms could be explained by a relatively high amount of transaction costs, the limited private profitability of the supported investments or by a combination of the two factors. In contrast, $RDP_{inv}$ in medium and large farms are associated with significant $ITE$ levels. Nonetheless, on average, the effect of $RDP_{inv}$ even for large farms remains far below the unity (i.e., 0.514 and 0.655 for short- and long-run, respectively).

 Results show that not all the CAP measures translate into an equivalent change in farm income. Given the intense pressures to reduce the financial resources allocated to the $CAP$, it is crucial to consider the overall efficiency of the policies, including their $ITE$. The analysis has provided evidence that $RDP$ measures other than $RDP_{aes}$ and $RDP_{inv}$ are the most efficient in terms of income transfer. However, these measures are dominated by $LFA$ payments, whose participation costs are negligible. $RDP_{inv}$ and $RDP_{aes}$ resulted in lower levels of $ITE$ compared to $DDP$: this is consistent with the fact that these measures pursue objectives other than income support, besides being characterized by higher participation costs. Finally, in line with theoretical predictions and part of the reviewed literature, the income transfer efficiency of $CDP$ is not significant.

## 2.6 Conclusions

This study is relyed on in-depth analysis to explain the Income Transfer Efficiency regarding dynamicity aspects and economic farm dimensions.

We have based our conclusions on the solid econometric framework, considering the time and individual effect and dynamicity of behaviour (i.e., persistence). We have shown that some $CAP$ measures do not translate equally into the same amount of changes in farm income.

This point is paramount, in particular, when these outcomes are sees related to the intensifying pressure to cut $CAP$ budget. For this motivation, the policymakers need to have a robust methodology to set their decisions and regarding, it is crucial to consider the overall efficiency of policies, including their ITE.

Preliminarily the analysis has shown that some measures (i.e., $DDP$ and $RDP_{other}$) are more income-transfer efficient than others. Reducing the share of the budget allocated to these measures is expected to result in a decline in the overall income transfer efficiency of $CAP$ support.

On the other hand, we cannot analyse the $RDP$ measure such as a unique set; indeed, the outcomes show that some measures are most efficient in terms of transfer than others, but this conclusion diverges if relying on the economic dimension of the farm. In particular, lower $ITE$ have been found to $RDP_{inv}$ and $RDP_{aes}$ in comparison, particularly with $DDP$.

Finally, compared to theoretical predictions and the reviewed literature, CDP income transfer efficiency is not significant.

The findings have consequences for potential CAP revisions. Even if with a lower level of resource would also improve the effectiveness of the CAP, and at the same time would increase the efficiency of all programs, not just their. The Multiannual Financial System plan's discussion indicates that the financial expenditure for the second pillar would decline rather than the first (Matthews, 2018). This is expected to have negative consequences on the achievements of agri-environmental objectives and on-farm investments. However, the impact on the farmers' income would primarily rely on how the required expenditure will be distributed.

In particular, considering that the level of $ITE$ is lower than unity leads to say that the increase in efficiency is tight, specifically for $CDP$ and $DDP$.

Regarding $RDP$, the outcomes show that the efficiency depends on the farm size, in particular, if we relied on $RDP_{aes}$ and $RDP_{inv}$.

These results are awe-inspiring, as it undoubtedly shows that the improvement of the $CAP$ must necessarily be based on the assessment of economic dimensions. We hypothesized that these effects depend on the cost of the transactions, the bureaucratic costs (to access some measures is mandatory to have technical support) or the minimum level of expense (in particular for $RDP_{inv}$ where the beneficiary must make a minimum amount of investments, this discrimination for small framer represents another hurdle that one sums with bureaucracy costs).

Conversely, but in line with our hypothesis that decreasing bureaucracy and costs of access and transition leads to an increase in $ITE$, the outcome found that $DDP$ is highly efficient for small companies compared to large ones. $DDP$ is characterised to have a lower level of bureaucracy cost and a few requests to satisfy (compliance costs) in comparison with $RDP$ measures.

Finally, it is crucial to spot the light on the "persistency" or the feature to some measure to maintain his effects in the following years. This effect is due to autoregressive coefficient ( or persistency) that shows how the larger farmers, over than receive a better condition to accessing some $RDP$ measure, thanks to his income that is not oscillating, can achieve a better cumulate response of the measure designed to increase the income. This evidence shows the pivotal role of the farm's economic dimension on the CAP's measures design

# Chapter 3. The Implementation of the Income Stabilization Tool in Italy

The IST is a measure not yet applied in Europe but suggested by the EU to reduce the problems of income volatility. Considering that it is still under development, the latest concept of the IST is used for designing the study, with two-level approach has been developed: first, we check if the tool is efficient in stabilizing income (section 3.1); second, we identify/explain how to build the tool effectively and efficiently (section 3.1);.

In section 3.1, it's investigated the potential impact of the income stabilisation tool (IST) in economic farmer outcomes and to Mutua Fund, mainly devoted by the European Common Agricultural Policy to reduce farmers' income risks and use Italian agriculture as a case study. The paper extends the existing literature by investigating the effects of two implementation issues: level of aggregation of mutual funds (MF); definition of farmers' contribution (i.e. premium) to MF. We use a simulation approach based on a FADN panel data set of 3421 farms over seven years to investigate effects on i) farm-level income variability, ii) the expected level and variability of indemnifications at the level of mutual funds, and iii) the distribution of affordability from this policy instrument across the farm population. We find that the introduction of the IST would lead to a significant reduction of income variability in Italian agriculture[27]. Our results support establishing a national mutual fund due to the high volatility of indemnification levels at more disaggregated (e.g. regional or sectoral) levels. Besides, our results propose that farmers' contribution to mutual funds, i.e. premiums paid, should be modulated according to farm size. This reduces the inequality of the distribution of benefits of such tool within the farm population.

Section 3.2 reported four different approaches to design IST and compared the results with a multi-objective analysis. Predicting indemnity levels is essential for insurance ratemaking but challenging: indemnity distribution is zero-inflated, not-continuous, right-skewed, and several factors can potentially explain it. We address these problems by using Tweedie distributions and three machine learning procedures. The objective is to assess whether this improves the ratemaking by using the prospective application of the Income Stabilization Tool in Italy as a case study. We look at the econometric performance of the models and the impact of using their predictions in practice. Some of these procedures efficiently predict indemnities, use a limited number of regressors and ensure the financial stability of the scheme.

Section 3.1 reports a paper published in Journal of Policy Modelling written in collaboration with Simone Severini and Robert Finger (ETH Zurich) (" Severini S, Biagini L, Finger R. Modeling agricultural risk management policies – The implementation of the Income Stabilization Tool in Italy. *J Policy Model* 2019;**41**:140–55").

---

[27]The amount of funding dedicated by the Member States is not considered in this analysis. The purpose of this survey concerns a self-financed IST hypothesis.

Section 3.2 reports a working paper accepted for 9th AIEAA Online Conference "Mediterranean agriculture facing climate change: Challenges and policies" , EAAE Webinar 3: "Challenges and opportunities of machine learning" and for XVI EAAE Congress "Raising the Impact of Agricultural Economics: Multidisciplinary, Stakeholder Engagement and Novel Approaches" written in collaboration with Simone Severini, Nadja El Benni (Agroscope – Switzerland) and Robert Finger (ETH Zurich – Switzerland) (" Severini S, Biagini L, El Benni N and Finger R. "Applications of Machine Learning for the Ratemaking of Agricultural Insurances").

## 3.1 Modeling Agricultural Risk Management Policies

### 3.1.1    Introduction

Farming is a risky business because forces beyond the control of farmers affect their income (Mishra and Sandretto 2002). More specifically, farmers' income reveals a high variability due to volatile product prices and in yields, as well as due to the occurrence of catastrophic risks such as natural disasters and animal/plant diseases. The resulting income instability negatively affects farmers' well-being and their decisions, their ability to expand operations and repay debt and, in turns, this can also have secondary effects on agribusiness firms and creditors (Mishra and El-Osta 2001; Mishra and Holthausen 2002; Mishra and Sandretto 2002; Vrolijk et al. 2009; Vrolijk and Poppe 2008). Thus, farm income and farm income distribution are of high importance for policymakers (El Benni and Finger 2013; Liesivaara and Myyrä 2016, 2017; Simone Severini and Tantari 2013). Several instruments have been proposed to support farmers in coping with (increasing) income risks (Diaz-Caneja *et al.* 2008; Meuwissen, Assefa and van Asseldonk 2013 for examples and overviews). Indeed, agricultural policies both in the US and European countries have provided disaster relief by *ad hoc* measures to cover (*ex-post*) catastrophic risks, product price stabilisation policies and direct support (e.g. the Common Agricultural Policy (CAP) direct payments(S. Severini, Tantari, and Di Tommaso 2016; Simone Severini, Tantari, and Di Tommaso 2016)).

Recent policy developments have been characterised by a shift away from direct public intervention towards the support of public-private partnerships that help farmers to cope with the risks they face. National governments aim to reduce the provision of *ad hoc* measures to cover catastrophic risks because this of the unpredictable nature of the costs of such measures and the growing financial constraints faced in recent years. Recent CAP reforms have reduced the role of price stabilisation measures. Moreover, next CAP reform steps will likely reduce the decoupled

-42-

support provided by direct payment policies. All these developments are increasing EU farmers' income risks, which is furthermore expected to be amplified by increasing risk exposure due to climate change (Reidsma et al. 2010). However, the recent reform of the CAP has introduced three different risk management measures that are based on the developments of public-private partnerships within the framework of the EU Rural Development Policy (RDP) [28]. First, insurance premium subsidies have been introduced (already 2009, (Liesivaara and Myyrä 2017)). Second, mutual funds are supported (Meuwissen, Assefa, and van Asseldonk 2013). Third, the so-called Income Stabilisation Tool (IST) has been introduced (see (El Benni, Finger, and Meuwissen 2016), for details).

This latter tool explicitly aims at smoothing the farm income variability between years. Hence, in comparison with yield insurances, it focuses not just on production risk but on the overall farm income offering, at least potentially, large scope for managing income risk.

The IST provides compensation to farmers who experience a severe income drop, i.e. income decreases larger than 30% from the expected income.

The interest of policy-makers for the IST is due to four major reasons. First, farmers protection under the IST focuses on the key variable of interest, i.e. income. This represents the economic wellbeing of a farm household much better than revenues of a single commodity and implicitly accounts for various correlations between prices and yields and across profits of different farm activities (Meuwissen et al. 2003; Simone Severini, Tantari, and Di Tommaso 2016). Second, IST is in agreement with WTO green-box requirements (Mary, Santini, and Boulanger 2013). Third, IST has the potential to cover also systemic risks (specifically price risk) that are not covered by purely commercial insurances hampering the principles of risk pooling (Meuwissen et al. 2003).Fourth, the tools support a public-private partnership because farmers must be organised in mutual funds (MF) and have to cover part of the indemnity costs (i.e. at least 35% of these) as well as management costs incurred by the MF.

There is now a large amount of explorative research on the farm-, sector- and country-level effects of IST. This literature focuses on actuarial evaluations of a potential income insurance, its governmental costs, potential beneficiaries within the farm population as well as conceptual studies on problems of adverse selection and moral hazard with such whole-farm income insurance tools (El Benni, Finger, and Meuwissen 2016; Dell'Aquila and Cimino 2012; Finger and El Benni 2014a, 2014b; Liesivaara et al. 2012; Liesivaara and Myyrä 2016; Mary, Santini, and Boulanger 2013; Pigeon, Henry de Frahan, and Denuit 2014)

---

[28] Regulation (EU) No 1305/2013 of the European Parliament and of the Council of 17 December 2013 on support for rural development by the European Agricultural Fund for Rural Development (EAFRD) (OJ L 347, 20.12.2013, p.487).

This research has shown that the introduction of such tool: i) stabilises farm-incomes (Finger and El Benni 2014a), ii) affects the income inequality within the farm population (Finger and El Benni 2014a), iii) the benefits from such tool might be highly heterogeneous across farm types (El Benni, Finger, and Meuwissen 2016), iv) causes highly volatile levels of indemnification payments, requiring large buffers (Pigeon, Henry de Frahan, and Denuit 2014) v) might cause large transaction costs (Liesivaara et al. 2012) vi) indemnification patterns are highly dependent on the calculation of the reference income (Finger and El Benni 2014b) vii) might cause moral hazard problems along value chains (Liesivaara and Myyrä 2016)

The conducted research, however, remained on conceptual levels, without explicit aspects on implementation issues. This observation is also due to the fact that, despite the potential appeal of this tool, IST has not been implemented yet in EU Member States. Now, however, first countries and regions declared interest in establishing IST: Italy, Hungary and the Spanish region Castilla y León (Bardaji and Garrido 2016). Important steps towards implementation that have not been addressed in empirical research so far are: i) the specification of aspects concerning the structure of mutual funds (MF) across sectors and space and ii) the specification of farmers' contribution to MF. These decisions could affect income stabilising properties, viability of mutual funds, income inequality in the agricultural sector and the distribution of benefits across space and farm types. These issues have, for instance in Italy, led to a postponement of the implementation of the IST (ISMEA 2015; Trestini et al. 2018; Trestini and Boatto 2015).

Opting against a single national mutual fund but focus on specific sectors and/or regions has several advantages. Already existing institutions (e.g. Producer Organisations) could be used to facilitate the establishment of a MF and increase the likelihood of an agreement among farmers. Moreover, such MF would have a more detailed knowledge about the participating farmers and could thus better manage moral hazard problems. Thus, there are substantially lower transaction costs expected for low aggregation levels. Finally, such disaggregation implies a limited redistribution of benefits (as well as costs) of the policy among different sectors and regions, which is an important aspect for the acceptance of such policy. A potential limitation when MF operate on a specific sector and specific regions, is that it may not be possible to effectively pool risks because the underlying risks are systemic. For example, when only a small region or a specific sector is addressed, all farms may be affected by the systemic weather risk or the same market risk. Under these circumstances, this causes large indemnification events within a MF, requires large buffers and/or re-insurance, which increases costs.

Another important issue is how farmers should pay the contributions, i.e. premiums, to the MF. Farms might contribute to the IST using flat-rate premium levels (El Benni, Finger, and Meuwissen 2016) or contribute according to income or risk levels. The choice on farmers' contribution will have implications for the costs and risk reducing effects of the IST.

We illustrate the effects of the specification of MF and farmers' contributions on the variability of farm income and the distribution of benefits of IST among the farm population using the example of Italian agriculture in a unique empirical analysis. Thus, we contribute an empirical analysis that feeds directly into ongoing policy debates in Italy on the implementation of the IST. Moreover, shedding light into these implementation aspects constitutes an important basis for policy decisions also in other countries.

The remainder of this paper is organised as follows. Section 2 describes methodology and data. Section 3 presents the results of the analysis. The final section 4 concludes and draws some policy implications.

### 3.1.2 Methodology

In our analysis, we use bookkeeping (i.e. farm accountancy data network, FADN) data for the entire Italian agricultural sector across a large set of farms and years that allows us to draw conclusions for the farm population at large and to investigate a large set of farm and farmers' characteristics that may influence the effects of the IST (Finger and El Benni 2014a). The methodology section first introduces the general framework of the IST; secondly presents the employed scenarios and thirdly addresses the assessment of the outcomes based on income variability and distributions of benefits.

**General framework of the IST.**

As in previous analyses, we assume participation in the IST to be mandatory (El Benni, Finger, and Meuwissen 2016; EC 2009; Finger and El Benni 2014a, 2014b; Liesivaara et al. 2012). Following the EU regulation, a farmer is indemnified if his/her income drops more than 30 per cent compared with the expected income level. For the i-th individual farm:

$I_{it}$ is the realised income at the t-th year

$E_{it}$ is the expected income at the t-th year and it is assumed to be the average of the realised incomes of the previous three years (see (Finger and El Benni 2014b) for discussions) as: $E_{it} = \frac{1}{3}\sum_{t-4}^{t-1} I_{it}$

$I_{Rit}$ is the reference income at the t-th year and it is calculated as: $I_{Rit} = \alpha \cdot E_{it}$ where α is equal to 0.7.

The indemnity paid in the t-th year for the i-th farm is:

$$Ind_{it} = \begin{pmatrix} 0 & if & I_i \geq I_{Rit} \\ \beta \cdot (E_{it} - I_i) & if & I_i < I_{Rit} \end{pmatrix} \quad (1)$$

where β is equal to 0.7. This partial compensation is supposed to further reduce moral hazard effects of the IST.

**Scenarios of the specification of the IST.**

Simulations are developed regarding the level of aggregation of the MF. In particular, it has been considered the case of: a) a single national-wide MF; b) MFs working on the 3 altimetry regions of Italy (Mountain, hill, plain); c) MFs working on 5 macro-regions of Italy (i.e. North-West, North-East, Centre, South, Islands); d) MFs working on 7 specific sectors (i.e. farms specialised in: fieldcrops; horticulture; permanent crops; grazing livestock; granivore livestock; mixed crop farms; mixed livestock, crops and livestock farms).

Considering different specifications of the MFs is important because different regions and types of farming face different risk profiles. For example, in the north of Italy, production risk due to hail is substantially higher than in the other regions. Furthermore, farms belonging to the considered regions generally differ in terms of structural characteristics and production orientation. For example, farms in the plains are larger than those in the other altimetry regions. Finally, farms belonging to the same type of farming face very similar market risks. Thus, mutual funds tailored to specific regions and types of farming are expected to face some systemic risks, which would imply large fluctuations of the paid indemnities over time.

The impact of the IST is assessed under two different assumptions:
a) IST is fully subsidised by the government (i.e. farmers do not pay neither indemnities, nor management costs of the MF)
b) Farmers pay contributions that are set to recover 35% of the indemnities paid by the MF[29].

The total indemnification paid by the MF in year t ($TInd_t$) is thus derived by means of a weighted sum of farm level indemnities:

$$TInd_t = \sum_{i=1}^{n} Ind_{it} \cdot w_i \qquad (2)$$

where $w_i$ are the statistical weights attached to each FADN farm (Indexed by $i$). These weights indicate how many similar farms can be found in the total population (EC, 2010) and are used to extrapolating the results to the entire country and to groups of farms[30].

Farm level contributions are calculated in two alternative ways. First, a flat rate contribution is chosen that represents premium calculation, for example, in the catastrophic crop insurance program in the USA (Shields 2015). As in Finger and El Benni (2014a) we simply divide the 35% of the total indemnification paid by the MF in each of the years by the sum of the weights of the sampled farms.

---
[29] The EU regulation allows to cover up to 65% of the paid indemnities by public funds.
[30] Criteria that define similarity include region, type of farm and economic size class (EC, 2010).

$$ContF_t = 0.35 \cdot TInd_t / \sum_{i=1}^{n} w_i \qquad (3)$$

$ContF_t$ is the flat rate contribution to be paid by the farmers to the MF. In this way, each farm pays exactly the same contribution in a specific year regardless its size and the probability to receive an indemnification.

Moreover, we also calculate a farm-specific contribution by distributing the total indemnification paid by the MF proportionally to the expected income of each farm as it follows:

$$ContE_{it} = (0.35 \cdot TInd_t / TE_t) \cdot E_{it} \qquad (4)$$

$ContE_{it}$ is for the contribution paid by each farmer in each year, and $E_{it}$ is expected income of farm i in year t. In this case, the contribution differs among farms even in the same year.

**Assessment of the effects of IST and test strategies**

The relative frequency of indemnification can be represented by mean of the ratio of farms receiving indemnification over the total number of farms. The relative extent of the indemnities paid can be expressed as an indemnity rate. This is the ratio of total paid indemnity over total expected income ($TInd_t/TE_t$) where $TE_t$ is the weighted sum of the expected income of all farms of the sample. The indemnity rate can be calculated also for specific groups of farms.

In order to conduct pairwise comparisons among the considered farm groups and configurations of MF, we used non-parametric bootstrap to derive confidence intervals. Following Payton et al. (2003), non-overlapping 83% confidence intervals can be used as indication for the rejection of the hypothesis of identical values in both groups at the 5% level of significance.

The analysis relies on 4 different income indicators (for each farm and year, indexes are omitted).

$I$          observed income (i.e. without IST)

$I_I = I + Ind$         (5)

$I_{IF} = I_I - ContF$         (6)

$I_{IE} = I_I - ContE$         (7)

The last three indicators refer to the application of the IST but under different rules. $I_I$ includes the received indemnification but does not subtract farmers' contribution to the MF. This is an hypothetical situation in which the policy covers all costs of the IST.

On the contrary, indicators $I_{IF}$ and $I_{IE}$ include the received indemnifications (as $I_I$) but are net of the farmers' contribution to the MF. This latter is paid according to the two considered criteria (Index *F* refers to *ContF* while index *E* refers to *ContE*) but both allow the recovery of 35% of the whole amount of the indemnities paid.

Income variability has been assessed by calculating the standard deviation (SD), the Median Absolute Deviation (from the median) (MAD), a more robust measure for dispersion, and Coefficient of Variation (CV) over the years 2011-2014 for each farm. Comparing the income variability of the 4 considered income indicators allows to assess the potential income stabilising effect of the IST under different implementation rules. To this end we compare CVs using non-parametric Wilcoxon tests (Wilcoxon 1945)

The distribution of the cost and benefits of the IST has been assessed only under the assumption that farmers pay contributions. This has been done by using the ratio between average indemnities received and the average contribution paid (over the considered four-year period) for each farm:

$$\text{DCB}_i = \frac{Avg(Ind)_i}{Avg(Cont)_i} \tag{7}$$

This ratio, that is zero for farms that have not received any indemnity, represents the financial benefit/cost ratio related to the IST[31]. Analysing the distribution of this indicator within the considered farms allows to assess the extent of the redistribution of the net benefits of the policy. An uneven distribution of this indicator suggests that a relatively large number of farms do not benefit from this policy while other farms enjoy a relatively high level of net benefits. This piece of information is important in the design of the policy because policy-makers are generally interested on how policy benefits are distributed among the population.

### 3.1.3   Data

Our analysis is based on a balanced sample of 3421 farms belonging to the Italian FADN consecutively in the year from 2008 to 2014 and not having negative average income[32]. The focus is on the Farm Net Value Added that measures the amount available for remuneration of the fixed production factors (work, land and capital) (EC 2010a). This indicator has been adopted by the European Commission and several analyses related to IST because it is the most comparable indicator between Member States and because it does not vary according to the relative importance of family own production factors (El Benni, Finger, and Meuwissen 2016; EC 2009; Pigeon, Henry de Frahan, and Denuit 2014). Income is measured on a per farm basis. All figures are deflated to allow comparability among data from different years by mean of the Eurostat Harmonised Index of Consumer Prices (HICP)[33].

---

[31] Another important benefit of the IST for risk adverse farmers is the reduction of the down side income risk generated by this tool.

[32] As in previous analyses, the exclusion of farms with negative average income (i.e. 84 farms over 3505 that is 2.4%) is motivated by the fact that it is not clear how such farmers should be handled and, in particular, how they should pay the contribution to the MF (Finger and El Benni, 2014a).

[33] Available at: http://ec.europa.eu/eurostat/web/hicp/data/database. Accessed on March 2017.

*3.1.4    Results and discussion*

**Frequency and expected level of indemnification**

We find that around 55% of the represented farms would receive at least one indemnification within the four considered years (Table 2) or around 21% per year in the considered period. This latter figure is higher than the one reported by El Benni et al. (2016) for Swiss agriculture.

More 30% of those farms receive only one indemnification and more than 16% exactly two indemnifications within the four years. On the contrary, only a very limited number of farms receive indemnifications in more than two of the considered years (Table 2).

**Table 1. Sampled and represented farms. Number of observations and frequency (%).**

|  | **Sampled farms** |  | **Represented farms** |  |
|---|---|---|---|---|
|  | N. of obs.s | Freq. | N. of obs.s | Freq. |
| **All observations** | 3421 | 100.0% | 186191 | 100.0% |
| **Altimetry regions:** | | | | |
| Mountain | 709 | 20.7% | 30806 | 16.5% |
| Hill | 1458 | 42.6% | 78154 | 42.0% |
| Plain | 1254 | 36.7% | 77231 | 41.5% |
| **Macro-regions (MR):** | | | | |
| Center | 383 | 11.2% | 14612 | 7.8% |
| Islands | 194 | 5.7% | 18583 | 10.0% |
| South | 760 | 22.2% | 50020 | 26.9% |
| North-West | 1251 | 36.6% | 49647 | 26.7% |
| North-East | 833 | 24.3% | 53329 | 28.6% |
| **Types of farming (TF)** | | | | |
| Specialised fieldcrops | 844 | 24.7% | 47142 | 25.3% |
| Specialised horticulture | 305 | 8.9% | 12392 | 6.7% |
| Specialised permanent crops | 1012 | 29.6% | 75824 | 40.7% |
| Specialised grazing livestock | 770 | 22.5% | 30975 | 16.6% |
| Specialised granivore livestock | 98 | 2.9% | 1848 | 1.0% |
| Mixed crops | 201 | 5.9% | 9529 | 5.1% |
| Mixed livestock, crops and livestock | 191 | 5.6% | 8481 | 4.6% |

*Source: Own elaboration on the Italian FADN sample.*

**Table 2. Beneficiaries of indemnities over the four considered years (2011 – 2014) in the whole sample of farms. Shares of represented farms (%).**

| Not indemnified | Indemnified farms | | | | |
|---|---|---|---|---|---|
| | At least in one year | of which: | | | |
| | | 1 year | 2 years | 3 years | 4 years |
| 44.8% | 55.2% | 31.8% | 16.6% | 6.1% | 0.7% |

*Source: Own elaboration on a constant sample of farms of the Italian FADN.*

With a single national MF, the overall amount of indemnities paid in the years 2011 – 2014 is around 9% of the expected income (i.e. indemnity rate) (second column of Table 3). Given that this figure refers to the overall represented population, this suggests that the IST could provide a not negligible financial support. However, this level differs among the considered regions and types of farming (Table 3). For example, indemnity rate is relatively high in the hill regions and in the islands of Italy. Differences are also found among farm types: farms specialised in horticulture have a relative level of indemnification higher than the average, while the opposite is true for specialised grazing livestock farms (Table 3). This confirms that different groups of farmers face different levels of income risk.

**Table 3. Variability of indemnity rates (Total indemnities paid over expected income - %) in the period 2011 – 2014 according to different configurations of the MF^.**

|  | Mean | Standard Deviation (SD) | Coefficient of Variation (SD/Mean): | |
|---|---|---|---|---|
|  |  |  | CV^^ | Confidence Intervals 8.5 – 91.5% |
| **National MF** | 9.3% | 0.7% | 0.079 | 0.078-0.081 |
| **MF by altimetry regions:** |  |  |  |  |
| Mountain | 8.2% | 1.0% | 0.121 | 0.117-0.124 |
| Hill | 10.7% | 0.5% | 0.048 | 0.047-0.049 |
| Plain | 8.7% | 1.2% | 0.135 | 0.135-0.139 |
| **MF by macro-regions (MR):** |  |  |  |  |
| Center | 7.8% | 3.3% | 0.388 | 0.373-0.392 |
| Islands | 14.5% | 3.1% | 0.223 | 0.221-0.229 |
| South | 8.6% | 0.8% | 0.098 | 0.096-0.100 |
| North-West | 9.7% | 1.2% | 0.120 | 0.117-0.121 |
| North-East | 8.0% | 0.9% | 0.106 | 0.104-0.109 |
| **MF by types of farming (TF)** |  |  |  |  |
| Specialised fieldcrops | 9.3% | 1.4% | 0.156 | 0.153-0.158 |
| Specialised horticulture | 14.9% | 1.9% | 0.127 | 0.125-0.129 |
| Specialised permanent crops | 10.0% | 0.9% | 0.094 | 0.094-0.097 |
| Specialised grazing livestock | 6.3% | 1.3% | 0.200 | 0.196-0.202 |
| Specialised granivore livestock | 13.2% | 6.1% | 0.437 | 0.427-0.444 |
| Mixed crops | 8.3% | 2.6% | 0.327 | 0.322-0.333 |
| Mixed livestock, crops and livestock | 9.1% | 2.9% | 0.313 | 0.310-0.321 |

*^Values derived by non-parametric bootstrap (n=1000). ^^The distributions of the CV in all groups are statistically different at 1% from that of the National MF according to Wilcoxon test. Source: Own elaboration on a constant sample of farms of the Italian FADN.*

The variability over time of the relative level of indemnifications (third and fourth column of Table 3) sharply and significantly increases in almost all cases when farms are grouped according to the considered dimensions. In two of the three altimetry regions, CV are significantly higher than the one calculated for the national MF case (Table 3). This is even more emphasised when farms are grouped by macro-regions and TF. Relevant is the case of the Center of Italy where, despite the low expected indemnity rate, this has a very large variability. Within the TFs, the variability is significantly higher in granivore livestock farms as well as the two mixed crops and mixed livestock groups of farms.

These results suggest that MF defined at a lower than national aggregation level can be less viable than a national MF in pooling the risk of all members. This depends on the fact that similar/contiguous farms face the same risks and that MF aggregate a lower number of farms.

However, under this specification, MF could better define the contribution of the farms making it more tailored to the peculiar extent of the income risk each group is facing. Hence, this reduces the extent of the distribution of policy benefits among farms. However, this latter result crucially depends also on how the contribution is calculated.

### 3.1.5 Farm contributions to the IST

The flat rate contributions (i.e. each farm of a group pays the same amount in a given year) are around 1400 Euro per farm and year under the assumption of the national MF and a subsidy rate of 65% of the paid indemnities. However, when MF are constructed according to altimetry regions or macro-regions, each group pays a different contribution. This clearly depends on both the expected relative level of indemnification and the size of the sampled farms. Regarding the altimetry regions, premiums would be highest in the plain region. Large differences can also be found among macro-regions with high contributions found in Islands and North-West of Italy. Even larger differences are found among TF with the highest levels in specialised granivore livestock farms.

When contribution is defined according to the average income level, there are large differences in the contribution paid by the farmers. This is true not just for the national MF, but also when the MFs are developed within the considered groups.

These results clearly suggest that the two considered approaches to calculate the contributions differ in terms of distribution of the cost of the IST among farms.

### 3.1.6 Impact on income variability

We find the IST to significantly reduce income variability. This is because this tool shifts to the left the distribution of the CV of the income among the considered farm sample (Figure 1) by reducing the frequency of the high levels of income variability.

Implementing the IST under the full subsidisation hypothesis ($I_I$) allows a reduction of the absolute level of income variability (i.e. SD and MAD) as well as in a significant reduction of the CV level (Table 4). Note that this finding is caused by two components. First, income variability is reduced. Second, the subsidy leads to an increase of mean and median income levels. Thus, our results confirm the findings of Finger and El Benni (2014a) that lowering the subsidisation rate reduces the income stabilising effect of the IST.

Moving from the full subsidy scenario to a contribution by farmers, only slightly reduces the income stabilising role of the IST (Figure 1). Under both scenarios how premiums are derived ($I_{IF}$ and $I_{IE}$), the absolute variability declines in comparison with the observed condition (i.e. without IST). Introducing the IST under the $I_{IF}$ and $I_{IE}$ specifications generates a large reduction of income variability compared to the no IST situation both expressed in absolute terms (i.e. SD and MAD) as well as in terms of relative income variability using the CV (Table 4). Furthermore, the mean and median income levels increase, reflecting that parts of the indemnities are recovered by means of the farmers' contributions.

**Figure 1. Distribution of the coefficient of variations (CV) of income among farms under different IST scenarios. National MF implementation.**

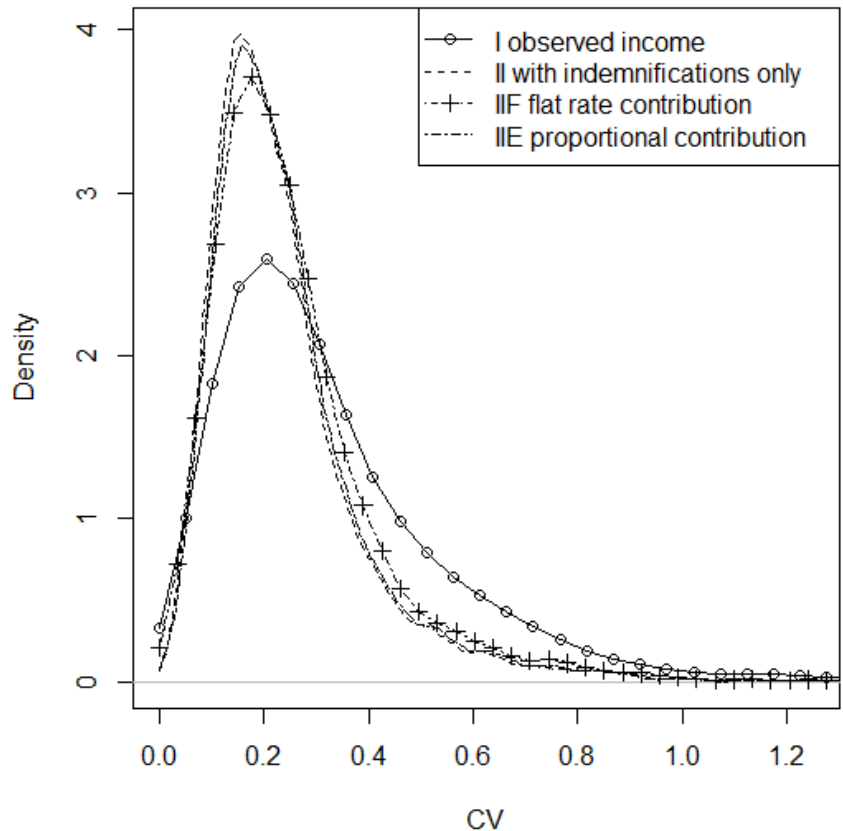

*Epanechnikov kernel with n = 512.*

*Source: Own elaboration on a constant sample of farms of the Italian FADN.*

However, the way farmers pay the contribution to MF can affect this result. When the contribution is proportional to the expected income ($I_{IE}$), the median CV declines slightly more than in the case of the flat-rate contribution ($I_{IF}$) (Table 4 and Figure 1). Hence, the way the contribution system is designed has implications on the income stabilising effect of the IST even considering exactly the same design of the indemnification mechanism. More specifically, the proportional contribution approach is found to be very effective to reduce income variability.

**Table 4. Variability of farm income over time under different IST scenarios. Application of the IST at the national level (i.e. Only one MF). Median values.**

| | Central values | | Absolute variability | | Coefficient of Variation (SD/Mean)^ | |
|---|---|---|---|---|---|---|
| | Mean | Median | Standard Deviation (SD) | Median Absolute Deviation (MAD) | | |
| **Observed income (without IST) (I)** | | | | | | |
| | 35769 | 35025 | 9438 | 6646 | 0.271 | |
| **Income with indemnities paid by the IST:** | | | | | | |
| - without any farmers contribution ($I_I$) | | | | | | |
| | 39132 | 37750 | 7736 | 5168 | 0.204 | *** |
| - with a flat rate contribution per farm ($I_{IF}$) | | | | | | |
| | 36858 | 35430 | 7721 | 5201 | 0.219 | *** |
| - with contributions proportional to expected income ($I_{IE}$) | | | | | | |
| | 38111 | 36673 | 7690 | 5139 | 0.211 | *** |

^ *Calculated on the single farm CVs. Median CVs with IST are statistically different at 1% according to Wilcoxon test (\*\*\*).*

*Source: Own elaboration on a constant sample of farms of the Italian FADN.*

The income stabilising role of the IST does not change according to how the MF are designed when the whole sample is considered (Table 5). However, this is not the case when specific sub-groups of farms are considered. For example, with flat rate contributions, the IST reduces the income variability of mountain farms stronger than that of farmers located in the hills (fourth column of Table 5). Similarly, when farmers pay a contribution proportional to their expected income (fifth column of Table 5), farmers located in the Islands experience a drop in variability from the baseline situation that is substantially higher than in the case of those located in the Center of Italy.

Table 5. Median of the CV of the income over time of the sampled farms under different designs of the MF.

| | Median values: | | | | Relative changes: | | |
|---|---|---|---|---|---|---|---|
| | Observed: | With IST: | | | With IST: | | |
| | | No contribution | With contribution:[a] | | No contribution | With contribution:[a] | |
| | | | Flat | Proportional | | Flat | Proportional |
| | $I$ | $I_I$ | $I_{IF}$ | $I_{IE}$ | $I_I$ | $I_{IF}$ | $I_{IE}$ |
| National MF | 0.271 | 0.204 | **0.219** | **0.211** | −24.6% | **−19.4%** | **−22.2%** |
| MF by altimetry regions: | | | **0.218** | **0.211** | | **−19.6%** | **−22.0%** |
| Mountain | 0.268 | 0.194 | 0.201 | 0.201 | −27.7% | −25.1% | −25.0% |
| Hill | 0.259 | 0.205 | 0.220 | 0.212 | −21.0% | −15.0% | −18.2% |
| Plain | 0.281 | 0.209 | 0.224 | 0.216 | −25.7% | −20.4% | −23.1% |
| MF by macro-regions (MR): | | | **0.218** | **0.211** | | **−19.6%** | **−22.1%** |
| Center | 0.257 | 0.192 | 0.215 | 0.217 | −25.4% | −16.6% | −15.6% |
| Islands | 0.291 | 0.204 | 0.222 | 0.198 | −29.6% | −23.5% | −32.0% |
| South | 0.256 | 0.205 | 0.215 | 0.207 | −19.7% | −16.1% | −19.0% |
| North-West | 0.286 | 0.209 | 0.227 | 0.210 | −27.0% | −20.7% | −26.8% |
| North-East | 0.266 | 0.200 | 0.214 | 0.213 | −25.0% | −19.7% | −19.9% |
| MF by types of farming (TF): | | | **0.219** | **0.211** | | **−19.4%** | **−22.2%** |
| Specialised fieldcrops | 0.267 | 0.206 | 0.219 | 0.212 | −22.9% | −18.2% | −20.5% |
| Specialised horticulture | 0.268 | 0.202 | 0.239 | 0.216 | −24.8% | −11.1% | −19.6% |
| Specialised permanent crops | 0.287 | 0.212 | 0.226 | 0.220 | −26.3% | −21.3% | −23.3% |
| Specialised grazing livestock | 0.264 | 0.192 | 0.205 | 0.197 | −27.6% | −22.5% | −25.6% |
| Specialised granivore livestock | 0.307 | 0.228 | 0.272 | 0.243 | −25.7% | −11.5% | −20.8% |
| Mixed crops | 0.289 | 0.203 | 0.214 | 0.209 | −29.9% | −26.1% | −27.6% |
| Mixed livestock, crops and livestock | 0.241 | 0.192 | 0.219 | 0.203 | −20.1% | −8.8% | −15.5% |

[a]Median values of the whole sample under the four configurations of the MF (i.e. figures in bold) are not statistically significant according to Wilcoxon test. This is true for both the flat and the proportional contribution cases.

Source: Own elaboration on a constant sample of farms of the Italian FADN.

## Distribution of the net financial benefits of the IST

The way contributions are calculated, strongly affects the distribution of the net financial benefit of the policy. Figure 2 shows the very different distribution of net gains from the IST under the two considered approaches used to calculate farmers' contributions to MF (i.e. flat and proportional rate).

**Figure 2. Distribution of the net financial benefits of the IST among farms in the period 2011-14. National MF.**

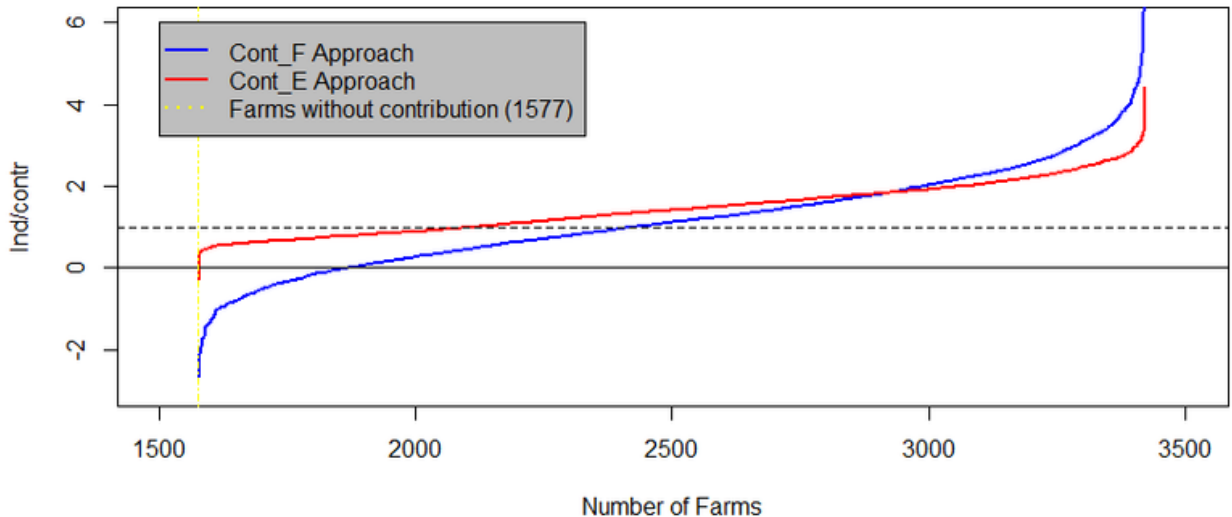

*Note: Ratio indemnities over contribution in the whole sample of farms (Ind/Cont). Flat rate contribution (Cont_F) and contribution proportional to expected income (Cont_E).*

*Source: Own elaboration on a constant sample of farms of the Italian FADN.*

In both cases, 1577 farms of the whole sample (3421) do not receive any indemnity in the four considered years (dotted yellow line in Figure 2). Thus, we observe a strong transfer component from farms with low income variability, receiving no indemnification under the IST, to indemnified farms. Yet, even several farms receiving indemnities have a negative net benefit because these indemnities do not compensate for the paid contributions (i.e. Ind/Contr lower than unity as represented by the horizontal dotted line in Figure 2). More specifically, only 1002 farms under the flat rate case have positive net benefits (i.e. Ind/Cont >1), while this number is 1304 under the proportional contribution case (Figure 2). This means that, under this latter case, 302 additional farms enjoy positive net financial benefits if compared with the flat rate case. This results can be explained by the fact that the flat rate contribution allows some farms to benefit of an extraordinary high level of Ind/Cont (i.e. those located on the right side of Figure 2). Thus, our findings confirm the results of El Benni, Finger and Meuwissen (2016) that flat rate contributions to the IST may generate uneven distributions of benefits of the IST.

### 3.1.7 Conclusions

This study offers the first empirical evidence of the impact on performance of IST, According to the outcomes, this tool, would significantly stabilise farm incomes in Italy.

It has shown that moving from geographical mutual funds (MF) disaggregation significantly increases the variability of the MF's total indemnities. This result is intuitive and is due to the pooling effect (Pigeon, Henry de Frahan, and Denuit 2014) and can be negatively reflected on an increase in the MF costs and, consequently, charged to farmers.

Another strand of analysis has focused on the type of farmers' contribution, founding that with a lower contribution rate, one found a reduction in the income stabilising effect. However, the findings demonstrate that the way farmers are paying is still important: a fixed payment has less impact on the income stabilisation effect than a value-added proportional payment. Consequently, a flat rate contribution generates a very uneven distribution of benefits across farms compared to proportional contribution.

One founded that the policymaker has to consider the level of aggregation of MF, relying on that a more extensive level is better than disaggregated, particular if we need to consider the transaction costs that necessarily increases.

Moreover, we need to highlight the importance of the design of the contributions. Modulation of this one can avoid an undesiderable effect of equity in economic impact, with the small farmer who may pay more than the stabilisation effect that receives with a consequent inequality decrease.

# 3.2 Applications of Machine Learning for the Ratemaking of Agricultural Insurances

## 3.2.1 Introduction

Risk is a fundamental issue in economics research, and insurance is one instrument used to cope with it. However, in general, the insured value in the world remains very low, and even the government policies used to increase participation through better information, subsidies and increasing trust have given only limited effect (Banerjee et al. 2019; Cai, de Janvry, and Sadoulet 2020; Cole et al. 2013). Insurances have received growing attention by researchers and policy-makers in the last decade (e.g., Diaz-Caneja et al. (2008) and Meuwissen, van Asseldonk, and Huirne (2008)). Recently, new insurance schemes such as the Income Stabilization Tool (IST) have been proposed in the European Union (EC 2017b; Meuwissen, Mey, and van Asseldonk 2018).

Accurately determining insurance premium levels is crucial for developing a successful insurance scheme in particular to prevent adverse selection and moral hazard problems (Glauber 2004; Ramirez, Carpio, and Rejesus 2011). The premiums should be significant enough to cover indemnities and other insurers' loading costs (Coble and Barnett 2013). However, high premiums are expected to make the scheme unattractive to farmers who have a quite elastic demand for insurances (Smith and Glauber 2012; Woodard and Verteramo-Chiu 2017). This calls to establish the premiums level, also accounting for the level of indemnification expected by each farmer (Goodwin and Mahul 2004). Thus, adequate ratemaking (i.e., the definition of premiums farmers have to pay, also referred to as pricing of insurance contracts) seems crucial to ensure both the financial sustainability of the scheme and the degree of satisfaction of participating farmers. This result cannot be obtained directly, but several farms and farmers' characteristics could be used to predict the expected level of indemnifications (El Benni, Finger, and Meuwissen 2016). Despite that predicting as accurately as possible the levels of indemnities is key to ensure the actuarial and financial performance of the Insurance (Goodwin and Mahul 2004), there is lack of studies proposing innovations in the ratemaking of agricultural insurance solutions.

Two central problems affect such exercise. First, the distribution of such indemnities is zero-inflated, and when indemnifications are paid, their distribution tends to be right-skewed, and thick-tailed estimation approaches requiring normal distributions cannot be applied. Second, many farms and farmers' characteristics could be potentially related to the probability and extent of indemnities. As this data needs to be collected coherently and efficiently, knowledge of the relevant variables is essential to reduce data collection costs. This aspect is coherent with the signal theory of Spence (1973, 2001); also, if the economist cannot see the real information that affects a particular value, he can use the variables to make a better decision and diminishes the effect of asymmetric information.

We propose a new approach to predicting possible future indemnities by addressing these two groups of problems encountered when assessing the indemnity levels of insurance schemes. First, we use the Tweedie distribution that is compatible with the distribution characteristics of indemnity. Some scholars have used this probability density function, in particular with applications in the actuarial field, with good results (see, for example, Jørgensen & Paes De Souza (1994) and Zhang (2013)). Tweedie distribution, in our knowledge, has not been used in the case of agricultural insurances.

Second, we use three Machine Learning (ML) procedures, i.e. LASSO, Elastic Net, and Boosting (Hastie, Tibshirani, and Friedman 2009; Hastie, Tibshirani, and Wainwright 2015; James et al. 2013; Storm, Baylis, and Heckelei 2020) to identify those farm and farmers` characteristics that are best suited to predict indemnities. In other words, the ML tools are used to denoise the signals and to discover only the essential information that affected the generation of our dependent variables (i.e., the Data Generating Process, DGP[34]). This is important in the practice of the implementation of a new insurance scheme because the results of the analysis inform decision-makers on which variables are indeed useful when designing insurance schemes as cost-efficiently as possible.

The general objective of the analysis is to assess whether the proposed approaches could improve the ratemaking. In addition to the previous literature, we are looking at the econometric efficiency of the models, but we also evaluate the effect of using model-based predictions in the practice of ratemaking by focusing on identifying the premiums based on a limited amount of information. We compare the various models' econometric performances looking at two different goodness-of-fit indicators and the amount of information required for the estimations. The rationale is that gathering and managing data is costly because data must be collected directly from farmers with collection costs that are expected to be correlated to the amount of information (Spence 1973). Hence it is helpful to select only the most essential variables.

Furthermore, we consider the implications of using the premiums suggested by the various models on the economic performances of the insurance. This aspect is essential to have an instrument that allows having a large enough number of contracts, ensuring satisfactory economic results and avoiding incurring relevant losses in some years. We add to previous analyses of agricultural income insurance schemes considering the compatibility of the premium with the actual income of each individual (Crocker and Snow 1986; Wilson 1977). This aspect is crucial because farmers consider it when deciding whether to participate in the scheme or not. Furthermore, we assess the distribution of the individual net premiums among the farm sample to assess how well the ratemaking is calibrated according to the specific risk profile of the policyholders (Crocker and Snow 1986; Miyazaki 1977; Rothschild and Stiglitz 1976; Spence 1978; Wilson 1977). In other words, the insurer would like to avoid the case in which the premium for some individual is lower than the expected indemnity and the opposite case because it may cause non-participation. Finally, the implications of using the results of the model on the economic results of the mutual fund are also assessed by considering the overall annual balance given by the differences between premiums collected and indemnities paid.

---

[34] Data Generating Process (DGP) describes the rules with which the data has been generated. "By this term, we mean whatever mechanism is at work in the real world of economic activity giving rise to the numbers in our samples, that is, precisely the mechanism that our econometric model is supposed to describe" (Davidson and MacKinnon 2003 page 32).

We use the perspective application of the EU Income Stabilization Tool in Italy and apply a large sample of individual farms using data from the whole FADN farm sample for the period 2008-18.

Our findings show that Tweedie is a distribution that is consistent with the indemnity space. The use of ML with Tweedie as the probability function can improve goodness-of-fit. In particular, ML reduces the collinearity problem by using a small number of regressors as it efficiently performs variable selection and, at the same time, enables to discover the proper Data Generating Process (DGP). The proposed approaches can also contribute to developing a targeted ratemaking, ensuring to reach a relatively large share of farmers to face a premium that is compatible with the level of their available income. Furthermore, it is shown that the ML models (even if using fewer regressors that the GLM model) allow having a ratemaking that calibrate better than other approaches to the specific risk profile of the policyholders.

The remainder of this paper is structured as follows. The following section offers a perspective on insurance ratemaking and the Income Stabilization Tool, putting our research within the literature context. The methods and data used in the analysis are discussed in section 3. Section 4 discusses the findings of the analysis, addressing objectively the econometric outcomes and how the estimates may be used in ratemaking. The last section summarises the study findings, pointing out the potentialities of the strategies being discussed to establish a practical implementation of insurance schemes. This also points to the methodological limits, discusses potential restrictions on its applications and recommends areas for further study.

### 3.2.2 *Background on insurance ratemaking and functioning of the IST*

**Insurance ratemaking**

Insurance can be defined as the exchange of money now for money payable contingent on the occurrence of certain events (Zweifel and Eisen 2012). The insurance contract specifies the premium ($Prem$), or the disbursement needed to buy the contract. In contrast, the indemnity *(Ind)* is the payment that the insurer reimburses to the policyholder after a specific event occurs. The indemnity is defined as $Ind = Loss - (a + d \times Loss)$, where the term into brackets is the level of a deductible that we define as a proportion of the loss where $d$ is the percentage of deductible and $a$ is a fixed level of deductible, established in the insurance contract. Hence, the insurance contract can be described as a two-dimensional vector $S = \{s \in \mathbb{R}^2 : (Prem, Ind)\}$ with *Prem>0* and *Ind>0.*

Ratemaking may be defined as *"The process of predicting future losses and future expenses and allocating these costs among the various classes of insureds"* (Vaughan and Vaughan 2014). In other words, ratemaking (aka insurance pricing) is intended here to determine what rates, or premiums, to charge for insurance. The prediction of the indemnities is an essential step of ratemaking and can be done based on the following:

$$E(Ind) = f(X_i) \qquad (2)$$

With $X_i$ is the policyholder's characteristics, i.e. individual farm and farmers' characteristics in our case (Bernard 2013).

According to Embrechts (2000), Insurance premiums can be calculated as follows:

$$\Theta_{i,t} = E(Ind_{i,t}) + \delta E(Ind_{i,t}) \qquad (3)$$

Where $\Theta_{i,t}$ representing the 'fair' insurance premium, $E(Ind_{i,t})$ is the expected indemnity of individual $i$ at time $t$. $\delta$ is the loading function that should be large enough to sufficiently secure margins in the mutual fund that can derive from ruin estimates of the underlying risk process. In our case, to simplify the analysis without loss of clarity and significant difference in assessing results, the loading cost is omitted.

The fundamental consideration that drives insurance ratemaking is that the premium should be high enough to bring forth sellers and low sufficient to induce buyers to enrol (Finn and Lane 1997). Hence it is crucial to choose a ratemaking that tends to satisfy condition (3) as much as possible. Furthermore, buyers will enrol only if the premium level is compatible with the resources available to the potential insured. In particular, it is common to assume that farmers participate to the insurance scheme only if their disposable incomes (*I*) allow to cover the premium (i.e., $I - Prem \geq 0$) (Wilson 1977; P. Zhang and Palma 2020). This second aspect is not generally considered in agricultural insurances while it is fully exploited in this analysis. Finally, the insurer is interested in evaluating the overall balance given by the differences between premiums collected and indemnities paid over a suitable number of years ($\Pi_{t=0}^{T}$):

$$\Pi_{t=0}^{T} = \sum_{t=0}^{T}\sum_{i=1}^{N}(Prem_{i,t} - Ind_{i,t}) \qquad (4)$$

However, the insurer is interested also in the yearly variation of the balance and, in particular, to the cases in which a negative balance large enough to threatening its financial position may occur (i.e. solvency).

Goodwin and Mahul (2004) show that the availability of data and the modelling techniques shape the insurance scheme and the ratemaking procedures. In contrast with the ratemaking of previously existing insurance products, where historical data are available, in the case of new insurance products, it is impossible to capture the impact of insurance on farmers' behaviour (Goodwin and Mahul 2004). Our case study falls in this latter situation because IST is a new insurance product.

### The functioning of the Income Stabilization Tool

Income volatility characterises the agricultural sector dramatically (EC 2009). To efficiently cope with this issue, the EU has looked at several instruments (Cafiero et al. 2007; Diaz-Caneja et al. 2008;

-61-

EC 2001, 2017b; Meuwissen, Assefa, and van Asseldonk 2013; Meuwissen, van Asseldonk, and Huirne 2008) and focusing on income insurance schemes through the development of the Income Stabilisation Tool - IST (EC 2010, 2011a, 2011b, 2013a, 2013b).

EC has designed IST to be WTO green box compatible to strengthened support to insurance instruments and mutual funds. IST is also coherent with other CAP instruments, in particular market instruments. IST is a Mutual Funds-based income farm insurance scheme and partially compensates for losses (it is not full insurance).

The trigger level is set at 70 per cent of the Olympic average of the last five years or the arithmetic average of the previous three-year income. If the current level of income is below the trigger level, the farmer can be indemnified. The maximum indemnity level is 70% of the difference between the trigger level and the current income value. Following Finger and Benni (2014), we have chosen the arithmetic average of the past three years of income to identify the trigger level.

IST provides compensation to farmers who experience a severe income drop, i.e. an income drop of more than 30% compared with the expected income level (Bardaji and Garrido 2016; EC 2017b; Meuwissen *et al.* 2018). In such a tool, farmers financially contribute to mutual funds similar to what farmers do when enrolling in an insurance scheme. The indemnification paid by the IST scheme is defined as follows:

$$Ind_{i,t} = \begin{pmatrix} 0 & if & I_{i,t} \geq I_{R\,i,t} \\ b\left(E(I)_{i,t} - I_{i,t}\right) & if & I_{i,t} < I_{R\,i,t} \end{pmatrix} \qquad (5)$$

where:

$I_{i,t}$ is the realised income of the $i_{th}$ farm at the $t_{th}$ year;

$E(I)_{i,t}$ is the expected income of the $i_{th}$ farm at the $t_{th}$ year and based on the average of the realised $I_{i,t}$ of the previous three years (see Finger and El Benni 2014b for discussions) as $E(I)_{i,t} = \frac{1}{3}\sum_{t-1}^{t-3} I_{i,t}$;

$I_{R\,i,t}$ is the trigger level of the $i_{th}$ farm and is defined as : $I_{R\,i,t} = a\,E(I)_t$ where parameter $a$ is set by the Regulation as 0.7.

The EC regulation also sets the parameter of deductible as 0.3; consequently, the MF pays b = 0.7 of the loss $\left(E(I)_{i,t} - I_{i,t}\right)$. The presence of deductible defined this instrument as "Partially Insurance". The regulator has introduced the deductible to consider the moral hazard (Cordier and Santeramo 2020). Note that the IST refers to farm income only and not to the overall farm household income. Farmers are asked to contribute to the mutual fund, but the mechanism is still unspecified (Liesivaara and Myyrä 2017).

Consequently, the space of indemnity $\overline{Ind}$ is defined, in IST, as:

$$\overline{Ind}: \{Ind \in \mathbb{R}: Ind(0 \cup min_{Ind}, +\infty)\} \qquad (6)$$

In words, *indemnity* is defined in the positive and continuous space but a closed between 0

(excluded) and the minimum value of indemnity.

Timing is essential because it has consequences in the model and the econometric framework. We consider two periods ($t_{-1}$ and $t$) (Figure 1). At the end of the first period, which we assume corresponding to the accounting year, the farmer can draft the balance sheet identifying the level of farm income and other relevant parameters. The mutual fund then gathers the data and, based on this, determines the premium that will be indicated in the contract. After this, farmers decide their production plan and only after this, an adverse event may occur. At the end of the period $t$, it will be possible to collect the data to assess the income losses, and if the drop of income exceeds the trigger level, the level of *Ind*.

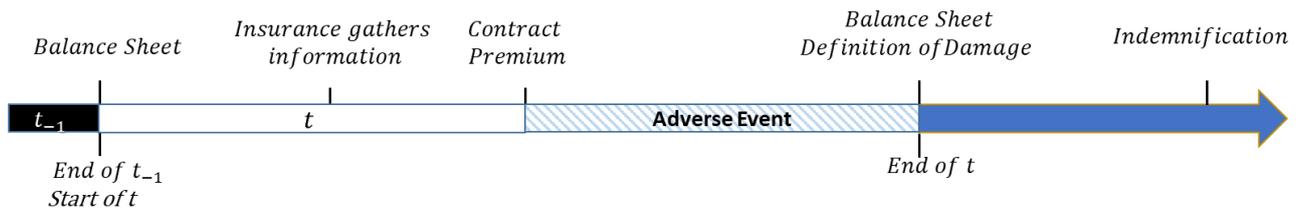

**Figure 1 - Timing of IST**

It is important to stress that it is not possible to use the current value of $X_{i,t}$ to estimate $E(Ind)$ because their levels are unknown when establishing the insurance contract. For this motivation, it is needed to use variables of the previous period ($X_{i,t-1}$) and to identify an estimation model $E(Ind_{i,t-1}) = f_{t-1}(X_{i,t-2})$ where $f_{t-1}(\cdot)$ represents the Data Generating Process (DGP) a time $t-1$. Then we use it to predict the current level that is: $\widehat{Ind_{i,t}} = f_{t-1}(X_{i,t-1})$

### 3.2.3    *Methodology and Data*

**Assumptions**

Our analysis is based on several assumptions. First, we assume that insurance and policyholders have the same information (i.e., no-asymmetric information). Second, we use the rational expectation hypothesis (Muth, 1961; Pesando, 1976). The past dependence of the conditional distributions is limited to the dependence on the most recent observation (i.e. Markov property) (Pesando 1976). Because of this, we assume that the $X_{i,t-1}$ represent the best information to address our problem. Note that our model considers individual covariates at time $t_{-1}$ only (Werner, Modlin, and Watson 2016) or Markovian assumption.

The Markov assumption eliminates the problems of reversible causality of some variables. We need to considerer the inference problem from an endogenous variable, like value-added, that enter directly into the estimation of Indemnity. We have to avoid the correlation with no-fixed variables like labour and the new information. This last effect is not filtered via Machine Learning but only through a theoretical assumption.

Stricly related to previous assumptions is the risk characteristic of farmers assumed neutral. This is not a very strong assumption because we make a sensitivity analysis with different affordability levels to make the outcome not particularly affected by this restriction. Considering the risk measure is obtainable only after applying this insurance instrument.

To avoid both effects, we use the level of variables at the beginning of the production process. Taking into account that we use accounting data, the figure at the start of the process coincide with the accountable value of $t_{-1}$ level (Bellemare, Masaki, and Pepinsky 2015; Fajgelbaum, Schaal, and Taschereau-Dumouchel 2017).

We also assume that only one insurance firm is present on the market and that it, being a mutual fund, tends to have zero profits. We consider that preferences, technology, behaviour, information and interactions remain constant throughout the year but varying between years (year-to-year approach[35] and *ceteris paribus*).

In contrast with previous analyses, we do not assume that IST is mandatory (EC 2009; Finger and El Benni 2014b, 2014a; El Benni, Finger and Meuwissen 2016; Severini, Biagini and Finger 2019). In particular, we consider different scenarios of income vs premium compatibility, as explained earlier. We also assume that farmers decide whether to buy or not the contract in a formula "take-it-or-leave-it".

The insurance industry should establish a premium before issuing the contract. This should account for loading costs. However, because we do not have information on the extent of such costs, we assume these are zero. The lack of loading cost does not affect the assessment of different ratemaking because it is only a proportional increase in premium.

---
[35] For lack of the space the discussion about the choice of year-to-year approach, in particular the comparison with the use of year fixed effect and the linkage with the signals theory of Spence is reported in Appendix.

Moreover, we assume that the mutual fund proposes a homogeneous contract to all farmers that is nor exchangeable between farmers. The last assumptions are needed to prevent another policyholder from using a contract designed for a farmer with different risk characteristics. Finally, we assume that the premium is the only parameter that can change from an individual to another.

### Modelling the distribution of indemnities

As illustrated, the space of indemnity has a peculiar probability density function represented in (6), which poses a very challenging problem for a maximum likelihood estimation.

We have compared other probability density function, but only the Tweedie (Jørgensen (1987) and Jørgensen and Paes De Souza (1994)) accounts for all characteristics of the space of indemnities (Table 1).

**Table 6 – Comparison between different Probability Density Functions with space of Indemnities**

| PdF | Support | Zero & $R^+$ | Zero Inflated | Tick Tailed | Asymmetric Distribution | Discontinuity Function (Deductible) | Continuous after Min Value of Indemnity |
|---|---|---|---|---|---|---|---|
| Normal | $-\infty \leq x \leq +\infty$ | No | Partially | No | No | No | Yes |
| Poisson | $\mathbb{N}$ | Yes | Yes | No | Yes | Yes | No |
| Negative Binomial | $\mathbb{N}$ | Yes | Yes | Yes | Yes | Yes | No |
| Inverse-Gaussian | $x \in (0, \infty)$ | Yes | Yes | Yes | Yes | No | Yes |
| Gamma | $x \in (0, \infty)$ | Yes | Yes | Yes | Yes | No | Yes |
| Tweedie | $x \in (0 \cup_{min} x^+ x, \infty))$ | Yes | Yes | Yes | Yes | Yes | Yes |

Tweedie distributions are Poisson-Tweedie mixtures belonging to the Exponential Dispersion Models. The probability density function of the Tweedie is defined as:

$$f(y; \mu, \phi) = a(y, \phi) exp\left\{\frac{[y\psi - \kappa\psi]}{\phi}\right\} \qquad (7)$$

Where $a(\cdot)$ are given functions, $\kappa(\cdot)$ is a cumulant function, $\psi$ is the canonical parameter and $\phi$ is the dispersion parameter with $\phi > 0$. According to Jørgensen (1987) and Jørgensen & Paes De Souza (1994): $\mu = E(y) = \kappa'(\psi)$ and $Var(y) = \phi V(\mu) = \kappa''(\psi)$, where $V(\mu) = \mu^p$. Moreover, $p \in (-\infty, 0] \cup [1, +\infty)$.

### Identification of the Data Generating Process and variable selection by Machine Learning

The main issue of the analysis is to estimate the Data Generating Process at time $t_{-1}$ or DGP $f_{t_{-1}}(\cdot)$ based on $X_i$ and $Ind_i$ to identify a valid set of regressors.

We use Machine Learning (ML) and compares it with the classic GLM approach. ML differs from the classical inference that assumes a given stochastic model generates the data. Instead, ML uses algorithmic models based on whether the DGP is unknown (Breiman 2001)[36].

To reach a correct estimation, we need to use a large amount of information, as our dataset is. However, many variables lead to an increase in the collinearity problem, an issue highlighted by Spence (1973) in his Signals theory.[37]

To reduce these multiple problems, or "dirty" dataset, it is necessary to tidy and clean them [38]. We need a methodology that eliminates redundant information and selects only the more essential and not correlated variables (E. W. Bond and Crocker 1991; Crocker and Snow 1986; Hoy 1982). Several standard variable selection methods can be used. For instance, El Benni, Finger and Meuwissen (2016) have used stepwise regression and Genetic Algorithm in the case of IST with good results. Still, these algorithms are not specifically designed to reduce the collinearity problem and are based on quasi-predefined DGP.

As referred by Breiman (2001), Hastie, Tibshirani and Friedman (2009), Varian (2014), Storm, Baylis and Heckelei (2019), and Efron (2020), ML can solve complex problems concerning collinearity, dimensions of the dataset, management of numerous variables and discover the real DGP in a very efficient way[39].

Four different models are utilised (Table 1), all with Tweedie as the Likelihood function. The baseline used to compare the performance of ML is the GLM model (defined by Efron (2020) as a "traditional prediction model"), including all available variables[40]. However, collinearity appears to be a relevant problem in our case study. Roughly half of the regressors used in the GLM show high collinearity reporting a VIF > 5 (Fox et al., (1992) and Fox & Weisberg, (2012)).

The ML tools are the following three: Least Absolute Shrinkage and Selection Operator - LASSO (R. Tibshirani 1996), Elastic Net - EN (Zou and Hastie 2005), and Boosting (Friedman 2001). We chose these ML tools because, differently from other methods, such as neural network, it is easy to assess their accuracy (i.e. the difference between predicted and observed values) and interpret the estimated coefficients. The ML can improve predictions and imposes limited restrictions on the functional forms. In particular, Boosting shows that the relation between regressor is not necessarily linear (Hastie, Tibshirani, and Friedman 2009). This aspect is practical when only little theoretical knowledge exists on the considered issue (Storm, Baylis, and Heckelei 2020).[41]

---

[36] A deeper discussion about the Machine Learning and the difference with Classical Inference methods can be found in Appendix

[37] A deeper discussion about the Spence Signals theory and his relation with ML can be found in the Appendix.

[38] Outliers detection in the initial dataset is performed by using Fox (1991, 2016) and Ziegel, Fox, & Park (1993) and observations with Cook's Distance D > cut-off (where the cut-off $4/n - k - 1$ with $n$ is the number of observations and $k$ is the number of regressors) are eliminated (Chatterjee and Hadi 2009).

[39] To avoid potential misunderstandings, we define "machine learning" using the definition of Efron (2020) (i.e., "Simple Prediction Algorithms"). However, we use the ther ML because it is widely used.

[40] Data are described in a section of the paper later on. The full list of these variables is available as supplementary material.

[41] ML approaches, the difference of this last one with inference model, and the tools used in this paper are discussed in deeply mode in appendix.

**Table 7. Definition of the utilised models**

| Acronym | Description | Type of Estimation | Main Reference |
|---|---|---|---|
| GLM | Generalised Linear Model with all 129 regressors | Maximum Likelihood Estimation | (Jørgensen and Paes De Souza 1994) |
| EN | Elastic Net: Regularised method with variable shrinkage parameter variables ($\alpha = 0.1 \div 0.9$) | Machine Learning | Zou & Hastie (2005) |
| LASSO | LASSO: Regularised method with shrinkage fixed parameter ($\alpha = 1$) | Machine Learning | Tibshirani (1996) |
| Boosting | An algorithm which converts weak learner to strong learners. | Machine Learning | Friedman (2001). |

**Model setting**

We use a specific algorithm to account for the characteristics that emerged from the IST evaluation (R Core Team 2020). The R-Package "TWEEDIE" (Dunn 2017) is used to obtain the best power parameter ($p_{opt}^*$). The residuals from the GLM-TWEEDIE estimation (Goldburd, Khare, and Tevet 2016) are used to remove some of the outliers (minus than 5%) by using the Chatterjee and Hadi (1986) cut-off. The dataset obtained in the previous step is used to estimate the $p_{opt}^{**}$. Furthermore, GLM without outliers is applied with $p_{opt}^{**}$. Finally, the $p_{opt}^{**}$ is used in GLM. We use the following R-Packages: for Elastic Net and LASSO the "HDtweedie" (Qian, Yang, and Zou 2013)) and for Boosting the "TDboost" (Yang, Qian, and Zou 2016).

We use "cross-validation" to find hyperparameters (the setting of ML algorithm) that minimise RMSE (McCullagh and Nelder, 1989; Fox 1991, 2016; James *et al.* 2014; Efron 2020).

Grouped regression has been used to handle the categorical regressors assuming multiple discrete values (i.e., levels) following Qian et al. (2016).[42]

The out-of-sample procedure is adopted to allow an exact evaluation of the models. Ten random groups of observations for every year are implemented. For every random group, there is a training set (containing 75% of the set), and as a test-set, we have used all the observations at $t$[43]. The estimation is implemented in the training set, while the test set is used to make the predictions.[44] The out-of-sample procedure also permits to confirm of the presence of overfitting cases (Efron 2020; Hastie, Tibshirani, and Friedman 2009)

The combination of out-of-sample and cross-validation can produce some issues defined in Breiman (2001) and recently discussed by Efron (2020)[45]. In particular, we can find multiple DGPs. In numerous case, DGP change for minimal parts, for example, can change in regressor with deficient "weight" in the prediction. It is essential to evaluate the core of selected variables. These are the

---

[42] Further explanations are reported in the Appendix
[43] For example, we use 75% of 2011 dataset to train-set and all 2012 observation as test-set. The motivation of this setting is explained in the theoretical framework – Timing of IST to obtain result compatible with the real availability data for Mutual Fund.
[44] The out-of-sample methodology, in the case of strong presence zero-inflated, can generate two unbalanced subsamples. To avoid this effect, we have adopted a procedure that maintains the proportionality between zero and non-zero values in the training-test sets.
[45] These include Rashomon's effect, Occam's effect and Bellman's effect. Further explanations are reported in the Appendix

information that represents the "signal" and not the "noise" in Spence's sense (Spence 1973, 1978).

The use of the Year-to-Year approach with out-of-sample procedure allows the discovery of the "transitory patterns" or the case where DGP describes with variables that are useful only in rare cases.[46]

### Econometric and economic assessment

The goodness-of-fit of the models is assessed based on the $R^2$ defined in Cameron and Windmeijer (1997) and the Root Mean Squared Errors (RMSE) (Hastie, Tibshirani, and Friedman 2009; P. Murphy 1991; Van der Paal 2014). The former is used because ML can produce non-linearity correlations between predicted and simulated indemnity values we use this specific $R^2$. The RMSE is also considered being another metric to verify the robustness of results that can be used in all considered models. We then analyse the number of selected regressors as well as how much they vary. This figure is important because it allows us to assess the stability of the estimated DGP. Finally, we compare both goodness-of-fit and the number of selected regressors to evaluate whether a trade-off exists between them in the considered case.

Both the goodness-of-fit metrics are not corrected for the degrees-of-freedom (DoF). Usually, the econometrics compare different models penalising the models with a high number of regressors (i.e. corrected-$R^2$ or AIC or BIC). Unfortunately, ML does not allow using this correction because the definition of GDF in the ML is still an open issue (R. J. Tibshirani and Taylor 2012). To remedy this deficiency, we use a graphical analysis jointly the number of regressors selected with the $R^2$.

The economic implications of using the predictions of the model for ratemaking are based on a few simple indicators. The first refers to the number of farms for which the compatibility between premium and indemnities is satisfied. We use the level of Value Added (*VA* - defined in the next section) of the past year as a proxy of disposable income that is considered by the farmer when deciding whether to buy the insurance or not. Considering that it is not possible to identify the suitable compatibility of each farmer, we have made a sensitivity analysis based on the following four levels of compatibility: $Prem \leq VA, Prem \leq 0.50 \times VA, Prem \leq 0.25 \times VA, Prem \leq 0.10 \times VA$.[47] This last level is very close to the expense for insurance premium in the whole sample of farms. In particular, note that the 95th percentile of the premium is around 9% of past year VA level in farms that paid the insurance premium.

We refer to the balance of the mutual fund considering both the multiannual value and its yearly figures.

The multiperiod financial balance sheet (2012-2018) of insurance, obtained with the formula (4), allows analysing regarding the capacity of MF to not incurred in bankruptcy or in an excessive

---

[46] Further details are available in Appendix

[47] We report in the main text only the compatibility level Premium ≤ 0.10 VA because it is similar to reality. Results for the other levels can be found in Appendix

increase of premium to cover the losses (i.e. solvency).

The economic assessment allows us to analyse the financial stability and suitability of the models. We compare the balance sheet of insurance in all years and its stability in the period of analysis considering different modelling policies

An evaluation only based on the total amount, without the analysis of the earning and losses classes, can hide some effects. We need to analyse $Prem_{i,t} - Ind_{i,t}$ considering that the insurance is a pooling scheme between policyholders who receive an indemnity against a premium always paid: the latter cover the former. We also need to consider that the farmers with a high positive value of $Prem_{i,t} - Ind_{i,t}$ are not likely to subscribe to the insurance in the next year.

It is crucial to assess some aspects of this analysis, such as the entity, the relationship, the distribution and last but not least, the different classes. In particular, regarding the classes, the analysis considers both the value of $Prem_{i,t} - Ind_{i,t}$ and the number of participants in the scheme for every class.

We classify the results in the following two main cases:

- ≥ 0 € and < 0 €

Furthermore, we consider the following more detailed classes:

- ≥ 100000 €; ≥ 250000 €, ≥ 500000 €, ≥ 750000 €, ≥ 1000000 €,
- ≤ -100000 €, ≤ -250000 €, ≤ -500000 €, ≤ -750000 € ,≤ -1000000 €

This analysis allows us to assess the sustainability of models, in particular regarding the objective of pooling characterised by the insurance scheme. Specifically considering how targeted is the ratemaking.

### Data

This analysis is based on a panel of individual farm accountancy data belonging to the Italian Farm Accountancy Data Network. The period of study regarding 2008–2018 with 118,748 farm observations. All economic values have been deflated using the Harmonized Index of Consumer Prices[48].

Consistently with the decisions of the Italian government, the farm income is defined in terms of farm value-added (ISMEA 2015; Mipaaf 2017). Value added is given by farm revenues and public payments (e.g., CAP direct payments) minus costs for external inputs.

First of all, we use the data to estimate IST indemnity. Because of IST's mechanism to calculate the indemnity, the value-added of the three years before time $t$ is used to find the trigger level. For this motivation, for the first three years, the trigger level is not obtainable, and consequently, the period from 2008 to 2010 is not usable for estimation. Besides, only farms with positive reference incomes are maintained because the EU regulations do not provide information on how the IST should consider negative incomes (0.48% of the total sample). These criteria bring the database to have 47.898 observations.

---

[48] HICP: Eurostat - https://ec.europa.eu/eurostat/web/hicp/data/database .

IST Indemnity simulated distribution is represented in Figure 2

**Figure 2 Density plot of indemnity**

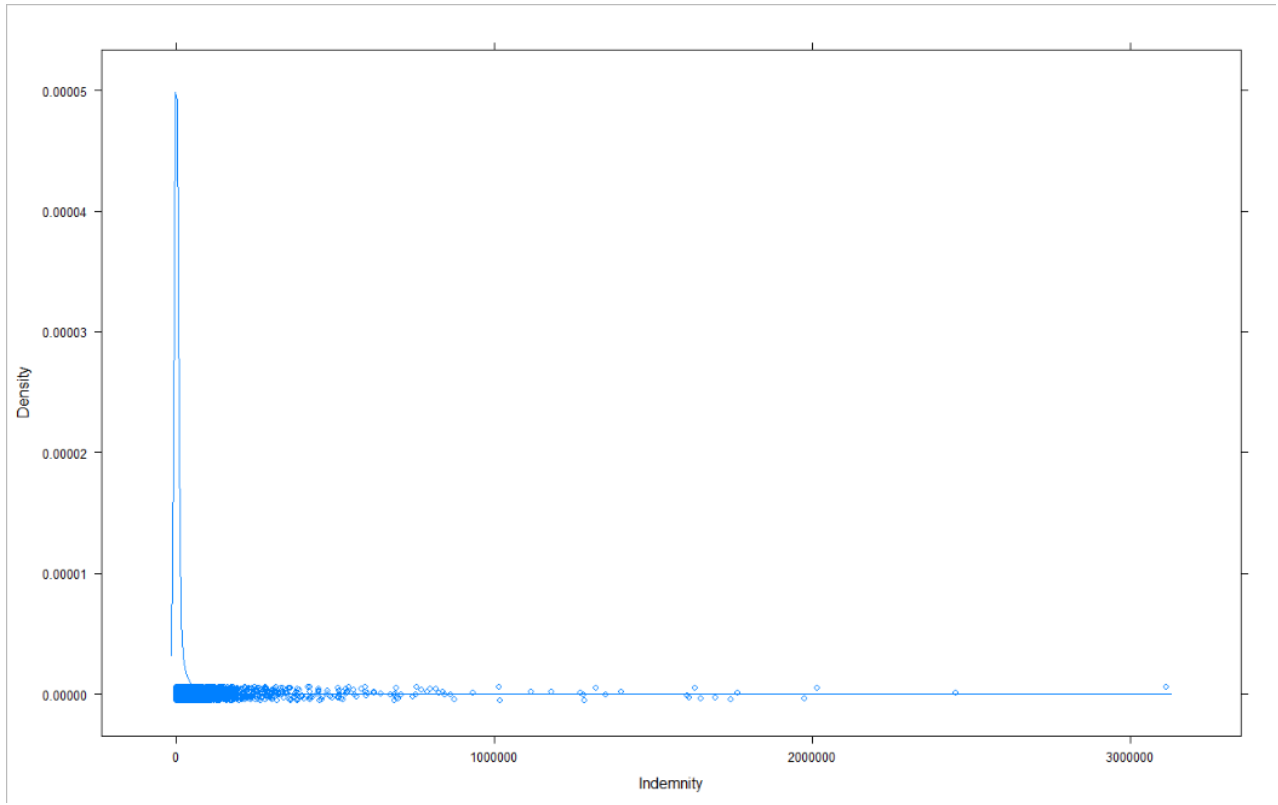

*Source: Own elaborations on Italian FADN data.*

The simulation shows how the expected indemnity level is not the same for all policyholders (Ohlsson and Johansson 2010), with a peculiar distribution is zero-inflated right-skewed and thick-tailed (Werner, Modlin, and Watson 2016) and also in line with space of indemnity.

An essential factor to evidence regarding the discontinuity of IST's distribution: the minimum value of Indemnity is € 236.40 larger than zero.

**Table 8 - Descriptive statistics of indemnity**

|  | n | mean | sd | median | mad | min | max | skew | kurtosis |
|---|---|---|---|---|---|---|---|---|---|
| Total Sample | 47898.00 | 7154.75 | 46644.17 | 0.00 | 0.00 | 0.00 | 3111487 | 26.83 | 1094.01 |
| Ind>0 | 10592.00 | 32354.45 | 94994.44 | 12377.22 | 12184.54 | 236.40 | 3111487 | 13.55 | 270.56 |

*Source: Own elaborations on Italian FADN data.*

The comparison between the sample used in this paper with the total sample is very similar[49]; for instance, the share of farms grouped by altimetry regions and types of farming. Overall, the sample represents the essential components of the farm population.

After these preliminary steps, it is necessary to define the factors potentially affecting the IST

---

[49] Data available on request.

indemnification level (i.e. $X$). As established in the theoretical framework, the level of IST indemnities is correlated to the factors that affect the income downside risk. We consider an extensive set of potential explanatory (El Benni, Finger, and Meuwissen 2016). We have not necessary to make a strong assumption to insert a variable or not because the scope of ML is whose to discover DGP eliminating the not necessary regressors. Nevertheless, we do not base our model only on a data-driven assumption. We take the critics of prof. Cox for the reflections of Breiman (2001), also corroborated to other studies (Athey and Imbens 2019; Box 1979; Efron 2020; Einav and Levin 2014; Mullainathan and Spiess 2017; Varian 2014) that considerer that a pure ML approach cannot make possible deep analysis in causal aspect. In other words, with a pure data-driven model like ML (i.e. without economic assumptions), we can make excellent predictions, but, at the same time, we cannot profoundly analyse the phenomena at stake. Our point of view is that we should consider some of the most relevant constraints are deriving from economic theory, and we cannot use a purely data-driven approach. Because of this, we use a structural model approach[50] (according to Hurwicz (1966) definition), balancing the data-driven and theory-driven model to define the components of regression, the assumptions and the constraints of the model (Iskhakov, Rust, and Schjerning 2020). The choice of regressors follows this approach.

Topographic, climatic, and socio-economic conditions affect the relative profitability of the farm enterprises, the availability and characteristics of the production factors, and, in turn, production possibilities. Farmers located in the mountain regions have been found to face higher (relative) income variability than farmers located in hill or valley regions (El Benni, Finger, and Mann 2012; Simone Severini, Tantari, and Di Tommaso 2016). Farm characteristics such as farm size, production characteristics, financial features have also been found to be relevant (e.g., Mishra and El-Osta (2001); Yee Ahearn Huffman (2004)). Large farms could better manage extreme events than small farms (El Benni, Finger, & Mann, 2012; Finger et al., 2018). Various farm enterprises, i.e., different types of farming, have been found to face different income variability (Simone Severini, Tantari, and Di Tommaso 2017). Similarly, production intensity affects production risk (see, e.g. (McBride and Greene 2009; Gardebroek, Chavez, and Lansink 2010; D'Antoni and Mishra 2012; Schläpfer, Tucker, and Seidl 2002). On-farm diversification is expected to decrease downside income risk (e.g., Di Falco, Penov, Aleksiev, & van-Rensburg, (2010)). Similar considerations apply to farm financial characteristics, such as cost flexibility, availability of liquidity, use of credit, and farm insurances (El Benni et al. 2012; Hardaker et al. 2015; Jodha 1981; Robison and Barry 1987). Farmers' characteristics, such as gender and age, have been found to influence risk preferences and risk perceptions (Hartog, Ferrer-i-Carbonell, and Jonker 2002; Menapace, Colson, and Raffaelli 2013; Sherrick et al. 2004). Agricultural policies can also affect farm income risk. In particular, in the EU, a large share of farm income is generated by CAP direct payments that have been found to reduce income variability, being a relatively stable income source (El Benni, Finger, and Meuwissen 2016; Finger and Lehmann 2012; Simone Severini, Tantari, and Di Tommaso 2016). Finally, income volatility experienced in the past is also a characteristic of the mutual fund that could predict the size of the indemnities.

---

[50] We use the definition of (Sims 2002) derived from Hurwicz (1966) "*A model is structural if it allows us to predict the effect of "interventions" — deliberate policy actions, or changes in the economy or in nature of known types. To make such a prediction, the model must tell us how the intervention corresponds to changes in some elements of the model (parameters, equations, observable or unobservable random variables), and it must be true that the changed model is an accurate characterization of the behaviour being modeled after the intervention.*"

Based on these considerations and the availability of FADN data, the variables used in the model refer to aspects such as farm size, farm production, and financial characteristics, farm policies. For crucial variables, such as income, the model includes the level and the variability over the previous three years as standard deviation[51]. Unfortunately, we cannot account for the off-farm incomes because the FADN does not provide this kind of information. Some descriptive statistics of the main characteristics of the farms in the sample are reported.

**Table 9. Descriptive statistics of the farms in the sample and differences between indemnified and not indemnified farms (In Italics Standard Deviations).**

| Description | Code | All farms | Indemnified farms | Not indemnified farms | Wilcoxon test Our Sample /All Sample |
|---|---|---|---|---|---|
| *Farm characteristics* | | | | | |
| Utilised agricultural land [Ha] | LND | 33.29 | 29.04 | 34.50 | 0.01 |
| | | *(57.20)* | *(51.85)* | *(58.58)* | |
| Livestock Units [LU] | LU | 49.35 | 40.51 | 51.86 | 0.45 |
| | | *(363.39)* | *(310.29)* | *(377.07)* | |
| Labour input [AWU] | LAB | 1.96 | 1.66 | 2.05 | 0.00 |
| | | *(2.47)* | *(1.96)* | *(2.58)* | |
| Total revenues [€] | REV | 141011.40 | 96812.38 | 153560.50 | 0.00 |
| | | *(433029.20)* | *(385837.60)* | *(444722.60)* | |
| *Farm production characteristics:* | | | | | |
| Livestock intensity [LU/Ha] | LU_I | 3.64 | 2.29 | 4.03 | 0.65 |
| | | *(105.05)* | *(37.47)* | *(117.34)* | |
| Machinery intensity [kWh/Ha] | MACHIN | 16.00 | 17.15 | 15.68 | 0.00 |
| | | *(30.69)* | *(31.47)* | *(30.46)* | |
| Labor intensity [UL/ha] | LAB | 0.28 | 0.27 | 0.28 | 0.10 |
| | | *(0.86)* | *(0.68)* | *(0.91)* | |
| Land productivity [€/ha] | VA | 7769.13 | 4310.73 | 8751.05 | 0.90 |
| | | *(52011.44)* | *(16263.52)* | *(58256.51)* | |
| Specialisation [Herfindhal Index] | H Index | 0.67 | 0.66 | 0.67 | 0.00 |
| | | *(0.25)* | *(0.26)* | *(0.25)* | |
| Other Gainful Activities [%] | OGA | 0.36 | 0.10 | 0.43 | 0.00 |

---

[51] We have not considered a longer interval to calculate the standard deviation because this would have resulted in a excessive reduction of the available sample size.

|  |  |  | (8.61) | (3.72) | (9.55) |  |
|---|---|---|---|---|---|---|
| *Farm financial characteristics:* | | | | | | |
| Fixed cost [€/ha] | FXCOST | 1642.36 | 1511.42 | 1679.53 | 0.00 |
|  |  | (7405.89) | (4713.10) | (8006.70) |  |
| Current over total costs [%] | CURCOST | 0.65 | 0.65 | 0.64 | 0.00 |
|  |  | (0.19) | (0.20) | (0.18) |  |
| Labour over total costs [%] | LBRCOST | 0.20 | 0.19 | 0.20 | 0.30 |
|  |  | (0.15) | (0.15) | (0.15) |  |
| Insurance premia over total costs [%] | INSURE | 0.02 | 0.02 | 0.02 | 0.00 |
|  |  | (0.04) | (0.04) | (0.04) |  |
| Relative amount of fixed capital [%] | FXK | 0.57 | 0.53 | 0.58 | 0.00 |
|  |  | (0.93) | (0.30) | (1.04) |  |
| Relative amount of debts [%] | DEBT | 0.02 | 0.02 | 0.02 | 0.00 |
|  |  | (1.21) | (0.09) | (1.37) |  |
| Relative amount of net worth [%] | NETK | 0.98 | 0.98 | 0.98 | 0.00 |
|  |  | (1.21) | (0.09) | (1.37) |  |
| *Farm policies* | | | | | | |
| Decoupled Direct Payments [€] | DDP | 47.31 | 47.28 | 47.32 | 0.00 |
|  |  | (240.72) | (272.54) | (230.90) |  |
| Coupled Direct Payments [€] | CDP | 48.66 | 36.54 | 52.09 | 0.00 |
|  |  | (332.34) | (261.16) | (349.85) |  |
| Rural Development Payments - Agroenvironmental [€] | RDP_AES | 56.73 | 58.51 | 56.23 | 0.00 |
|  |  | (193.89) | (159.58) | (202.57) |  |
| Rural Development Payments - Less Favoured Area [€] | RDP_LFA | 35.34 | 36.76 | 34.93 | 0.00 |
|  |  | (189.79) | (127.10) | (204.10) |  |
| Rural Development Payments for Investments [€] | RDP_INV | 30.15 | 31.47 | 29.78 | 0.00 |
|  |  | (994.90) | (1588.64) | (744.55) |  |
| *Other farm characteristics:* | | | | | | |
| Number | N_FARMS | 47898 | 10592 | 37306 |  |
| Individual farms | INDIV | 41742 | 9352 | 32390 |  |
| Gender of the holder | MALE | 38090 | 8353 | 29737 |  |
| Young holder | YOUNG | 4699 | 946 | 3753 |  |
| Organic farms | ORGAN | 5455 | 1246 | 4209 |  |
| Plain regions | PLAIN | 10967 | 2391 | 8576 |  |
| Hill regions | HILL | 21028 | 4519 | 16509 |  |
| Mountain regions | MOUNT | 15903 | 3682 | 12221 |  |

*Source: Own elaborations on Italian FADN data.*

### 3.2.4 Results

**Econometric evaluation**

As already said earlier, we look at the goodness-of-fit of the models (referring to $R^2$ and $RMSE$) and the number of regressors.

To consider the overfit, we compare $R^2$ in the training-sets (Figure 3) with $R^2$ in test-sets (

Figure 4) using all the resample in all the years. Overfitting is dramatically noticeable for GLM; this is observable when one compares $R^2$ in the training-sets, which is very high, with a low value resulting in the test-sets. This finding is consolidated in the literature that found that the $R^2$ for GLM with too many parameters is (Breiman 2001). ML is less affected by this problem showing substantial equivalent performances between training-sets and test-sets (Figure 3 and Figure 4). Boosting has a less evident effect: the $R^2$ in the training-sets (Figure 3) is way higher than the $R^2$ in test-sets, on average.

As expected, the overall goodness-of-fit of models is not exceptionally high; GLM has the worst performance with the $R^2$ in test-sets close to zero. In contrast, most of the $R^2$ in test-sets in the considered ML approaches range between 0.1 and 0.3 depending on the model.

Another essential aspect to consider is the variability of the goodness-of-fit because a low variability suggests a relatively high reliability of the model under different conditions (i.e. years and sub-samples). In other words, for the same average goodness-of-fit, it should be preferred a model with limited variability of econometric performances. This aspect also concerns the multiple DGPs that is generable, as indicated in the previous section. To most stable selection correspond a more validity of estimation of real DGP (Rashomon's effect in Breiman, 2001).

In this regard, Boosting, while having an average goodness-of-fit similar to that of the other two ML approaches, has a lower variance in comparison with them (

Figure 4). This result suggests that Boosting is more reliable than the different considered methods in the considered case study.

**Figure 3 - $R^2$ on Training-Sets**

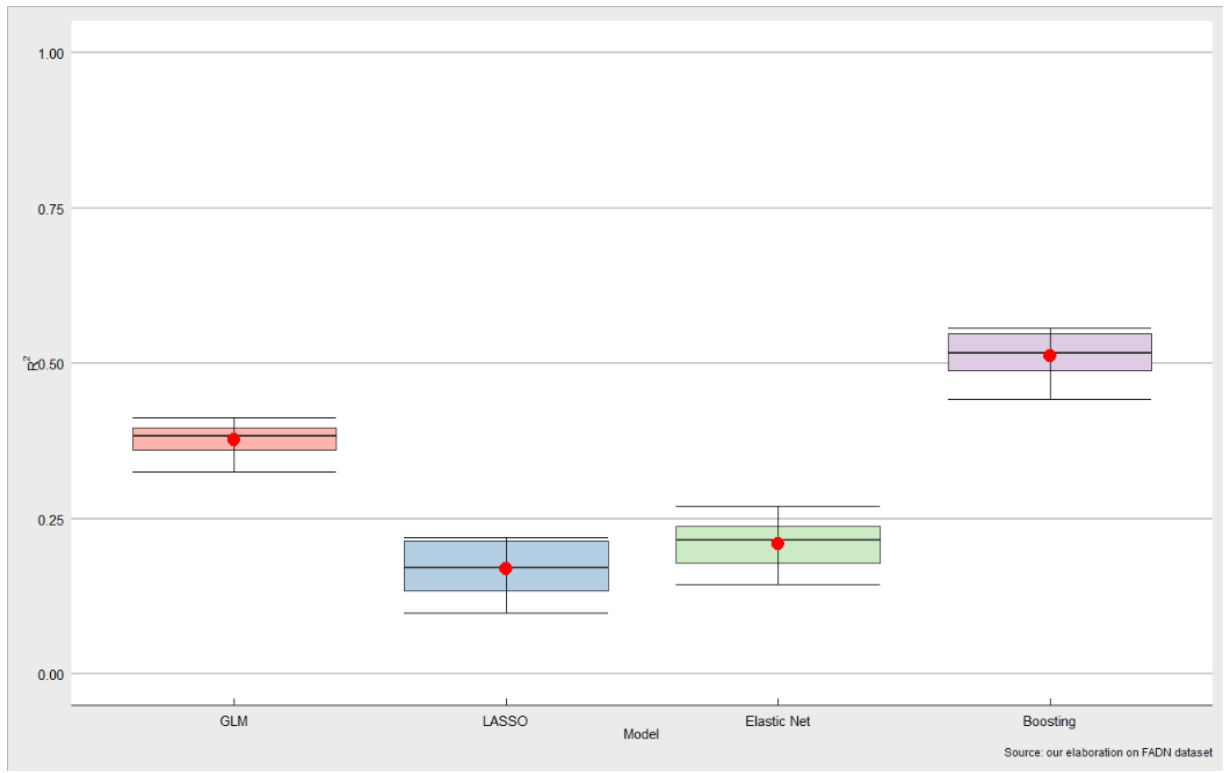

# Figure 4 - $R^2$ in Test-Sets

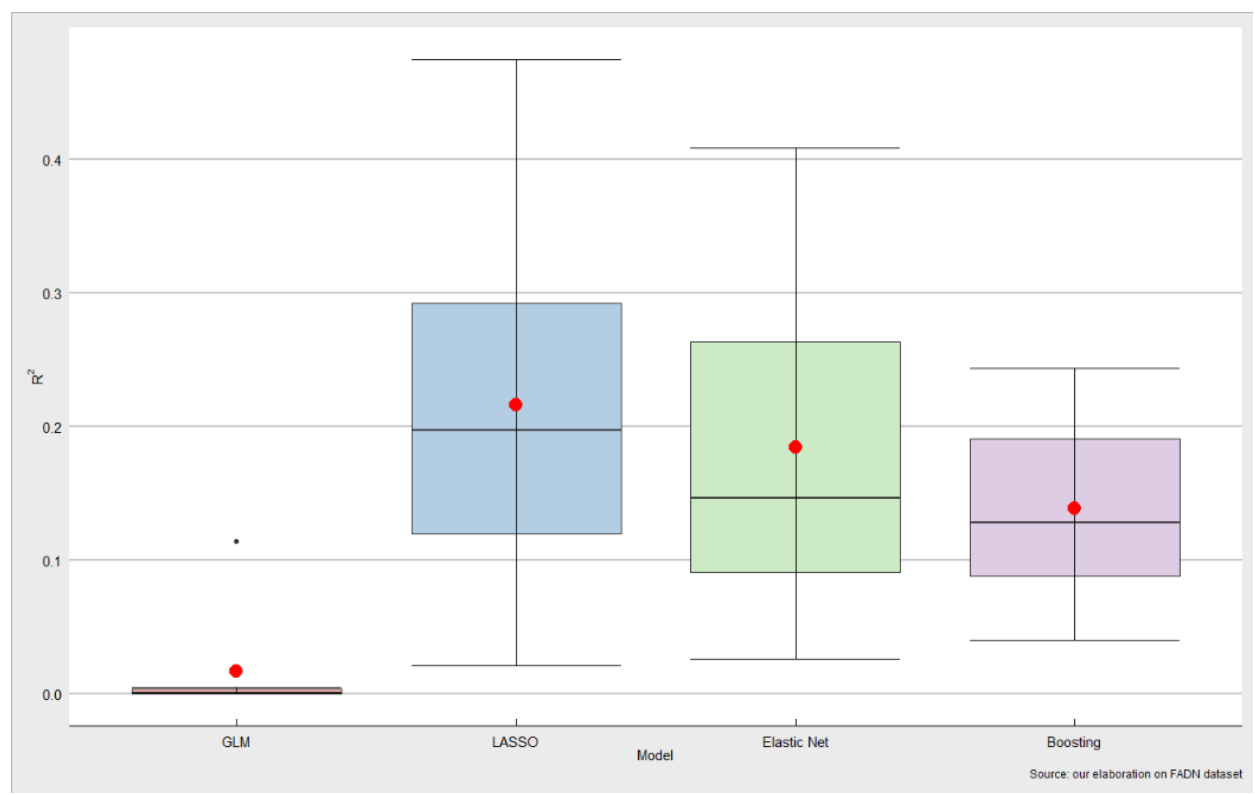

## Table 10 – Mean $R^2$ of the test-sets in all years (sd in parenthesis in italics)

| Model | 2012 | 2013 | 2014 | 2015 | 2016 | 2017 | 2018 | Mean |
|---|---|---|---|---|---|---|---|---|
| GLM | 0.000 | 0.114 | 0.005 | 0.000 | 0.001 | 0.000 | 0.002 | 0.017 |
|  | *(0.000)* | *(0.121)* | *(0.016)* | *(0.000)* | *(0.003)* | *(0.000)* | *(0.003)* | *(0.020)* |
| LASSO | 0.474 | 0.292 | 0.116 | 0.123 | 0.197 | 0.021 | 0.292 | 0.216 |
|  | *(0.076)* | *(0.179)* | *(0.024)* | *(0.100)* | *(0.162)* | *(0.017)* | *(0.241)* | *(0.114)* |
| EN | 0.408 | 0.379 | 0.100 | 0.082 | 0.148 | 0.027 | 0.147 | 0.184 |
|  | *(0.228)* | *(0.039)* | *(0.016)* | *(0.071)* | *(0.113)* | *(0.025)* | *(0.126)* | *(0.088)* |
| Boosting | 0.187 | 0.243 | 0.117 | 0.195 | 0.128 | 0.040 | 0.060 | 0.139 |
|  | *(0.086)* | *(0.080)* | *(0.054)* | *(0.049)* | *(0.052)* | *(0.023)* | *(0.026)* | *(0.053)* |

*Standard deviation in parenthesis in italics - Source: Own elaborations on Italian FADN data.*

The stability of the $R^2$ in test-sets over the years can be better appreciated in reading Table 10. Goodness-of-fit varies significantly over the years for LASSO and EN (e.g. ranging from 0.021 to 0.474), while for Boosting it goes less from 0.040 to 0.243.

We also report the Root Mean Squared Errors (RMSE because, as said earlier, it is very often used in the analyses using ML) (Hastie, Tibshirani, and Friedman 2009; P. Murphy 1991; Van der Paal 2014). The results confirm how indicated for $R^{2,}$ in particular, the very poor goodness-of-fit of the GLM (Table 11). However, it suggests that Boosting has the lowest level of RMSE in many years compared to the other two ML approaches. Furthermore, the level of RMSE for Boosting is way more stable over the years compared to the other two ML approaches, where the level of RMSE wanders over the years.

**Table 11 – Mean RMSE of the test-sets in all years**

| Model | 2012 | 2013 | 2014 | 2015 | 2016 | 2017 | 2018 | Mean |
|---|---|---|---|---|---|---|---|---|
| GLM | 8.00E+15 | 4.70E+16 | 2.10E+13 | 4.80E+22 | 7.90E+78 | 4.30E+57 | 6.50E+102 | 9.30E+101 |
|  | *(2.30E+16)* | *(1.50E+17)* | *(5.90E+13)* | *(1.50E+23)* | *(1.60E+79)* | *(1.30E+58)* | *(2.00E+103)* | *(2.80E+102)* |
| LASSO | 5.44E+05 | 3.90E+07 | 5.33E+04 | 1.02E+08 | 6.09E+04 | 9.23E+05 | 5.48E+04 | 2.03E+07 |
|  | *(1.56E+06)* | *(1.23E+08)* | *(6.40E+03)* | *(2.22E+08)* | *(7.77E+04)* | *(2.42E+06)* | *(6.70E+04)* | *(4.99E+07)* |
| EN | 1.90E+08 | 3.50E+07 | 6.40E+04 | 3.80E+09 | 8.10E+06 | 6.40E+06 | 5.80E+25 | 8.20E+24 |
|  | *(5.90E+08)* | *(1.10E+08)* | *(2.10E+04)* | *(1.20E+10)* | *(2.50E+07)* | *(2.00E+07)* | *(1.80E+26)* | *(2.60E+25)* |
| Boosting | 4.98E+04 | 5.88E+04 | 5.27E+04 | 3.70E+04 | 3.58E+04 | 3.26E+04 | 3.32E+04 | 4.28E+04 |
|  | *(2.65E+03)* | *(2.58E+03)* | *(3.03E+03)* | *(3.57E+03)* | *(1.04E+03)* | *(8.68E+02)* | *(1.71E+03)* | *(2.21E+03)* |

*Standard deviation in parenthesis in italics - Source: Own elaborations on Italian FADN data.*

The number of selected regressors is the other considered aspect (Figure 5). This feature is essential because fewer variables are necessary to reduce the costs related to data gathering and processing. While GLM uses all the 129 regressors, all ML approaches perform variable selection. However, they differ in this regard: EN and, even more, LASSO are very parsimonious models selecting from 4 to 15 regressors and 10 to 24 regressors, respectively, depending on the years. In contrast, Boosting is less parsimonious because it selects around 50 (42 ÷ 54) of the available variables as regressors (Occam's effect in Breiman, 2001 ).

**Figure 5 – Selected regressors in all years and models.**

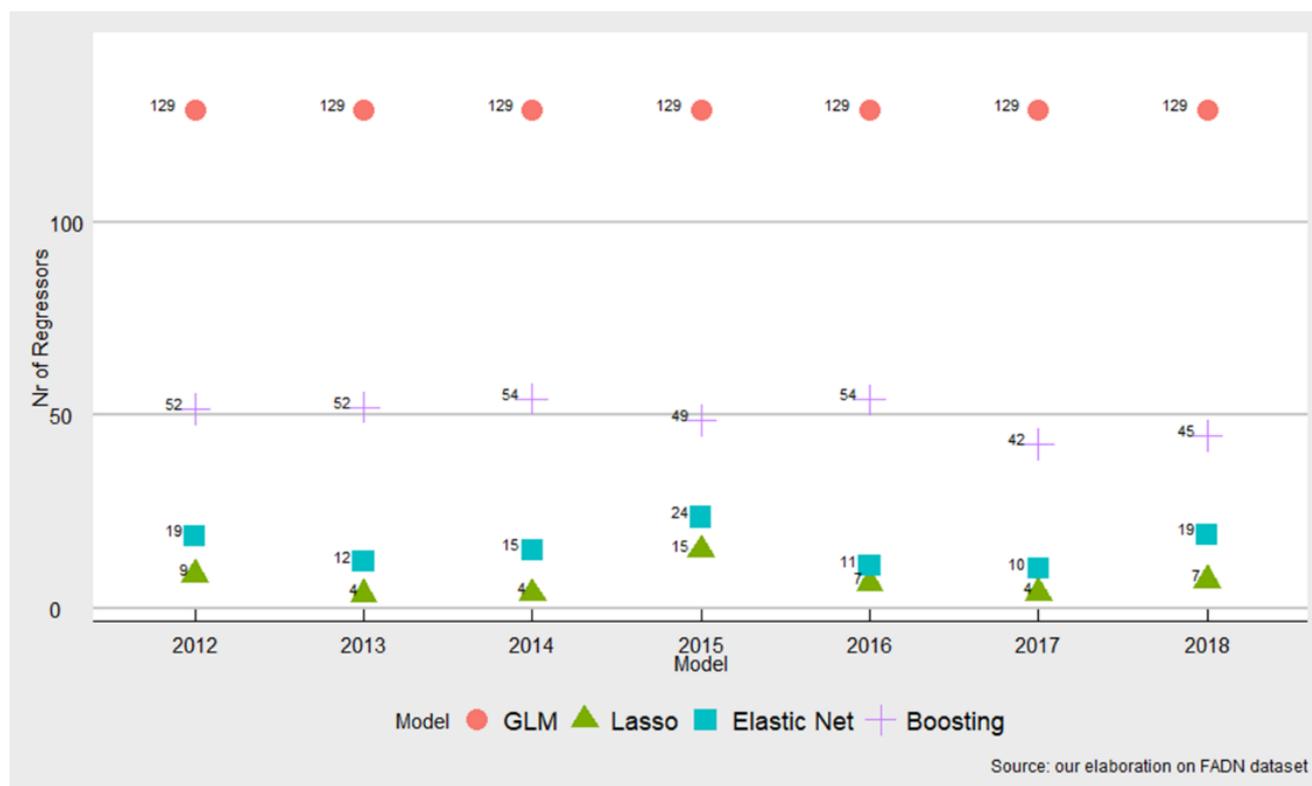

It is also important to note that the number of selected regressors is more or less the same in the considered years. This characteristic is particularly evident in Boosting. This result demonstrates the relative stability of DGP identified by using the ML approaches and, in particular, using Boosting.

However, it is essential to evaluate the frequency of selection to assess whether the selected regressors are the same in the different considered cases (resamples and years) to fully assess the stability of the ML models (Bellman's Effect in Breiman, 2001). Furthermore, the analysis of the selected regressors can explain to discover of the variables that better describes the DGP (

Table 12).

First of all, the stability of DGPs obtained by Boosting is more extensive than in the case of EN and LASSO. This result is in line with theory: more selective ML tools have a lower probability of finding the real DGP. In particular, the variables identified by LASSO are less frequently selected (i.e. lower selection rate) than EN and Boosting: the variables have a frequency of selection always lower than 50%. EN is less selective than LASSO and has a higher selection rate (i.e., the highest selection rate is roughly 75%). In the case of Boosting, we have ten variables that are always selected and many others with a very high selection rate compared to EN and LASSO. This is a peculiarity of the estimation of the DGP obtained by ML.

We assumed to have only one GDP per year on all farms. This assumption is motivated by the need to keep the analysis simple and to get easily interpretable results. The trade-off between complexity and interpretability is evaluated, considering the stability of the models. In this regard, Boosting seems to perform better than the other ML models.

**Table 12 – First 50 variables selected per model: percentage of selection in all sample (10) and all years (7) for a total of 70 samples[52].**

| LASSO | | | EN | | | Boosting | | |
|---|---|---|---|---|---|---|---|---|
| Coef | n | % on total | Coef | n | % on total | Coef | n | % on total |
| AVG_VA | 38 | 54% | sd_VA | 53 | 76% | AVG_VA | 70 | 100% |
| sd_VA | 32 | 46% | AVG_VA | 48 | 69% | CUR_COST_LAB_sd | 70 | 100% |
| TR_sd | 28 | 40% | FNVA_sd | 43 | 61% | FNVA_sd | 70 | 100% |
| L_IMM_L1 | 21 | 30% | L_IMM_L1 | 43 | 61% | K_CIRC_L1 | 70 | 100% |
| L_IMM_sd | 19 | 27% | TR_sd | 43 | 61% | K_CIRC_sd | 70 | 100% |
| LAB_COS_sd | 18 | 26% | AWU_sd | 35 | 50% | L_IMM_L1 | 70 | 100% |
| KW_M_L1 | 17 | 24% | LAB_COS_sd | 34 | 49% | L_IMM_sd | 70 | 100% |
| BREED_sd | 15 | 21% | KW_M_L1 | 33 | 47% | NET_K_sd | 70 | 100% |
| AWU_sd | 13 | 19% | L_IMM_sd | 31 | 44% | sd_VA | 70 | 100% |
| FNVA_sd | 12 | 17% | K_CIRC_sd | 29 | 41% | TR_sd | 70 | 100% |
| Land_PROD_Index_sd | 11 | 16% | BREED_sd | 26 | 37% | FNVA_L1 | 69 | 99% |
| COST_PLUR_sd | 10 | 14% | CUR_COST_rat_L1 | 21 | 30% | TR_L1 | 69 | 99% |
| K_CIRC_sd | 10 | 14% | Land_PROD_Index_sd | 21 | 30% | K | 68 | 97% |
| CUR_COST_rat_L1 | 8 | 11% | COST_PLUR_sd | 20 | 29% | VA_L1 | 68 | 97% |
| MECC_sd | 8 | 11% | ALT.3 | 19 | 27% | CUR_COST_LAB_L1 | 67 | 96% |
| NET_K_L1 | 8 | 11% | K_CIRC_L1 | 18 | 26% | LAB_COS_L1 | 66 | 94% |
| ALT.3 | 7 | 10% | MECC_RENTAL_COST_sd | 17 | 24% | AWU_L1 | 65 | 93% |
| FIX_COST_L1 | 7 | 10% | MR.NOC | 17 | 24% | CUR_COST_0 | 65 | 93% |
| Land_LEASE_sd | 7 | 10% | AWU_L1 | 15 | 21% | NET_K_L1 | 65 | 93% |
| MECC_RENTAL_COST_sd | 7 | 10% | ToF.5 | 15 | 21% | Total_COST | 65 | 93% |
| MR.NOC | 7 | 10% | FIX_COST_L1 | 14 | 20% | FIX_COST_L1 | 63 | 90% |
| ToF.5 | 7 | 10% | GEND | 14 | 20% | EXTRACAR_L1 | 62 | 89% |
| CDP_L1 | 6 | 9% | FAWU_L1 | 13 | 19% | LAB_COS_sd | 60 | 86% |
| FIX_COST_sd | 6 | 9% | Land_LEASE_sd | 13 | 19% | MECC_sd | 60 | 86% |
| ToF.3 | 6 | 9% | MECC_sd | 13 | 19% | DDP_L1 | 58 | 83% |
| ToF.7 | 6 | 9% | NET_K_L1 | 13 | 19% | Land_RENTAL_COST_sd | 57 | 81% |
| VA_L1 | 6 | 9% | ToF.3 | 13 | 19% | MECC_L1 | 57 | 81% |
| FXK_rat_sd | 5 | 7% | KFIX_L1 | 12 | 17% | PSA_L1 | 57 | 81% |
| HHI_L1 | 5 | 7% | CDP_L1 | 11 | 16% | FIX_COST_sd | 56 | 80% |
| KFIX_L1 | 5 | 7% | DEPREC_L1 | 11 | 16% | Land_RENTAL_COST_L1 | 54 | 77% |
| LU_L1 | 5 | 7% | EXTRACAR_sd | 11 | 16% | LU_L1 | 52 | 74% |
| AgriT_L1 | 4 | 6% | FIX_COST_sd | 11 | 16% | LU_sd | 52 | 74% |
| GEND | 4 | 6% | Land_RENTAL_COST_L1 | 11 | 16% | COST_PLUR_L1 | 51 | 73% |
| HHI_sd | 4 | 6% | VA_L1 | 11 | 16% | KFIX_L1 | 47 | 67% |
| LU_INT_Index_L1 | 4 | 6% | CUR_COST_LAB_sd | 10 | 14% | PSA_sd | 47 | 67% |
| MR.INS | 4 | 6% | FNVA_L1 | 10 | 14% | AWU_sd | 46 | 66% |
| MR.NOR | 4 | 6% | FXK_rat_sd | 10 | 14% | Insurance_L1 | 46 | 66% |
| ORGANIC | 4 | 6% | HHI_L1 | 10 | 14% | FXK_rat_L1 | 45 | 64% |
| PSA_L1 | 4 | 6% | MECC_L1 | 10 | 14% | Land_PROD_Index_sd | 44 | 63% |
| ToF.8 | 4 | 6% | ToF.7 | 10 | 14% | BREED_sd | 43 | 61% |
| EXTRACAR_sd | 3 | 4% | COST_PLUR_L1 | 9 | 13% | MECC_RENTAL_COST_sd | 43 | 61% |
| FAWU_L1 | 3 | 4% | DEPREC_sd | 9 | 13% | DEPREC_L1 | 42 | 60% |
| INSURE_rat_L1 | 3 | 4% | PSA_L1 | 9 | 13% | KW_M_L1 | 42 | 60% |
| Land_RENTAL_COST_L1 | 3 | 4% | LBR_COST_rat_L1 | 8 | 11% | OWN_UAA_L1 | 41 | 59% |
| NET_K_sd | 3 | 4% | OWN_UAA_sd | 8 | 11% | COST_PLUR_sd | 40 | 57% |
| OGA_sd | 3 | 4% | RDP_A_AES_sd | 8 | 11% | FAWU_L1 | 39 | 56% |
| RDP_A_AES_L1 | 3 | 4% | HHI_sd | 7 | 10% | CDP_L1 | 37 | 53% |
| RDP_A_AES_sd | 3 | 4% | INSURE_rat_L1 | 7 | 10% | Insurance_sd | 36 | 51% |
| RDP_A_LFA_sd | 3 | 4% | MR.INS | 7 | 10% | LIABILITY_sd | 36 | 51% |
| RDP_O_sd | 3 | 4% | NET_K_sd | 7 | 10% | MACH_L1 | 34 | 49% |

*Source: Own elaborations on Italian FADN data.*

This analysis shows that some variables represent the core of DGP in the case of our case study. These are, among others, the average Value Added and its standard deviation (AVG_VA, sd_VA), the standard deviation of the circulating capital (K_CIRC_sd), the level of liquidity available on-farm and its standard deviation (L_IMM_L1 and L_IMM_sd), as well as the standard deviation of the total farm revenues (TR_sd) and the relative importance of labour costs (LAB_COS_L1). These results allow us to conclude that the core of DGP depends on the economic or financial aspect.

---

[52] The full table is reported in the online appendix.

Considering that Boosting has more stability than the other ML tools, we analyse the result of this one.

The relevance of CAP for IST is less important in comparison with other variables, and only DDP and CDP have relatively high stability of selection.

Note that the selected variables are in most of the variables that are affected by farmers' behaviour (i.e., signals as Spence (1973)) and not stable characteristics of the farm and farmers characteristics (i.e., indices as  ). This theory is at the base of the year-to-year approach that we used in the estimation.

Additionally, note that some of the levels of the selected variables (e.g. current costs, Value Added, liquidity and labour) can be more easily and (less costly) changed than other variables such for example, amount of land, fixed costs and level of indebtedness. These aspects are coherent with the theory (Spence 1973, 1978, 2002)

Finally, it is essential to assess whether a trade-off exists between the number of selected regressors and goodness-of-fit in the considered ML approaches.

**Figure 6 – Comparison of different models using $R^2$ and the number of selected regressors**

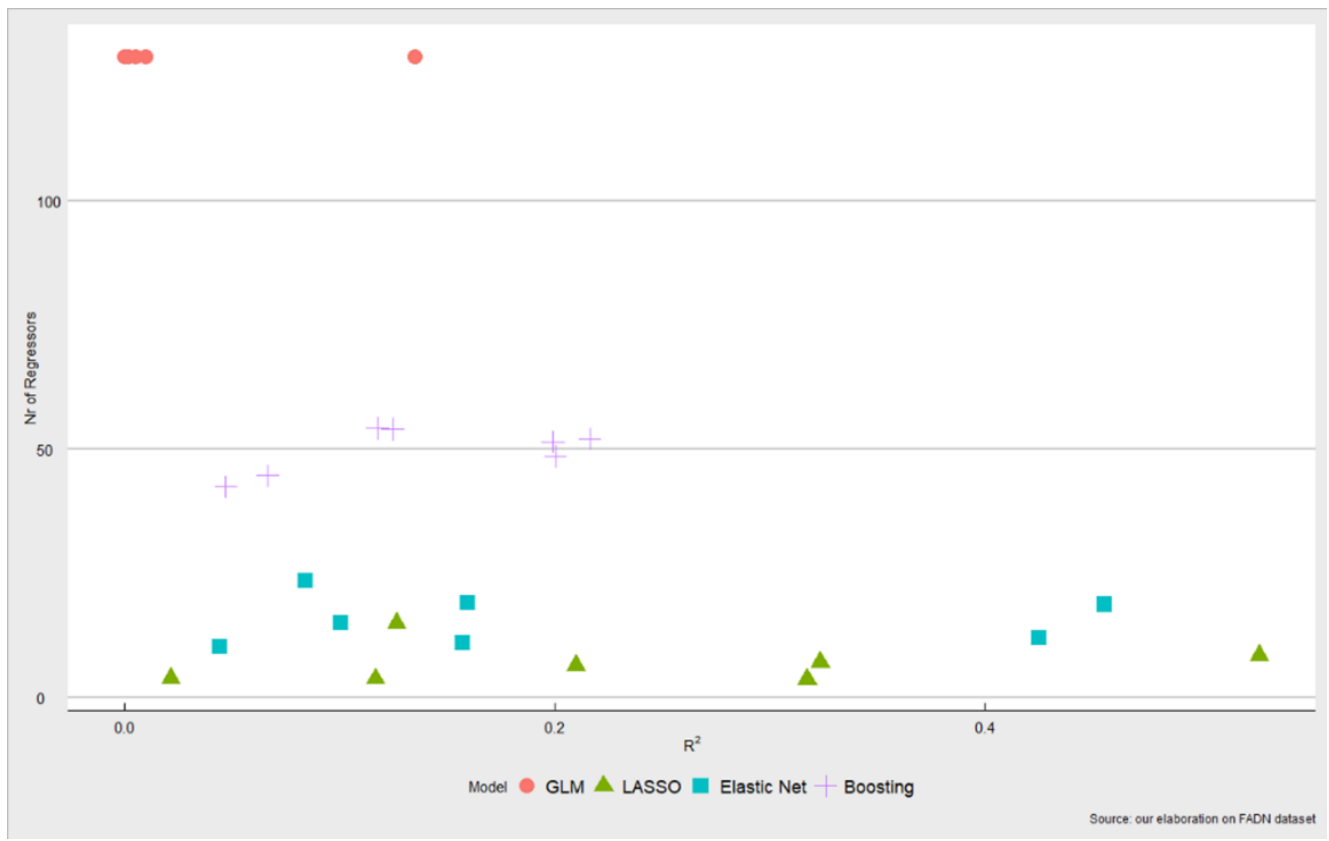

Figure 6 shows that a high number of regressors (i.e. much information) do not necessarily correspond to a high goodness-of-fit. This result is quite counterintuitive if considered based on the classical inference tools where generally a high number of independent variables corresponds to a high value of non-correct $R^2$.

### Economic evaluation

The use of estimates for the definition of premiums is assessed according to three main dimensions already discussed in the methodology section: i) compatibility of premium with farm income, ii) overall balance of mutual fund and solvency of scheme, iii) distribution of the individual difference between premium and indemnity.

The share of farms with a premium compatible with their income varies according to the considered threshold. When an extensive level of compatibility is considered (i.e., the premium should not exceed the available farm income), all four models allow having a considerable share of the farms (around 93% or above) for which premium is compatible with their income (Figure 7). However, the percentage of farms with a premium compatible with its income level firmly declines when more tight compatibility constraints are considered. In particular, less than 53% of farms considered have a premium that does not exceed 10% of their available income. This figure suggests that not considering the income compatibility of the ratemaking could lead to a relevant overestimate of the degree of farmers' participation in the new insurance scheme.

What is more important is that the four considered models provide very different results regarding the share of farms with a compatible premium when considering reasonably tight constraints. In particular, when the constrain is set at 10% of the farm income, the use of the Boosting results allows a more considerable degree of compatibility than the other ML approaches. This result suggests that using Boosting ensures a more considerable participation, *ceteris paribus*. Note that the use of GLM results could provide, in theory, an even larger share of farms with compatible premium. However, how it will be shown shortly, this is because the resulting ratemaking, in this case, is not financially sustainable: the premiums on average are too low to ensure a sound economic balance for the mutual fund.

**Figure 7 - Number of compatible observations under different levels of premium vs income compatibility (Premium as a share of farm Value Added) for all year and the four models.**

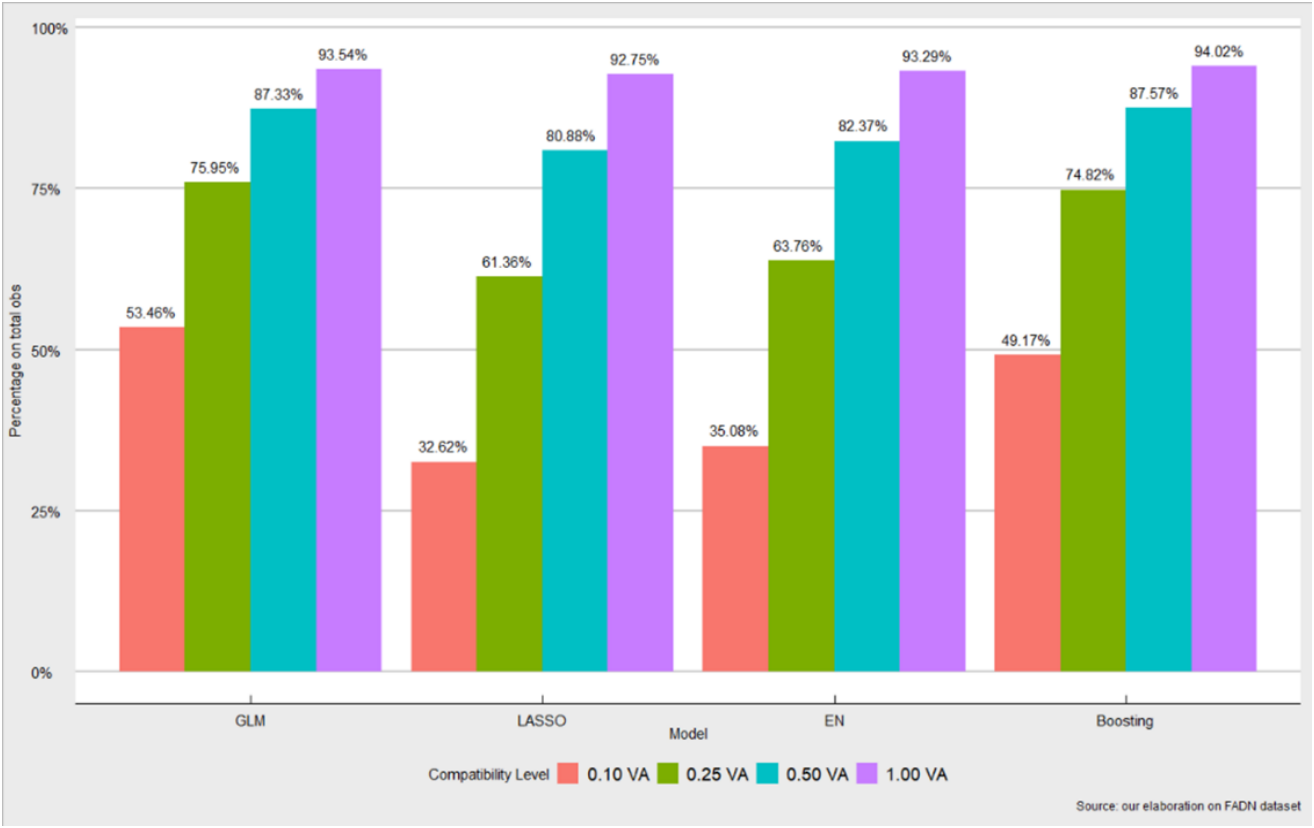

It is crucial to assess the multiannual balance of the mutual fund to ensure the economic sustainability of the scheme, i.e., the sum of the premiums minus indemnities over a reasonably long time (Figure 8).

As already said, using the results of the GLM for defining the premiums yields very negative consequences for the mutual fund: the amount of indemnities exceeds that of the received premiums. Note that this is true for all four considered compatibility levels. The multiannual balance is more satisfactory using the results of the ML models, at least when reasonably levels of compatibility are considered. In particular, balance is very close to zero when compatibility is set at 10% of the farm income with all three models. However, the LASSO and EN perform better than Boosting in this regard. This result is, even more, the case when slightly less tight compatibility constrains is applied (i.e., premium < 25% of farm income). This aspect suggests that the ML approaches, particularly EN and LASSO, allow reaching more satisfactory results in terms of the overall multiannual balance of the new scheme. Also, in this case, it is worth mentioning how results change when applying various compatibility constraints.

**Figure 8 - Multiannual balance of the mutual fund (Sum of Premium - Indemnities in the period 2012-2018) for different levels of premium vs income compatibility in the four models.**

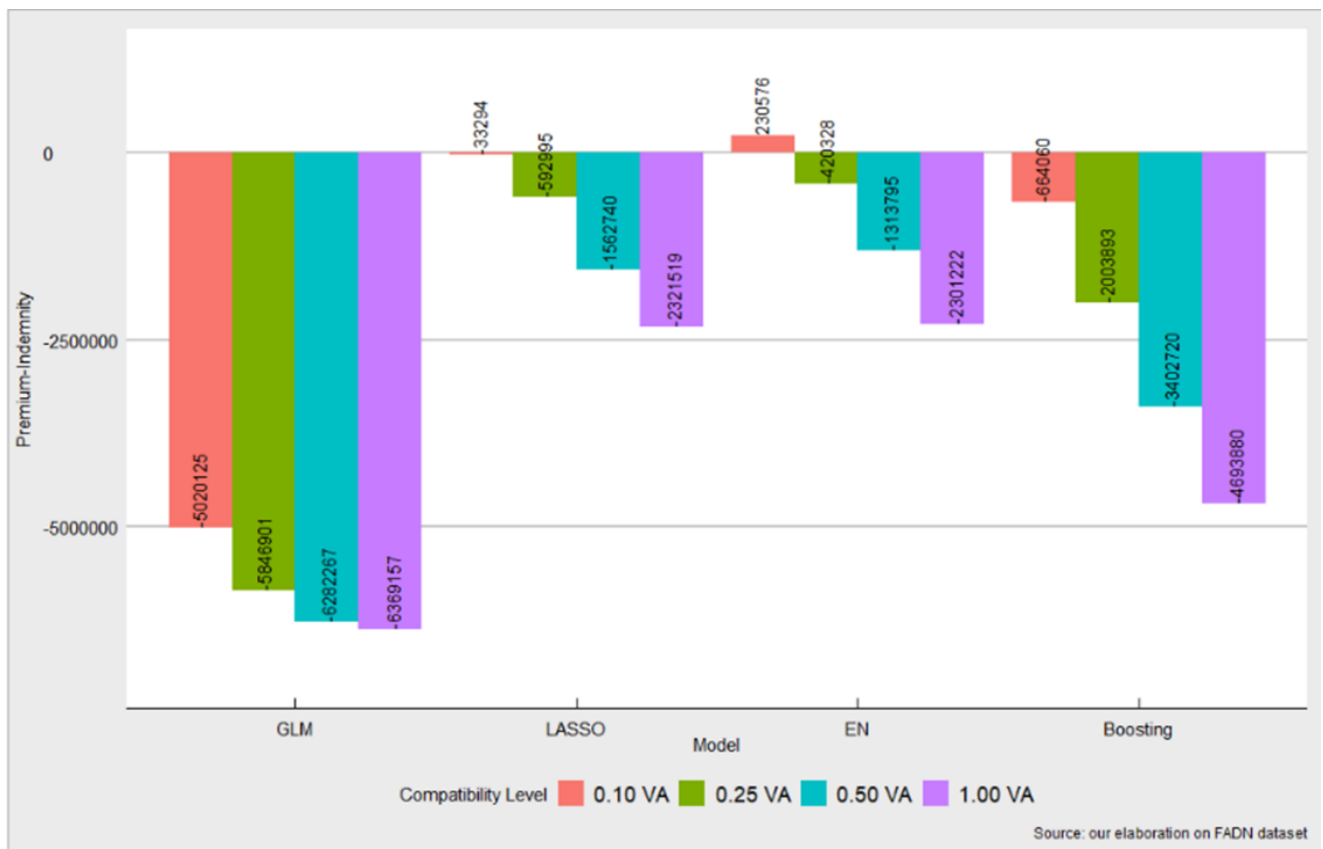

Apart from considering the multiannual balance, it is essential also to consider the evolution of the annual balances to identify whether in some years the balance reaches a significant negative level that could cause solvency problems for the mutual fund. This information is essential because it may suggest the mutual fund to have a high enough level of precautionary capital to meet claims and earning capacity.

Figure 9 provides the annual balances concerning the tighter level of compatibility level (i.e. premium < 10% of farm value-added) for the four models. These results confirm the low economic sustainability of the ratemaking provided by directly applying the results of the GLM model. In terms of the evolution of the annual balances, the three ML models provide similar results. The worst year is in all cases 2013: in all three cases, the balance is very negative and slightly more limited in EN. Furthermore, note that the Boosting yield a negative annual balance also for 2014. This could generate solvency problems.

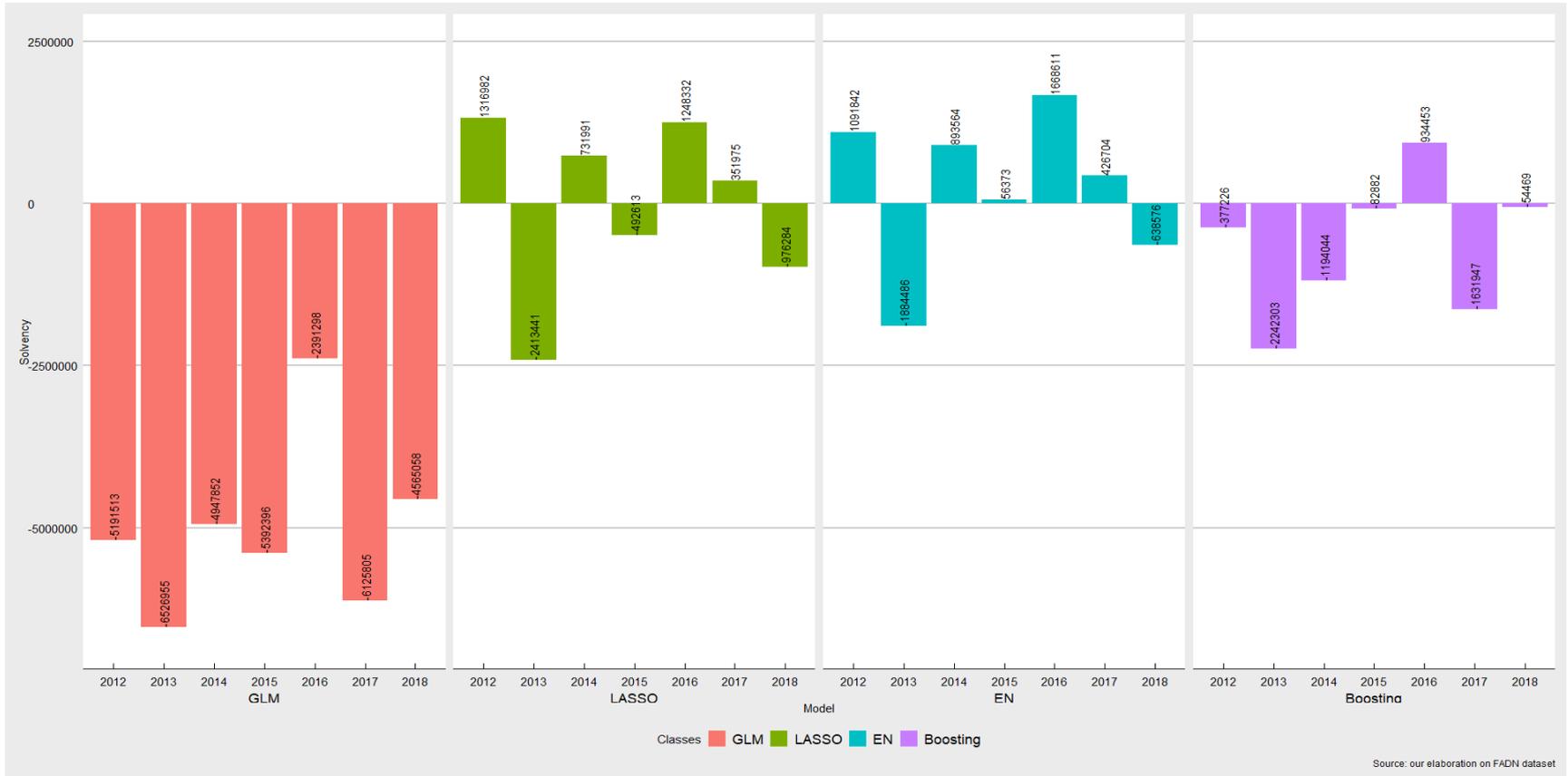

Figure 9 - Solvency per Year and model: Premium ≤ 0.10 VA.

Finally, it is crucial to examine the distribution of the difference between the premium and indemnity of each individual. We do so considering both different levels of premium vs income compatibility and some classes of $Premium - Indemnity$ or net premium (Table 13).

**Table 13 - Premium - Indemnities classes in the four models and hypothesis of Income Compatibility. Multiperiod assessment (2012-2018)**

| Level of Premium Compatability | Model | Premium-Indemnity | | | | | | | | | |
|---|---|---|---|---|---|---|---|---|---|---|---|
| | | Total | ≥0€ | ≥100k€ | ≥250k€ | ≥500k€ | ≥1000k€ | <0€ | ≤-100k€ | ≤-250k€ | ≤-500k€ | ≤-1.000k€ |
| 1.00 VA | GLM | -445840990 | 1007063903 | 30783312 | 9087101 | 4178688 | 3017438 | -1.453E+09 | -546937539 | -351696943 | -221668900 | -129013657 |
| 1.00 VA | LASSO | -162506300 | 1482155651 | 2142289 | 401001 | 0 | 0 | -1.645E+09 | -763949292 | -439568861 | -243390387 | -94744502 |
| 1.00 VA | EN | -161085546 | 1396054244 | 8592109 | 4633242 | 3267779 | 3267779 | -1.557E+09 | -655380280 | -346502625 | -168230017 | -52056948 |
| 1.00 VA | Boosting | -328571580 | 1032804790 | 15525480 | 4641342 | 785316 | 0 | -1.361E+09 | -413417525 | -227133708 | -124593220 | -47321985 |
| 0.50 VA | GLM | -439758702 | 843963721 | 9995656 | 2049352 | 654491 | 0 | -1.284E+09 | -470203077 | -305895291 | -200250986 | -119960730 |
| 0.50 VA | LASSO | -109391800 | 1278872622 | 699920 | 0 | 0 | 0 | -1.388E+09 | -619918436 | -347475336 | -188714511 | -67779411 |
| 0.50 VA | EN | -91965635 | 1208988265 | 1666985 | 278514 | 0 | 0 | -1.301E+09 | -510191837 | -255428887 | -116276108 | -28766474 |
| 0.50 VA | Boosting | -238190424 | 883284366 | 6237371 | 1841088 | 0 | 0 | -1.121E+09 | -286828412 | -152836290 | -79967970 | -27043030 |
| 0.25 VA | GLM | -409283048 | 655199052 | 4382088 | 1382391 | 654491 | 0 | -1.064E+09 | -395049549 | -270752355 | -183415103 | -112282874 |
| 0.25 VA | LASSO | -41509635 | 959979347 | 0 | 0 | 0 | 0 | -1.001E+09 | -414169895 | -219100139 | -110143595 | -33040422 |
| 0.25 VA | EN | -29422993 | 918634022 | 564061 | 278514 | 0 | 0 | -948057015 | -338966006 | -157322838 | -66513498 | -9974213 |
| 0.25 VA | Boosting | -140272486 | 695600638 | 2270953 | 681913 | 0 | 0 | -835873124 | -175806444 | -92308349 | -48580071 | -11873842 |
| 0.10 VA | GLM | -351408767 | 385797653 | 371784 | 0 | 0 | 0 | -737206419 | -302970685 | -228118448 | -164087506 | -105039460 |
| 0.10 VA | LASSO | -2330581 | 504703370 | 0 | 0 | 0 | 0 | -507033951 | -194274833 | -94043520 | -45274931 | -9788896 |
| 0.10 VA | EN | 16140326 | 494911218 | 0 | 0 | 0 | 0 | -478770892 | -149322190 | -60471587 | -22573604 | -3075054 |
| 0.10 VA | Boosting | -46484176 | 415583534 | 554507 | 399537 | 0 | 0 | -462067710 | -68244084 | -35567743 | -19246953 | -5856786 |

*Source: Own elaborations on Italian FADN data.*

A graphical analysis allows to draw interesting considerations (

Figure 10).

**Figure 10 – Distribution of different Premium-Losses classes – Compatibility level $P \leq 0.10 \times VA$ – Multiperiod assessment (2012-2018)**

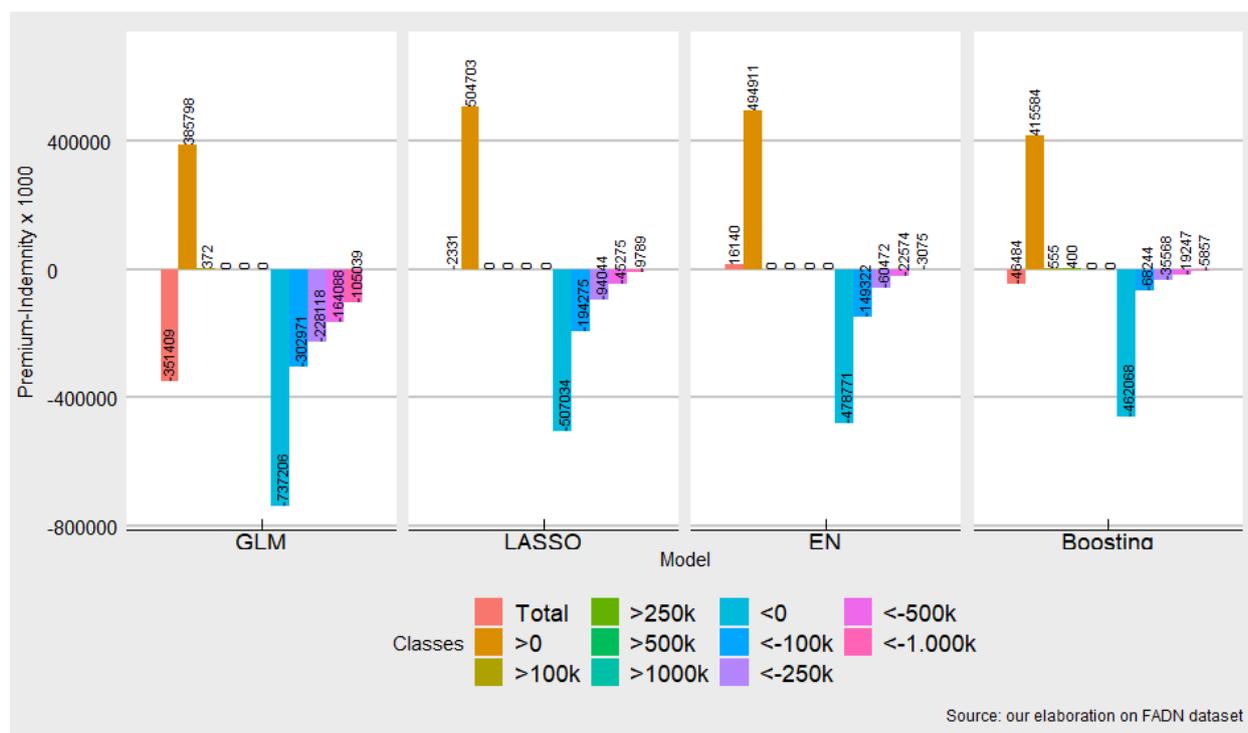

First, having a high occurrence of negative Premium-Indemnity cases is not a good outcome because it reflects the losses of the insurer and negatively affects its financial stability. Regarding this aspect, GLM has the worst performance also because of the prominence of highly negative cases. In comparison, ML tools have better performance. In particular, Boosting has better performance guaranteeing that the insurer does not take excessive risks using its results in the ratemaking.

Boosting seems to have a better performance considering the number of cases within the relevant classes of net premiums (Figure 7).

**Figure 11 - Number of farms withing different Premium - Indemnities classes in the four models. Premium vs income compatibility level: Premium ≤ 0.10 VA. The annual average of the period 2012 – 2018**

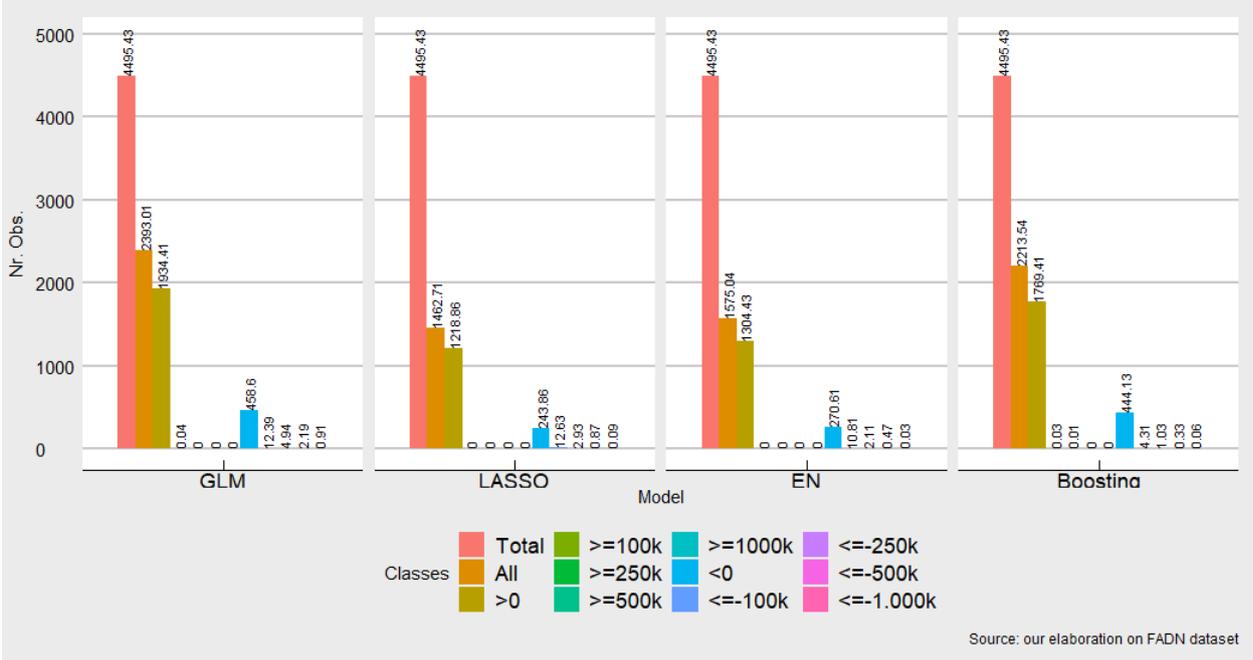

The comparison between Figure 10 and Figure 11 shows how the models predict a high number of farms to positive Premium-Indemnity difference, that is the consequence of the fact the farmer has no adverse event and then paying only the premium.

Conversely, negative values of net premiums are concentrated in roughly 20% of the total sample. This result is in line with the distribution of simulated Losses.

Boosting has an outstanding performance with a low level of Premium-Indemnity Premium lower than -100k and a minimal number of subjects with this characteristic.

The level of difference ensemble to the number of excessive loss for MF demonstrated that Boosting, overall, allows also a low expensive recognition, validation and calibration of future ratemaking.

This evaluation is synthesised in Table 14 and can justify a stop-losses threshold.

**Table 14 – Comparison between the average solvency and the number of compatible farms in different models and classes.**

| Model | Income Comp. 0.10VA | ≥0€ | of which: ≥100k€ | <0€ | of which: ≤-100k€ | ≤-250k€ | ≤-500k€ | ≤-1000k€ |
|---|---|---|---|---|---|---|---|---|
| *Premiums - Indemnities (Euro)* | | | | | | | | |
| GLM | -5,020,125 | 5,511,395 | 5,311 | -10,531,520 | -4,328,153 | -3,258,835 | -2,344,107 | -1,500,564 |
| LASSO | -33,294 | 7,210,048 | 0 | -7,243,342 | -2,775,355 | -1,343,479 | -646,785 | -139,841 |
| EN | 230,576 | 7,070,160 | 0 | -6,839,584 | -2,133,174 | -863,880 | -322,480 | -43,929 |
| Boosting | -664,060 | 5,936,908 | 7,922 | -6,600,967 | -974,915 | -508,111 | -274,956 | -83,668 |
| *Number of farms* | | | | | | | | |
| GLM | 2,393.01 | 1,934.41 | 0.04 | 458.60 | 12.39 | 4.94 | 2.19 | 0.91 |
| LASSO | 1,462.71 | 1,218.86 | 0.00 | 243.86 | 12.63 | 2.93 | 0.87 | 0.09 |
| EN | 1,575.04 | 1,304.43 | 0.00 | 270.61 | 10.81 | 2.11 | 0.47 | 0.03 |
| Boosting | 2,213.54 | 1,769.41 | 0.03 | 444.13 | 4.31 | 1.03 | 0.33 | 0.06 |

*Source: Own elaborations on Italian FADN data.*

### 3.2.5 Conclusions

The combination of Tweedie distribution with machine learning approaches guarantees an undoubted increase in prediction performance in insurance scheme such as IST. The space of indemnity is very strictly related to Tweedie pdf characteristics and not creates prediction values that are not credible ( f.e., negative and/or only discrete value): this pdf overcome many issues that are not faced by other approaches such as OLS, GLM with Poisson or negative binomial.

The robustness of predictions is, without doubt, one point-of-strength of ML that, in concomitance with the selection ability, has demonstrated a great capacity to solve the trade-off between the number of regressors and the accuracy of forecasting.

We use three ML tools with LASSO and EN featuring a very parsimonious approach while Boosting selected more variables but reaches a better goodness-of-fit performance.

In this study, we have not focused only on the econometrician aspect of the model but also on economic evaluation.

An insurance scheme is robust when it guarantees the reach of the main objective, which is equal to zero profit for MF in IST. The achievement of this goal is not without challenge.

To obtain a suitable insurance scheme, we declined the economic problem in a twofold aspect: good for ML and good for the policyholder.

First of all, how we can discern the "good" fo MF? We assess the balance sheet evaluating in the long term; to evaluate it, we use the sum of indemnities reimbursed in front of premiums received at the end of the simulation period. This can allow evaluating a compressive balance sheet that should be, in the optimal case, zero.

Nevertheless, this is not the only parameter to evaluate. We have to need to assess the sustainability of MF with a particular focus on the year-to-year economic performance: high unbalance years between Premium and Indemnity (and then more distance concerning zero profit goal) can put on the knees the MF with the only possibility to cease the future implementation of the scheme and the go bankrupt.

To overcome the issues of policyholder, usually, it has been used the mandatory assumption. We have reputed this aspect not veritable and the assumption strongly heroic. For this motivation, we have made the exercise to relax mandatory to verify the effect.

This assessment is carried out to evaluate Premium's affordability level compared with the farmer capacity to pay.

Though this approach seems naïve, thanks to it, we obtain a prime evaluation of the supply-side, never considered in other papers investigating the IST issues relaxing the heroic assumption of mandatory insurance.

Finally, we can found that Boosting is powerful tools in comparing the economic outcomes, followed by EN and LASSO. GLM can put the MF in a difficult sustainable economic position and request a large amount of precautionary capital to face the magnitude of imbalance between Premium and Indemnity and a less capacity to achieve a good level of affordability ML approach.

More in general, the analysis has provided one of the few empirical applications of machine learning in agricultural economics. The obtained results seem to confirm that it will see more of these techniques tailored and applied to economic, as stated by Storm, Baylis, and Heckelei (2020). In particular, some of the considered models seem able to efficiently select from many farm and farmers' characteristics on which the explanatory variables of interest may depend.

# Chapter 4.     Final conclusions

This thesis addressed issues concerning both the income transfer efficiency (ITE) of several CAP measures and the effects of IST (Income Stabilisation Tool) in income stabilization. The analyses were developed at the farm level, using the FADN database, but applying three different methodologies.

As far as the evaluation of ITE is concerned, the study thoroughly applied the dynamic panel approach of Blundell and Bond, the SYS-GMM, considering the fixed effects. This econometric strategy allows exploiting the potential of the panel dataset, reducing both dimensionality and omitted variables problems, besides eliminating some source of bias, such as the components of autoregressive error and collinearity problems. To the best of my knowledge, this study is the first that uses a dynamic SYS-GMM to evaluate the ITE of specific CAP measures, taking into account that they can vary according to the economic size of farms. This analysis revealed that the various CAP measures have a different ITE and that it changes according to the economic size of the farm.

The IST is a tool not yet implemented in Europe but proposed by the EU to reduce the problems of income instability. Considering that it is still under construction, the latest definition of the IST is used for developing the analysis, with a two-level approach has been developed: first, we check whether the tool is efficient in stabilizing income; second, we define/describe how to design the tool effectively and efficiently.

Concerning the first level (Section 3.1), was simulated the application of IST. Since the FADN is a representative dataset for the Italian farms, it permits us to verify whether the IST application is an effective stabilisation tool at the national level, hence using the weights assigned to each farm to have comparable results with the universe of Italian farms. Considering that the actual EU regulation does not indicate the size of the Mutual Funds, it was investigated whether different levels of aggregation (both spatial and typological) had different stabilization effects. Following this assessment, an initial verification was carried out on two hypotheses of contribution to MF: constant for each farm and proportional to the added value. This was necessary to understand if the variable contribution was sustainable and fair compared to the fixed one.

The second level of analysis refers to the design of IST (Chapter 3.2). The contribution used in the previous simulation was replaced with a premium based on the expected indemnity. This approach is much more rational than the first because the premium is established according to risk parameters. First of all, it estimated the expected indemnity using many potential explanatory variables already indicated in the literature. This is affected by two main issues: on the one hand, the distribution of the indemnity is characterized by a peculiar shape; on the other hand, concerning the variables influencing the indemnity, many are irrelevant while others not very significant and some others collinear. Two tools have been used to overcome these drawbacks: the machine learning approach using a Tweedie distribution. This study is the first applying machine learning (ML) in agricultural insurance economics to the best of my knowledge. The ML allows one to select the most relevant variables for estimating the indemnity and reduce collinearity. The Tweedie distribution is the second innovative element, which is coherent with the indemnity space. The second part of the analysis concerned the evaluation of the economic impact of this insurance-like design. It is not enough to have a premium similar to the actual indemnity. However, it is also necessary to verify at the same time two economics goals: i) the premium must be compatible with the economic capacity of the individual company (i.e. if the premium is too high compared to his income, the entrepreneur does not participate in IST) ii) the long-term profit of the MF must aim for zero, without too high annual fluctuations between premium income and indemnities paid.

The following section explains the above in detail, addressing the ITE and IST issue.

## 4.1 The role of Common Agricultural Policy in enhancing farm income: A dynamic panel analysis accounting for farm size in Italy

Results show that not all the CAP measures translate into an equivalent change in farm income. Given the intense pressures to reduce the financial resources allocated to the $CAP$, it is crucial to consider the overall efficiency of the policies, including their $ITE$. The analysis has provided evidence that $RDP$ measures other than $RDP_{aes}$ and $RDP_{inv}$ are the most efficient in terms of income transfer. However, these measures are dominated by $LFA$ payments, whose participation costs are negligible. $RDP_{inv}$ and $RDP_{aes}$ resulted in lower levels of $ITE$ compared to $DDP$: this is consistent with the fact that these measures pursue objectives other than income support, besides being characterized by higher participation costs. Finally, in line with theoretical predictions and part of the reviewed literature, the income transfer efficiency of $CDP$ is not significant.

The results have relevant policy implications for future CAP Reforms. Even maintaining the same overall budget, a shift of financial resources from more to less efficient income-transfer measures will reduce the overall effect of $CAP$ in terms of income- enhancement. The current debate on the Multiannual Financial Framework proposal seems to anticipate that the financial budget for the second pillar will decrease more that for the first pillar in relative terms (Matthews, 2018). This is expected to have negative consequences on the achievements of agri-environmental objectives and on-farm investments. However, such a proposal on the enhancement of the farmers' income will largely depend upon how the available budget will be allocated between pillars and, hence, among measures. The results of the analysis have shown that some measures (i.e., $DDP$ and $RDP_{other}$) are more income-transfer efficient than others so that reductions of the share of the budget allocated to these measures is expected to result in a decline of the overall income transfer efficiency of $CAP$ support.

Furthermore, most of the considered policy measures have displayed an $ITE$ lower than unity supports the idea that there is room for improvement, specifically for measures such as $CDP$ and $DDP$, which are intended to support farm income. Generally speaking, the results point to a significant economic pressure exerted by the cost of participating in policy measures, hampering a more efficient transfer of policy support to farmers' income.

For $RDP$-related payments, the income transfer efficiency increases with farm size. $RDP_{aes}$ and $RDP_{inv}$ are examples of measures resulting in negligible $ITE$ for small farms. These results suggest that these measures' participation costs can enormously differ between farms of different size. In particular, participation costs can be non-linearly related to farm size, as a part is fixed and independent from the amount of support. Nevertheless, according to the current $CAP$ configuration, farms located in the same region and participating in the same measure obtain the same unitary support (e.g. euro/ha), regardless of their structural characteristics.

In contrast, $DDP$ has shown a positive marginal impact on the income of small farms roughly equal to that of larger ones. This relates to the diverse nature of specific compliance costs these groups of farms face and the simplified procedure to apply for direct payments that both favour small farms.

Finally, adding to the previous literature, it has identified the long run cumulated effects of a change in support over subsequent years, highlighting differences across farm size. It finds more significant LR effects for large farms. Furthermore, the fact that the $ITE$ of the $RDP$ measures is vast for large than for small farms, and that $DDP$ $ITE$ is more significant for large than medium farms, suggests that there may be economies of scale in terms of policy-participation costs. Both considerations suggest that farm size should be taken into consideration when designing policy measures. Finally, reducing participation costs could be an effective solution for pursuing both efficiency and equity objectives, especially regarding $RDP$ measures for small farms, via a further simplification of the $CAP$. Another way to improve the welfare of small farms is to increase the persistency of their income because this increases the $LR$ income effects of the policy. This could be pursued by further extending the relevant risk management strategies and tools, including those supported by the $CAP$, aimed at stabilizing farm income.

### 4.1.1 Modelling Agricultural Risk Management Policies - The Implementation of the Income Stabilization Tool in Italy

According to the analysis results, the income stabilization tool recently introduced by the CAP would lead to a significant stabilization of farm incomes for Italian agriculture. This analysis provides the first empirical evidence on IST impacts and potential management costs.

It has shown that moving from a national to a regional/sectorial mutual funds (MF) increases the variability of the total indemnities paid by the MF significantly. More specifically, the relative variability of indemnification rates can be up to five times higher if moving to smaller aggregation levels of MF. Hence, such MF focusing on specific regions and sectors would face highly volatile levels of indemnification payments, requiring large buffers (Pigeon, Henry de Frahan, and Denuit 2014) and/or reinsurance. This would result in an increase in the MF costs and, consequently, charged to farmers. On the contrary, a nation-wide MF seems to face a limited variability of the total amount of indemnifications over time. This implies that public expense for this tool will also be less variable (more stable) over time. It was found that the specification of farmers' contribution, i.e. premiums to the IST, affects its income stabilizing effect. For flat-rate contributions, regional/sector-specific MF seems to be more effective than a national MF in terms of income risk reduction. However, this is not the case when contributions are paid according to their expected income. The level of the contributions that farmers pay affects the income stabilizing effect of the IST. Lowering the contribution rate reduces the income stabilising effect of the IST. However, the results have shown that the way farmers pay is essential in this regard: a flat rate approach has seemed less significant than a contribution proportional to the average farm income level in terms of income stabilization.

According to which contributions are designed, the way has relevant implications on how the financial benefits of the policy are distributed among the farm population. For instance, it was found that a flat rate contribution generates a very uneven distribution of these benefits across farms. However, this phenomenon is strongly reduced when a proportional contribution is used.

These results support some policy implications. First, developing MF with a high level of aggregation (e.g. national level) may be desirable to avoid highly volatile total paid indemnifications. Second, this will not reduce the effectiveness of the IST as an income stabilising tool. Of course, this kind of design of the IST might cause enormous transaction costs. Hence, this trade-off should be carefully considered when designing the MF. Third, it seems very appropriate to modulate farmers' contributions to the MF according to farm size, avoiding more straightforward approaches such as flat-rate contributions. This modulation decreases the inequality of the distribution of financial benefits of the policy within the farm population and improves the risk-reducing effects of the IST.

Finally, the analysis has not considered some other aspects of significant policy relevance that should be explored in future analyses on the IST. These include, among others, the potential implications of the application of the IST on the income inequality within the farm population (Finger and El Benni 2014a) and the factors affecting the distribution of the benefits derived by this tool (El Benni, Finger, and Meuwissen 2016).

### *4.1.2 Applications of Machine Learning for the Ratemaking of Agricultural Insurances*

Tweedie distribution and machine learning approach to efficiently predict insurance schemes' indemnities (see Chapter 3.2).

While it uses the perspective application of the IST for Italy as a case study, the proposed methodology could be potentially applied to other insurance products providing valuable information for ratemaking. Indeed, the obtained results allow us to draw some general considerations.

Models relying on Tweedie distributions can adequately account for the very peculiar distribution of indemnities, which are zero-inflated, right-skewed, and thick-tailed. The analysis has also shown that this class of distributions can be satisfactorily used with different models, including GLM and the three considered ML techniques.

Models based on ML have allowed reaching relatively high goodness-of-fit compared with the classical GLM even if using way fewer regressors. Indeed, these approaches satisfactory account for the overfitting problem that affects the GLM model. This is done by selecting only a limited set of regressors. We use three ML tools that differ in this regard: while LASSO and EN have shown very parsimonious, Boosting has selected around 40% of the potentially available regressors. The capacity to reach a satisfactory economic performance with a limited number of regressors could be advantageous for ML approaches. Focusing on a sub-set of the whole set of potential explanatory variables allows reducing the costs related to gathering and processing information.

The analysis has also tested the economic implications of using the estimates of the considered models in ratemaking. The first implication is that the study of a new insurance scheme (i.e., to be introduced) should be done accounting for the compatibility of premium and the insured's financial resources. Using reasonably tight compatibility rules strongly reduces the share of individuals potentially willing to participate and, therefore, the size of the market. In our case, assuming that the premiums must not exceed 10% of the farm value-added, the number of subjects potentially interested in joining the new regime has been reduced on average to less than 50%. However, some of the models considered Boosting have excellent performance, allowing for a wider audience of potential adherents than the other two ML-based models.

The analysis has shown that using the estimates of ML models allows the insurer to have a sound multiannual balance between collected premiums and paid indemnities. This is not the case when the estimates of the GLM are used. Similarly, the ratemaking based on the forecast of the ML models allows having only a few years in which the balance is very negative, requiring to have precautionary funds or other tools to cope with solvency problems.

Finally, while the three ML models perform well in terms of multiannual balance, they differ in how targeted the ratemaking is. In the considered case study, the Boosting seems to perform better than the other models. It allows having only a limited number of cases in which the extent of very negative the net premiums (i.e., premium - indemnity) is significant (lower than - 100 thousand Euro) and only one case when this is very large (lower than - 250 thousand Euro). In other words, the use of the estimates obtained by Boosting models reduced the cases in which premium is set at a too low level compared to the risk profile of the farms.

The analysis has some limitations that should be taken into consideration. First, it has an ex-ante nature because the IST has not been applied so far, limiting the possibility of studying farmers' behaviour. In particular, the recursivity of the problem at stake when the IST is introduced is not considered. Second, the analysis heavily relies on the data from the previous period. However, the proposed approach can be easily extended to account for the history of the considered individual farms. Finally, it is essential to recall that the lack of enough and reliable data that mutual funds face can severely constrain their capacity to use the proposed methodology. These considerations suggest that there are several important areas for further research. When the IST is applied, it will be possible to use factual data and develop ex-post analyses that could provide insights on farmers' behaviour accounting and the interaction effects with other existing risk management tools and strategies.

Despite these limitations, the findings are highly relevant to policy and industry. The here proposed approach allows determining farm-specific expected indemnifications in insurance schemes. This can contribute to effective ratemaking based on a clear, transparent, and methodologically sound approach. Furthermore, having an efficient strategy for variable selection can limit the effort and cost associated with gathering and processing information. Thus, the analysis is expected to broaden the development of efficient agricultural insurance options and foster more resilient farming systems (Meuwissen et al. 2019).

More in general, the analysis has provided one of the few empirical applications of machine learning in agricultural economics. The obtained results seem to confirm that it will see more of these techniques tailored and applied to economic, as stated by Storm, Baylis and Heckelei (2019). In particular, some of the considered models seem able to efficiently select from many farm and farmers' characteristics on which the explanatory variables of interest may depend.

# Appendix

## Appendix to Chapter 2

### A. Econometric specification

The Farm Net Income ($FNI$) of the $i_{th}$ farm in the $t_{th}$ year can be represented in implicit form as[53]:

$$FNI_{i,t} = f(G_{i,t}, X_{i,t}) + G_{i,t} + X_{i,t} + \tau_t + (\eta_i + m_{i,t}) \tag{A.1}$$

Where $G = \{CDP, DDP, RDP_{aes}, RDP_{inv}, RDP_{other}\}$; $X = \{LV, Non\ LV, FAML, Leverage, Price-ratio, CER\ ratio, F\&V\ ratio\}$; $\tau_t$ is the year-specific intercept; $\eta_i$ is the individual farm-specific effect and $m_{it}$ is the idiosyncratic error with $m_{it} \sim MA(0)$.

Considering that $FNI_{i,t} = E(FNI_{i,t}) + m_{i,t}$ one can write[54]

$$FNI_{i,t} - E(FNI_{i,t}) + m_{i,t} = 0 \tag{A.2}$$

$E(FNI_{i,t})$ can be written as:

$$E(FNI_{i,t}) = E\left[f(G_{i,t}, X_{i,t}) + G_{i,t} + X_{i,t} + \tau_t + \eta_i\right] \tag{A.3}$$

$$FNI_{i,t} - E(FNI_{i,t}) = FNI_{i,t} - E\left[f(G_{i,t}, X_{i,t}) + G_{i,t} + X_{i,t}\right] + \tau_t + \eta_i + m_{i,t} \tag{A.4}$$

The farmer uses past information to optimise the combination of factors. In this case, we assume the presence of an autoregressive error $v_{i,t}$. The formula (A.4) becomes

$$FNI_{i,t} - E(FNI_{i,t}) = FNI_{i,t} - E\left[f(G_{i,t}, X_{i,t}) + G_{i,t} + X_{i,t}\right] + \tau_t + \eta_i + m_{i,t} + v_{i,t} \tag{A.5}$$

As in Blundell and Bond (2000), it is possible to specify autoregressive as $v_{i,t} = \rho\, v_{i,t-1} + \varepsilon_{i,t}$ with $\varepsilon_{i,t} \sim MA(0)$ and $\rho$ is the autoregressive coefficient.

This term can be obtained from the difference between the equation at time $t$ less the same equation at time $t-1$[55], following a Cochrane-Orcutt Transformation (Cochrane and Orcutt 1949)[56], as:

$$v_{i,t} = \rho\, v_{i,t-1} + \varepsilon_{i,t} = \{FNI_{i,t} - E[f(G_{i,t}, X_{i,t}) + G_{i,t} + X_{i,t}] + \tau_t + \eta_i + m_{i,t}\} =$$
$$\rho\{[FNI_{i,t-1} - f(G_{i,t-1}, X_{i,t-1}) + G_{i,t-1} + X_{i,t-1}] + \tau_{t-1} + \eta_i + m_{i,t-1}\} \tag{A.6}$$

Accordingly, farm income is specified as:

---

[53] The prices of input and output are included in $X$.

[54] The error is assumed $m_{i,t} \sim N(0, \sigma^2)$, the transition of $m_{i,t}$ from one part of equality to another is therefore irrelevant to the sign.

[55] At time t-1 the E(.) is omitted because the farmer has the value without error. The individual effect $\eta_i$ remains the same in the time by definition (time invariant).

[56] When lagged variables are included, the model may suffer from a large bias towards randomness (Cochrane and Orcutt 1949). If $\rho\, v_{i,t-1}$ tends to or equals zero, the estimator is much more unbiased. Such assumption is later verified through the No-Linear Common Factor.

$$FNI_{i,t} = \{E[f(G_{i,t}, X_{i,t}) + G_{i,t} + X_{i,t}] - \rho[FNI_{i,t-1} - f(G_{i,t-1}, X_{i,t-1}) + G_{i,t-1} + X_{i,t-1}]\} - \{(\tau_t - \rho\tau_{t-1}) + [\eta_i(1-\rho)] + m_{i,t} - \rho m_{i,t-1} + \rho v_{i,t-1} + \varepsilon_{i,t}\} \quad (A.7)$$

$$FNI_{i,t} = g(G_{i,t}, X_{i,t}) - \rho[FNI_{i,t-1} - g(G_{i,t-1}, X_{i,t-1})] + \{(\tau_t - \rho\tau_{t-1}) + [\eta_i(1-\rho)] + m_{i,t} - \rho m_{i,t-1} + \rho v_{i,t-1} + \varepsilon_{i,t}\} \quad (A.8)$$

$FNI_{t-1}$ leads to decisions at time $t$, and it may entail certain degrees of stickiness between different periods (see Hirsch and Gschwandtner, (2013) for more details over the topic of economic outcomes' persistency). Lagged variables for the two different farm capital assets account for the long-term nature of the optimal capital structure decisions (Tamirat, Trujillo-Barrera, and Pennings 2018). Thus, the dynamic nature of the model provides more accurate results, deepening the understanding of how explanatory variables (including CAP subsidies) affect farm income over time. In case of no autoregressive error, $\rho v_{i,t-1} = 0$ we have:

Considering $\varepsilon'_{i,t} = \rho v_{i,t-1} + \varepsilon_{i,t}$, this equation identifies the autoregressive error (note that when $v_{i,t-1} = 0$, $\varepsilon'_{i,t} = \varepsilon_{i,t}$, as $\rho v_{i,t-1}$ is minimised by the GMM estimator see Appendix B).

The income function used in the analysis is the following:

$$FNI_{i,t} = g(G_{i,t}, X_{i,t}) - \rho[FNI_{i,t-1} - g(G_{i,t-1}, X_{i,t-1})] + \{(\tau_t - \rho\tau_{t-1}) + [\eta_i(1-\rho)] + m_{i,t} - \rho m_{i,t-1} + \varepsilon'_{i,t}\} \quad (A.9)$$

The dynamic nature of the farm income (i.e., decisions are taken at $t-1$ based on the result in $t-2$) justifies the use of $FNI_{t-2}$, which is also motivated by the fact that such $AR(2)$ specification results as the best fitting lag order (for a similar specification, see Baldoni et al., 2017). Hence, the definitive specification of the empirical model estimated in the analysis takes the following explicit form:

$$FNI_t = \rho FNI_{i,t-1} + \rho^2 FNI_{i,t-2} + \sum_{k=1}^{5} \gamma_k G_{k,i,t} - \sum_{k=1}^{5} \rho\gamma_k G_{k,i,t-1} + \sum_{n=1}^{7} \gamma_n X_{n,i,t} - \sum_{n=1}^{7} \rho\gamma_n X_{n,i,t-1} + \{(\tau_t - \rho\tau_{t-1}) + [\eta_i(1-\rho)] + m_{i,t} - \rho m_{i,t-1} + \varepsilon'_{i,t}\} \quad (A.10)$$

or

$$FNI_t = \alpha_1 FNI_{i,t-1} + \alpha_2 FNI_{i,t-2} + \sum_{k=1}^{5} \beta_k G_{k,i,t} + \sum_{k=1}^{5} \gamma_k G_{k,i,t-1} + \sum_{j=1}^{7} \delta_j X_{j,i,t} + \sum_{j=1}^{7} \zeta_j X_{j,i,t-1} + \tau_t^* + \eta_i^* + \varepsilon''_{i,t} \quad (A.11)$$

Where $\tau_t^* = \tau_t - \rho\tau_{t-1}$; $\eta_i^* = \eta_i(1-\rho)$; $\varepsilon''_{i,t} = \varepsilon'_{i,t} + m_{i,t} - \rho m_{i,t-1}$

(A.10) is defined as the "restricted model" where the regressors of $FNI_{i,t-2}, G_{k,i,t-1}, X_{n,i,t-1}$ (i.e $\rho^2, \rho\gamma_k, \rho\gamma_n$) are not directly estimated by the regression. Motivated by Cochrane and Orcutt (1949) transformation, and relying on fewer variables: $\rho, \gamma_k, \gamma_n, \tau_{t-1}, \eta_i, m_{i,t-1}$ (as $\rho^2, \rho\gamma_k, \rho\gamma_n$ are imposed by restrictions).

(A.12) [57] is defined as the "unrestricted model" as all the coefficients are directly estimated via the $SYS-GMM$ model, which does require the researcher to specify the relationship existing among regressors and dependent variable.

Whether non-linear common factor restrictions (i.e. $\alpha_2 = \alpha_1^2$; $\gamma_k = \alpha_1\beta_k$; $\gamma_k = \alpha_1\delta_j$) are fulfilled in the estimated (A.11). Non-linear common factors are tested through the Wald Test, indicating if significant differences exist with (A.10) (Blundell, Bond, and Meghir 1996) (see Section B for further details).

B. Verifying non-linear common factor restrictions

The GMM, by construction, minimises the distance between the two models through the two-step estimator (see Hansen, 1982). If non-linear common factors restrictions are verified, the model has no autoregressive errors.

Following Blundell et al. (1992), we have tested this difference through the Wald Test. The results (Table B.1) show that for two (Total sample and Small farms) out of four models, the test does not reject the null hypothesis than the Restricted and Unrestricted model are statistically identical. In other words, for both Medium and Large farm models, results need more cautious interpretation due to more unstable estimates, as the autocorrelation error is not completely eliminated but just minimised.

*B. Wald Test results for the non-linear Common Factor restrictions. Probability values.*

**Table B.1 - Wald Test results for the non-linear Common Factor restrictions. Probability values.**

| Model | p-value |
|---|---|
| Total | 0.000 |
| Small | 0.023 |
| Medium | 0.558 |
| Large | 0.000 |

*C. Specification of the econometric model*

Four groups of specification tests served as the basis for a robust and comprehensive application of the $SYS-GMM$ and its structure (Table C.1). According to the econometric specification of the model in (9), the F-test (Wooldridge 2010) indicates the model suffers from both individual and time effects, recommending the use of the two-way fixed effects. The purpose of the two-ways fixed effects, besides, to taking into account the specific common effects of time, prevents individuals' heterogeneity from being correlated at the error term

---

[57] (A.11), that is the equation (1) reported in the main text

(Wooldridge 2010).

Moreover, Hausman (1978) test has been used to specify whether a fixed or random effect (FE and RE, respectively) model should be estimated (Wooldridge 2010). Results indicate that $FE$ model should be preferred against the $RE$. The last row of Table presents the result from the Panel covariate-Augmented Dickey-Fuller ($Panel\ cADF$) (Hansen, 1995) test for unit-root in panel dataset. This points to the stability of the model by rejecting the null of the presence of unit-root. Whether the panel contains a unit-root is important since when the dependent variable is non-stationary, introducing own-lags in the model would result in a spurious regression (Mary 2013b).

**Table C.1 – Specification tests of the four models (Total sample, Small, Medium and Large farms)**

|  | Total Sample | Small Farm | Medium Farm | Large Farm |
|---|---|---|---|---|
| Poolability (F-Test)[a] | | | | |
| - *Individual Effect* | 2.081*** | 2.293*** | 2.038*** | 1.986*** |
| -*Two-ways Effect* | 2.102*** | 2.293*** | 2.043*** | 2.020*** |
| FE vs. RE[b] | 12197*** | 2641*** | 2883*** | 2635*** |
| Panel cADF test[c] | -116.925 (0.0001***) | -72,308 (0.0001***) | -86,702 (0.0001***) | -53,090 (0.0001***) |
| Significance codes for p-values: *** ≤ 0.01; ** ≤ 0.05; * ≤ 0.10. | | | | |
| [a] F Test (H$_0$: non-significant individual effect) | Individual Effect | Individual Effect | Individual Effect | Individual Effect |
| [a] F Test (H$_0$: non-significant two ways effect) | Two-ways Effect | Two-ways Effect | Two-ways Effect | Two-ways Effect |
| [b] Hausman Test | FE | FE | FE | FE |
| [c] Hansen covariate ADF (H$_0$: Unit Root) | The panel does not contain a Unit Root | | | |

*Source: Authors' elaboration on Italian FADN data.*

All estimated models account for Two-Ways $FE$ and were evaluated through the two-step approach, which is asymptotically more efficient than the one-step estimation procedure, albeit downward biased (Blundell and Bond 1998). Furthermore, the Windmeijer (2005) correction to the two-step covariance matrix has been applied to obtain robust standard error estimation when using finite sample and to account for heteroscedasticity.

*D. Short and Long-Run Elasticity of CAP measures on farm income (FNI). Total sample, Small, Medium and Large farm models.*

The elasticities are estimated as in Olper et al., (2014), at the sample mean, and are specified as follows:

$$Short-run\ elasticity: \frac{\partial lnG}{\partial lnFNI} = (\beta + \gamma)\frac{\bar{G}}{\overline{FNI}} \tag{D.1}$$

$$Long-run\ elasticity: \frac{\partial lnG}{\partial lnFNI} = \left(\frac{\beta+\gamma}{1-\alpha_1-\alpha_2}\right)\frac{\bar{G}}{\overline{FNI}} \tag{D.2}$$

Where, using the notation of formula (1), $(\beta + \gamma)$ is short-run effect and $\left(\frac{\beta+\gamma}{1-\alpha_1-\alpha_2}\right)$ is long-run effect as specified in Pesaran (2015), $\bar{G}$ is the sample mean value of a specific subsidy, and $\overline{FNI}$ the sample mean level of $FNI$.

**Table D.1 – Short- and Long-run elasticity estimates.**

|  | Short-Run | | | |
| --- | --- | --- | --- | --- |
|  | *Total* | *Small* | *Medium* | *Large* |
| $CDP$ | 0.007 | -0.001 | 0.025 | 0.006 |
| $DDP$ | 0.173 | 0.228 | 0.139 | 0.144 |
| $RDP_{aes}$ | 0.007 | 0.004 | 0.002 | 0.007 |
| $RDP_{inv}$ | 0.006 | 0.000 | 0.006 | 0.006 |
| $RDP_{other}$ | 0.029 | 0.036 | 0.051 | 0.019 |
|  | Long-Run | | | |
|  | *Total* | *Small* | *Medium* | *Large* |
| $CDP$ | 0.009 | -0.001 | 0.027 | 0.008 |
| $DDP$ | 0.221 | 0.250 | 0.149 | 0.184 |
| $RDP_{aes}$ | 0.009 | 0.004 | 0.002 | 0.009 |
| $RDP_{inv}$ | 0.007 | 0.000 | 0.006 | 0.008 |
| $RDP_{other}$ | 0.038 | 0.040 | 0.054 | 0.024 |

*Source: Authors' elaboration on Italian FADN data*

## Appendix to chapter 3

This appendix, after having provided minimal background information on Machine Learning (ML). Furthermore, it discusses the fact that ML does not provide CI for the estimated coefficients. Finally, it describes how the dummy variables are introduced in the models.

*Introduction*

Econometric analysis is based on different kinds of tools. A more simply are "descriptive statistics" used to analyse only observed economic data, not evaluating all the population and, making some assumptions, the "inference" to allows us to embrace our results in all the population.

In particular, the inference is the main instrument in econometrics. Indeed econometricians need focusing on properties of data and deductions regarding the distribution of probability. Finally, this approach allows to testing hypothesis and deriving estimates, that permits to verify the theory put at the base of an economic model and verify the outcomes.

Data Generation Processes (DGPs) are the basis of inference. DGP can be defined as a process in which a set of input variables $x$, called independent variables, is associated with a function generating an output y (or dependent variable).

Econometricians usually use a stochastic DGP or $y = f(x, \beta; \varepsilon)$ where output $y$ deriving from a function $f(\cdot)$ of input o predictor variables $x$, parameter $\beta$, and random noise $\varepsilon$.

The analysis in economics is based on inference using a stochastic DGP referring to a theoretical framework, defined in a theoretical framework, with the primary goal of making inference rather than obtaining a large predictive capacity of the model.

At the end of the 20$^{th}$ century, the increase in processing capacity thanks to the advent of ever more powerful computers created a new branch of statistics: Machine Learning (ML).

The first definition of ML is attributable to Samuel (1959), which he explained it as "*a field of study that gives a computer the ability without being explicitly programmed*".Iskhakov, Rust and Schjerning (2020 page 2) define ML with an extensive formulation as: "*the scientific study of the algorithms and statistical models that computer systems use to perform a specific task without using explicit instructions and improve automatically through experience.*"

This new statistical toolbox has also generated a deep rift between classical statisticians, linked to inference and those who instead saw in ML new possibilities for developing statistics.

Breiman (2001) with the famous essay "Statistical Modeling: The Two Cultures" highlighted this division by precisely defining the terms of the clash.

ML tools have different objectives than those used in statistical inference. First of all, the objective of ML is the ability to predict with respect to inference (many tools are biased by definition). Secondly, ML has more ability to make mapping or reduction of dimensionality of problems. Thirdly, we do not need to establish a DGP preliminarily: ML can deduce the DGP from data (data-driven). With the possibility of discovering new DGPs that may not have been previously defined in the literature or the theoretical context ( i.e. *a priori assumed*). Last but not least, ML is very efficient in the use of computational resources.

The critiques against the use of ML regards, first of all, the lack of mathematical models used at the base of inference: this concept have been defined by Cox (2001) *"Abandoning mathematical models comes close to abandoning the historic scientific goal of understanding nature"* and also supported by (Efron 2020; Parzen 2001). Secondly, the lack of confidence intervals and the possibility to use the marginal effects, that do not allow to extrapolate information on the relationship between the dependent and independent variable on the coefficient (Lee et al. 2016; Leeb and Pötscher 2006b, 2006a; Taylor and Tibshirani 2018). Thirdly, ML is usually not BLUE estimator but only Best Linear or Un-linear Estimator (BLE o BUE) because it is generally biased (van de Geer 2016). Finally also Breiman (2001) itself introduced some issues: i) the possibility to discover multiple DGPs (defined as Rashomon's effect); ii) the necessity to reduce the complexity of the model or to consider the trade-off between accuracy and interpretability of model (called Occam's effect); iii) In an economic model, to improve accuracy, it is necessary to increase the number of variables. This request leads to an increase in the complexity of the model and also in difficulty in estimating. A model with hutter amount of variables is often unresolvable. This inability to find a precise solution to complex problems using many variables is called by Breiman (2001), borrowing it from Bellman (1957), "curse of dimensionality" or Bellman's effect.[58]

Currently, the ML approach is not widely used in economics (Athey and Imbens 2019) mainly because economists base their analyses more on inference than on the predictive capacity of the model. So why can the economist be used ML?

Scholars are faced with a central question: should one use a very selective ML tool or one that uses more regressors but more precise? Currently, there is no single answer; a compromise of complexity and interpretability can be a good one (Efron 2020).

The fracture reported in Breiman (2001) is longer to be healed. As reported in recent works, this mending can only take place when statisticians and econometricians become aware that these tools are complementary and not opposed. (Athey 2017; Athey and Imbens 2019; Charpentier, Flachaire, and Ly 2018; Efron 2020; Einav and Levin 2014; Iskhakov, Rust, and Schjerning 2020; Kleinberg et al. 2017; Mullainathan and Spiess 2017; Rust 2019; Varian 2014).

*Machine Learning: some crucial issues*

**The lack of confidence intervals in ML**

A prominent critique against the use of ML refers to the lack of confidence intervals. This problem is common to all the algorithms used to select the variables (Leeb and Pötscher 2006b, 2006a).

According to Mullainathan and Spiess (2017) and Athey and Imbens (2019), the presence of confidence intervals is functional to a specific scope. In particular, the trade-off between accuracy (that be almost sure better in Machine Learning by construction) and analysis of confidence interval (that is a point of strength in econometric models) have should be articulated clearly in the goals of their investigation. At the same time, the researcher will have to specify why some specific properties of algorithms and estimators may or may not be relevant.

---

[58] To decrease the "Curse of dimensionality" it is necessary to reduce the size of the model while maintaining the characteristics of the phenomenon (in statistics it is called "mapping"). This problem can be solved with mathematical tools such as in Dynamic Programming but also through ML which allows you to select only the variables necessary to estimate the DGP..

For example, in a model like IST, is more critical the level of collinearity, that is very low in ML or the confidence interval, deriving from Maximum Likelihood Estimation (MLE)? And in this last case, the confidence interval is functional to the scope of paper or to answer a specific research question or not?

This trade-off will interest the researchers in the next future almost surely the econometricians cannot discovery a substitute of confidence intervals for ML.

According to Leeb and Pöscher (2006) and Leeb and Pötscher (2006, 2008) make inference after selection is difficult. In particular, the use of bootstrap can obtain peculiar results (e.g., some regressors selected by LASSO can have negative lower bounds and positive upper bounds, this is caused by the multiples DGPs effects). To try to overcome this problem (Lee et al. 2016; R. J. Tibshirani et al. 2016) proposed to use a conditional approach leading to a truncated normal reference distribution. Notwithstanding this implementation, the definitive solution to this problem has not yet been achieved.

Mullainathan and Spiess (2017 Figure 2 page 97) used the comparison between multiple DGPs to obtain a comparison similar to that found in econometric inference for confidence intervals.

We used this procedure in our study. We considered the number of times a variable is selected in all years of simulation and all tests. We can say that, if a variable is chosen in all tests and in all years, its significance is almost certain. These results give an idea of the strength of the variables in establishing the DGP.

*A brief explanation of ML tools used*

**LASSO and Elastic Net**

LASSO and Elastic Net[59] are ML methods based on "regularisation" or "shrinkage" methodology (Hastie, Tibshirani, and Friedman 2009). These allow us to obtain threefold results: select variables, enhance the accuracy of prediction and reducing collinearity.

In the beginning, the shrinkage methodology has been used for the Ridge Regression or Tikhonov regularisation: while the Least Squared seeks to minimise the sum of squared residual $\|Ax - b\|_2^2$ the Ridge Regression adds a Regularisation term $\|\Gamma x\|_2^2$. The objective of the Ridge Regression (Loss Function) is minimising the Mean Square Error (MSE) of $\|Ax - b\|_2^2 +$

---

[59] Elastic Net regression is part of the Shrinkage Regression Family (R. Tibshirani 1996). This, unlike the classical statistical prediction, aims to find a function that gives a "good" prediction of $y$ as a function of $x$ where "good" means it minimizes or maximize some objective of the inference as RSS, Deviance, AUC, AIC, BIC (Varian 2014).

$\|\Gamma x\|_2^2$ (with $\|\Gamma x\|_2^2$ is Euclidean norm defined as $l_2$)).

The LASSO regression uses the following modification of the Tikonov term: $\|Ax - b\|_2^2 + |\Gamma x|$ (with $|\Gamma x|$ also defined as $l_1$).

Elastic Net uses an intermediated value of the regularisation term that lays in between Euclidean norm and Absolute norm (between $l_1$ and $l_2$).

We can rewrite Loss Function as minimisation of MSE of : i) The RIDGE = $L_{\text{Ridge}} = RSS + \lambda \sum_{j=1}^{p} \beta_j^2$; ii) The LASSO= $L_{\text{Lasso}} = RSS + \lambda \sum_{j=1}^{p} |\beta_j|$; iii) The EN is given with a combination of Ridge and LASSO thanks a coefficient $\alpha$ : $L_{\text{Elastic Net}} = \frac{RSS}{2n} + \lambda \left( \frac{1-\alpha}{2} \sum_{j=1}^{p} \beta_j^2 + \frac{\alpha}{2} \lambda \sum_{j=1}^{p} |\beta_j| \right)$. $\lambda$ denotes a tuning parameter indicating the strength of the penalty term, and it is the hyper-parameter of Shrinkage Tool. $\lambda$ is set via Cross-Validation[60].

**Boosting**

Boosting is a machine learning tool with the primary objective to reduce distortion and variance by using a "weighting" methodology that transforms weak predictors into strong predictors.

The central hypothesis of "boosting" is the division into strong and weak regressors. While the former provides a high contribution to the explanation of the model, the latter is not very important for prediction. Boosting methodology "weighs" the regressors differently based on their explanatory power.

The algorithm performs many simulations to obtain the correct weight. Each of these simulations is used as a basis for the next phase (learning method). For example, the first step uses a constant weight for each regressor, but only a few of them have high predictive power.

Necessarily, it happens that two or more weak regressors together have the same explicative power than a single strong regressor. The ability to convert a mediocre regression into one that works extremely well through a learning algorithm is one of the strengths of Boosting[61].

There are several implementations of the Boosting approach. We use the Gradient Boosting algorithm (Friedman 2001).

The Loss Function in Boosting is defined as $\hat{t}(x) = argmin \sum_{i=1}^{N} L(\hat{y}_i, y_i)$.

With $\hat{y}_i$ = predicted value, $y_i$ the observed value and i is the observations and $L$ a function that we can also use to obtain MSE, RMSE or Huber (Adaptative) Loss, Entropy or Exponential Loss. We report the meta-algorithm used in Gradient Boosting.

---

[60] We can use the EN Loss Function, in general, to make also RIDGE with $\alpha = 0$ and LASSO with $\alpha = 1$ and $\alpha = (0,1)$ Elastic Net.

[61] For more details see Bühlmann & Hothorn (2007), Friedman (2001), Hastie et al., (2013) and Yang, Qian, & Zou (2017).

1. In the beginning, one starts with a causal tree and do the first stage where the results $F_0 = L_0(\hat{y}_i, y_i)$.
2. In the second stage, one takes $y_i - F_0(X_i) = h_0(x)$ with $F_0(X_i) = \hat{y}_i$ or the prediction function using the regressors matrix $(X_i)$ with $X_i = \{x_{1,i}, \dots, x_{m,i}\}$. It uses MSE[62] the first stage $L_{MSE} = \frac{1}{2}(y_i - F_0(X_i))^2_i$ and use the negative gradient $r_{i,m}(x) = -\left[\frac{\partial L_{MSE}}{\partial F(X_i)}\right]_{F(X_i) = F_{m-1}(X)}$ to find the so-called pseudo-residuals. With $M = \{1, \dots, m\}$ is the number of iteration
3. One fits the leaner $h_m(x)$ using the training-set $h_m(x) = \{(x_i, r_{i,m})\}_{i=1}^n$
4. One computes multiplier $\gamma_m$ to solving the problem

$$\gamma_m = \underset{\gamma}{argmin} \sum_{i=1}^n L(y_i, F_{m-1}(x_i) + \gamma\, h_m(x_i))$$

5. One updated the model $F_m(x) = F_{m-1}(x) + \gamma_m h_m(x)$
6. And a the end one obtains the Output $F_M(x)$

**Some considerations**

The comparison between Boosting and shrinkage methods, especially in economics, deserves reflection.

Boosting allows for very high accuracy without the need for a mathematical model definition. Conversely, with EN and Lasso, one must first define the functional form of the regressors.

The choice of one or the other approach depends very much on the assumptions made in the theoretical model and the research questions. If you intend to have high accuracy without first setting the functional shape of the regressors, Boosting certainly has the best performance. On the other hand, if it is necessary to verify a more structural model, based therefore on economic theory, assumptions and constraints, surely LASSO and EN allow better evaluations.

There is no perfect ML tool valid for all economic analyses, but it must be chosen from time to time based on research needs.

*Treatment of dummies with multiple categories: the grouped approach*

The grouped regression has been used to handle the categorical regressors assuming multiple discrete values (i.e., levels).

The "shrinkage" methods (but this also applies to other methods used to select regressors) can choose only specific categories of categorical variables[63].

---

[62] In this example, we use MSE, but the other loss functions are used the same way.

[63] To explain this issue, let's take the case of the macro-regions in Italy which have five categories, North-West (NOR), North-East (NOC), Center (CEN), South (MER), Islands (INS). To avoid the dummy variable trap, we use one of the categories as contrast (for example NOR), and we use the others with values 0 or 1. Without selection, the regression allows us to compare all other categories with NOR. In the case of variable selection, we may have that NOC is not selected. This effect implies that the contrast is no longer NOR but NOR + NOC, which does not allow

We must then "group" these categories and create a constraint to select or not the "group" and not a single class.

This method is called "Grouped Regression" (Bühlmann and van de Geer 2011; Qian, Yang, and Zou 2016).and has been applied for all categorical variables with classes higher than two.

---

a clear interpretation of the model. To avoid this issue, we can impose a selection of all or none of the MR categories to the algorithm.